\documentclass{article}
\usepackage[dvips]{graphicx}
\usepackage{epsfig}
\usepackage[mathscr]{euscript}
\usepackage{amsfonts}
\usepackage{amssymb}

\newcommand{\Tr}{\;\mbox{Tr}\;}

\begin{document}
\title{Matrix geometries and Matrix Models}
\author{Rodrigo Delgadillo-Blando$^a$\footnote{rodrigo@stp.dias.ie}, 
Denjoe O'Connor$^{b}$\footnote{denjoe@stp.dias.ie}\\
$^a$Department of Mathematical Physics, NUIM Maynooth, Ireland \\
$^b$School of Theoretical Physics, DIAS, Dublin, Ireland}
\date{\today}
\maketitle

\abstract{ We study a two parameter single trace $3$-matrix model with
  $SO(3)$ global symmetry. The model has two phases, a fuzzy sphere
  phase and a matrix phase. Configurations in the matrix phase are
  consistent with fluctuations around a background of commuting
  matrices whose eigenvalues are confined to the interior of a ball of
  radius $R=2.0$.  We study the co-existence curve of the model and
  find evidence that it has two distinct portions one with a
  discontinuous internal energy yet critical fluctuations of the
  specific heat but only on the low temperature side of the transition
  and the other portion has a continuous internal energy with a
  discontinuous specific heat of finite jump. We study in detail the
  eigenvalue distributions of different observables. }

\section{Introduction}

One of the most striking and interesting features of multi-matrix
models is the phenomenon of emergent geometry. The notion of classical
geometry changes drastically within the context of matrix models; the
geometry is no longer a basic concept which exists a priori but
instead it emerges dynamically as a consequence of the reordering of
degrees of freedom. This is in many ways exciting because matrix
models can lead to new ways of thinking about the structure of the
space time.

The interest in matrix models has grown since it was suggested that
they might provide a non-perturbative definition for M theory
\cite{BFSS}.  Several kinds of matrix models have been proposed for
this purpose \cite{BFSS,IKKT,BMN}.  The IIB matrix model (IKKT model)
\cite{IKKT} is one of these proposals; it is a large $N$ reduced model
\cite{reduced} of ten-dimensional supersymmetric Yang-Mills theory and
the action is a matrix regularised form of the Green-Schwarz action of
the IIB superstring.  It is postulated that it gives a constructive
definition of type IIB superstring theory.

Finite dimensional matrix models have also been used to 
regulate field theories \cite{GKP,WW,fuzzyactions,st:phi4,presnajder:phi4,denjoelowd,st:randomgauge} and diverse fully non-perturbative 
numerical studies have been performed, see for instance \cite{landi,nishi:numeS2,xavier,bad:simu,panerophi4,prlROY,longpaper}.

The pure commutator action, which we refer to as the Yang-Mills 
matrix model, is also particularly interesting
since in $d=10, 6$ and $4$ dimensions it corresponds to the bosonic part of the
IKKT model. The model is well defined in dimensions $d>2$ and matrix
size $N>3$ \cite{integralsYM}.  In dimension $d=2$ the model, with a
quadratic term is added to stabilise it, is exactly solvable
\cite{hoppe,kazakov}. These bosonic Yang-Mills matrix models in
different dimensions have been considered as possible realizations of
emergent geometry/gravity \cite{steinemgravity}.

Numerical studies of pure Yang-Mills matrix models were performed in
\cite{reduced} for different dimensions (numbers of matrices) and more
recently in \cite{OConnorThMp,FilevOConnor} it is argued that
perturbation theory around a background of commuting matrices gives a
good approximation to the 3-dimensional model. In \cite{FilevOConnor}
based on a two loop computation it is predicted that the eigenvalues
of the background commuting matrices are uniformly distributed
within a ball of radius $R\sim 1.8$ which is in broad agreement with
our findings here.

Our starting point is an action in which the basic objects are simple
Hermitian matrices; no geometrical background is assumed a
priori.  The model describes the statistical
fluctuations of matrices with prescribed energy
functional. The geometry arises as a condensate around which the
system fluctuates.  Generically, multi-matrix models can undergo
transitions between different geometries and phases with no
geometrical content.  Given the novelty of these phenomena it is worth
studying the simplest model that exhibits such phenomena in detail.

The simplest model in which a geometry has been shown to be emergent
is a 3-matrix model consisting of the trace of the square of the
commutator of the matrices (a Yang-Mills term) plus the epsilon-tensor
contracted with the trace of the three matrices (a Myers
term)\cite{myers}. This model was studied in
\cite{nishi:numeS2,bad:simu,prlROY,longpaper} and is a static bosonic 
subsector of the BMN model \cite{BMN}.
 
The model exhibits a geometrical phase for sufficiently large
coupling to the cubic Myers term, with the geometry being that of a
fuzzy sphere \cite{madore,hoppe}; a non-commutative \cite{connes} 
version of the commutative sphere \cite{bal:year1}.

At a critical coupling, which can be traded for a critical
temperature, a phase transition occurs and the condensed geometry
evaporates. In the geometrical, low temperature phase, small
fluctuations around this condensate correspond to a $U(1)$ gauge
and scalar field multiplet \cite{st:randomgauge,PRY1,longpaper}.

The model can be extended by adding appropriate potentials in order to enhance
the range of parameters in which the fuzzy sphere phase is stable
\cite{longpaper}.  The general phase diagram for a class of such models was
predicted in \cite{longpaper}, one of the purposes of this paper is to
check these predictions in detain in a non-perturbative study. We find
that indeed the phase diagram is well predicted by theoretical 
expressions presented in \cite{longpaper}.

Manifolds in higher dimensions can emerge from more general
matrix models with essentially the same structure and
phenomenology as can be seen in \cite{PRY:phaseSXS,RY:uvSXS,bad:gaugeCP2}.

In the current paper we study, in detail, the high temperature phase of the 
model and establish that this phase is consistent with the three matrices 
fluctuating  around commuting matrices where the eigenvalues 
of the commuting matrices are confined to the interior of a 
solid three dimensional ball. The fluctuations are still significant 
as the commutator of pairs of matrices is itself a peaked (almost triangular)
distribution.

We concentrate on the effect of adding a quadratic, 
mass like perturbation to the 3-matrix model. 
We find that below a critical temperature any negative massive 
perturbation induces a transition from the matrix phase 
to the fuzzy sphere phase. For higher temperatures a more negative quadratic 
coupling is necessary. The phase diagram, which is one of the principal 
results of this paper, is shown in figure 3. 

\begin{itemize}
\item We find the ground states of the system are characterised by
  either the $N$ dimensional irreducible representation of $SU(2)$
  (the fuzzy sphere phase) or a continuum spectrum (matrix phase) for
  one of the matrices. These characterize the two phases of the
  system. Though meta stable states other than the irreducible fuzzy
  sphere are present in the system they never correspond to the true
  ground state, they do however appear stable in the large $N$ limit
  where tunneling is suppressed.
\item We study the phase diagram (see figure 3) as a function of the
  two parameters $\tau$ and $\tilde\alpha$ and locate the co-existence
  curve with some precision.
\item The co-existence curve rapidly asymptotes to the special line
  $\tau=\frac{2}{9}$ where the energy functional defining the model
  becomes a complete square.
\item We found evidence for two distinct types of transition in the
  emergent geometry scenario: for $0<\tau<\frac{2}{9}$, as the
  transition is approached from the fuzzy sphere phase with fixed
  $\tau$, the system has a divergent specific heat with critical
  exponent $\alpha=\frac{1}{2}$ while crossing the phase boundary for
  fixed $\tilde\alpha>4.02$ there appear to be no critical
  fluctuations and the transition is one with a continuous internal
  energy and discontinuous specific heat.
\item We find that a useful description of the matrix phase is in
  terms of fluctuations about a background of commuting matrices whose
  eigenvalues are concentrated in a three dimensional ball of radius
  $R=2.0$.
\end{itemize}

The paper is structured as follows: Section 2 describes the model and the
predictions of the phase diagram from the effective potential \cite{longpaper} 
in the fuzzy sphere phase. Section 3 describes the matrix phase and 
the consequences of a background of commuting matrices with eigenvalues 
uniformly distributed within a ball. Section 4 describes our numerical
results and section 5 contains our conclusions.

\section{The model}

We begin by considering an action functional built as a single trace
quartic polynomial of 3 Hermitian matrices with $SO(3)$ symmetry.  There are
four available invariants:
\begin{eqnarray}
\Tr D_aD_bD_aD_b,\;\;\;\; \Tr (D_a^2)^2, \;\;\;\; \Tr i\epsilon_{abc}D_aD_bD_c,\;\;\;\; \Tr D_a^2 .
\end{eqnarray}
In this paper we restrict our study to the two parameter model given by 
the action
\begin{equation}\label{mu-action}
S[D]=\frac{\tilde{\alpha}^4}{N}Tr\left[-\frac{1}{4}[D_a,D_b]^2+\frac{i}{3}{\epsilon}_{abc}[D_aD_b]D_c+\tau D_a^2\right]\ ;
\end{equation}
where stability of the model requires the $\tau>0$. For $\tau<0$ the action 
for commuting matrices is unbounded from below.
The most general model would include in addition $Tr(D_a^2)^2$ with an 
extra coupling.

The action (\ref{mu-action}) is invariant under unitary 
transformations $U(N)$, $D_a\to U D_a
U^{\dagger}$ and global $SO(3)$ rotations of the matrices. 
The modes $c_a=Tr(D_a)$ decouple from the others
and we therefore choose to work with traceless matrices,
$Tr\; D_a=0$.  The parameters of the model are $\tau$ and
$\tilde{\alpha}$. $\tilde{\alpha}$ can be identified either as the
Yang-Mills coupling constant $g^2 \equiv \tilde{\alpha}^{-4}$ or
the temperature $T\equiv \tilde{\alpha}^{-4}$ \cite{longpaper}.

Saddle points of the action, derived from the condition $\delta S=0$,
are given by solutions of
\[ [D_b,iF_{ab}]+2\tau D_a=0, \qquad \mbox{with} \quad 
F_{ab}=i[D_a,D_b]+\epsilon_{abc}D_c\] 
and include the trivial solution $D_a=0$ and $D_a=\phi J_a$ where $J_a$ are
representations of $SU(2)$, not-necessarily irreducible.  
For $D_a=\phi J_a$, then $F_{ab}=-(\phi^2-\phi)\epsilon_{abc}J_c$ 
we get an algebraic equation for $\phi$ given by
\begin{equation}
(\phi^3-\phi^2+\tau \phi) =0
\end{equation} 
the explicit solutions are
\begin{equation}
\phi=\{\phi_0,\phi_-,\phi_+\}=\left\{ 0, \frac{1+\sqrt{1-4\tau}}{2},\frac{1-\sqrt{1-4\tau}}{2} \right\}.
\end{equation}

The first corresponds to $D_a=0$ and is the ground state of the system for 
$\tau>\frac{1}{4}$, is a local minimum for $\frac{1}{4}>\tau>0$ and 
though it appears to be a local maximum for $\tau<0$, the model 
has no ground state for such value of $\tau$.
The second solution gives 
the local minima $D_a=\phi_- J_a$, for $\tau<\frac{1}{4}$. 
These minima $D_a=0$ and $D_a=\phi_-J_a$ are separated by a potential barrier 
whose highest point is localised at the local maximum $\phi_+$.
As $\tau$ approaches zero $\phi_{-}\rightarrow 1$, which
corresponds to the case for $m^2=\tau =0$ studied in \cite{nishi:numeS2,prlROY,longpaper}. For arbitrary $\tau$ has been also studied by other authors \cite{nishi-mass,longpaper}. 

\begin{figure}[ht]
\begin{center}
\includegraphics[width=7.5cm,angle=0]{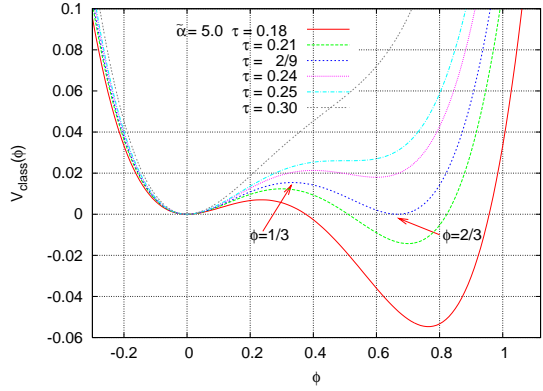}
\caption{Classical potential for different values of $\tau$ for fixed $\tilde{\alpha}=5$.}
\end{center}
\label{fig-figclaspotplot} 
\end{figure}

For the configurations $D_a=\phi J_a$ the action 
\begin{equation}\label{classpotmu}
S[D]=V_{class}(\phi)=\frac{2}{N}\sum_{i}n_iC_2(n_i)\;\tilde{\alpha}^4\left(\frac{\phi^4}{4}-\frac{\phi^3}{3}+\frac{\tau\phi^2}{2}\right)
\end{equation} 
where $C_2(n_i)$ is the Casimir of the representation of dimension $n_i$
and $\sum_in_i=N$. For $\tau<\frac{2}{9}$ the potential $V_{class}(\phi_{-})<0$ 
and the configurations $D_a=\phi_-J_a$ is of lower energy 
than $D_a=0$, which is zero. The configuration with 
minimum energy is then given by maximising the sum of Casimirs, 
this is achieved by $J_a=L_a$, where $L_a$ is
the irreducible representation of dimension $N$. 
For $\tau=\frac{2}{9}$ we have $V_{class}(\phi_{-})=0$ and
the configurations $D_a=0$ and $D_a=\phi_{-}J_a$, with any 
representation $J_a$, become degenerate. 
Figure \ref{fig-figclaspotplot}
shows the potential (\ref{classpotmu}) for different $\tau$.

Therefore the classical prediction is that for $\tau<\frac{2}{9}$, 
and any value of $\tilde\alpha$ the ground state is $D_a=\phi_{-}L_a$ 
and small fluctuations around this configuration have the
geometrical content of a Yang-Mills and scalar multiplet on a
background fuzzy sphere.  

The parameter domain $\frac{2}{9}<\tau<\frac{1}{4}$ is of special
interest.  The classical analysis above suggests that the fuzzy sphere
is unstable here, however, since the potential still has a local
minimum and in the $N\rightarrow\infty$ limit all representations such
that $\frac{1}{N}\sum_in_iC_2(n_i)\rightarrow0$ become degenerate with
the $D_a=0$ configuration one might wonder if one of these
configurations gives the ground state of the system when fluctuations
are included or perhaps as suggested by the analysis of
\cite{longpaper} and discussed in the next section, the fuzzy sphere
phase is stabilised by fluctuation. The answer as we will see, is that
numerical simulations do not support the assertion that the fuzzy
sphere is stabilised in this parameter range, but rather that the
coexistence curve between the fuzzy sphere phase and the matrix phase
asymptotes to the line $\tau=\frac{2}{9}$.
 
For $\tau=\frac{2}{9}$ the model is indeed special, the action is always 
positive  semi-definite and can be written in the form
\begin{eqnarray}\label{F2}
S[D]=\frac{\tilde{\alpha}^4}{N}Tr \left(\frac{i}{2}[D_a,D_b]+\frac{1}{3}\epsilon_{abc}D_c\right)^2,
\end{eqnarray}
from which we see that $D_a=0$ and the orbit 
$D_a=U\frac{2}{3}L_aU^{\dagger}$ have zero action.  

\subsection{Quantum corrections and critical behaviour}

In the previous section we described the classical potential predictions 
for the phase transition by choosing $D_a=\phi L_a$ and
studying the potential for $\phi$.  This analysis suggests 
that $\phi$ plays the role of order parameter for the transitions 
which are taking place in the model. We will now take into account 
the quantum fluctuations.
The computation of the quantum effective action and the quantum effective
potential was carried out in detail \cite{PRY1,prlROY} for a more
general $SO(3)$ globally invariant matrix model. The results there 
correspond to the present case by setting $m^2=0$. 
 
Using the standard background field method around the classical
configuration $D=\phi L_a$ one finds the effective potential in the
large $N$ limit is given by

\begin{equation}
\label{V_eff}
\frac{V_{\rm eff}}{2C_2}=\tilde{\alpha}^4\left[\frac{\phi^4}{4}
-\frac{\phi^3}{3}+\tau\frac{\phi^2}{2}\right]+\log \phi^2
\end{equation} 
As discussed in 
\cite{PRY1,prlROY} the phase diagram including fluctuations
is obtained from the minimum of this effective potential.  
Due to the log term the effective potential (\ref{V_eff}) is not
bounded from below near $\phi=0$ and since in its derivation it was 
assumed that we were expanding around $D_a=\phi L_a$ it is only 
valid where such a ground state exists.

The minimum is therefore one of the roots of the polynomial
$\phi\frac{\partial V_{\rm eff}(\phi)}{\partial\phi}=0$ which gives the equation
\begin{equation}\label{V'=0}
{\phi}^4-{\phi}^3
+\tau{\phi^2}+\frac{2}{\tilde{\alpha}^4}=0,
\end{equation}
and determines $\phi$ in $D_a=\phi L_a$ at quantum level. 
Explicitly the minimum is given by
\begin{eqnarray}\label{phisol}
{\phi}=\frac{1}{4}+\frac{1}{2}\sqrt{\frac{1}{4}-\frac{2\tau}{3}+d}+
  \frac{1}{2}\sqrt{\frac{1}{2}-\frac{4\tau}{3}-d+\frac{1-4\tau}{4\sqrt{\frac{1}{4}-\frac{2\tau}{3}+d}}},
\end{eqnarray}
with the definitions
\begin{eqnarray}
&&q=1-\frac{8\tau}{3}+\frac{\tilde{\alpha}^4\tau^3}{27},\;\;\;p=q^2-\frac{(\frac{8}{3}+\frac{\tilde{\alpha}^4\tau^2}{9})^3}{\tilde{\alpha}^4},\;\;\;\;
d=a^{-\frac{4}{3}}\bigg(\big(q+\sqrt{p}\big)^{\frac{1}{3}}+\big(q-\sqrt{p}\big)^{\frac{1}{3}}\bigg).
\end{eqnarray}

\begin{figure}[ht]
\begin{center}\label{fig:figVeff}
\includegraphics[width=7.0cm,angle=0]{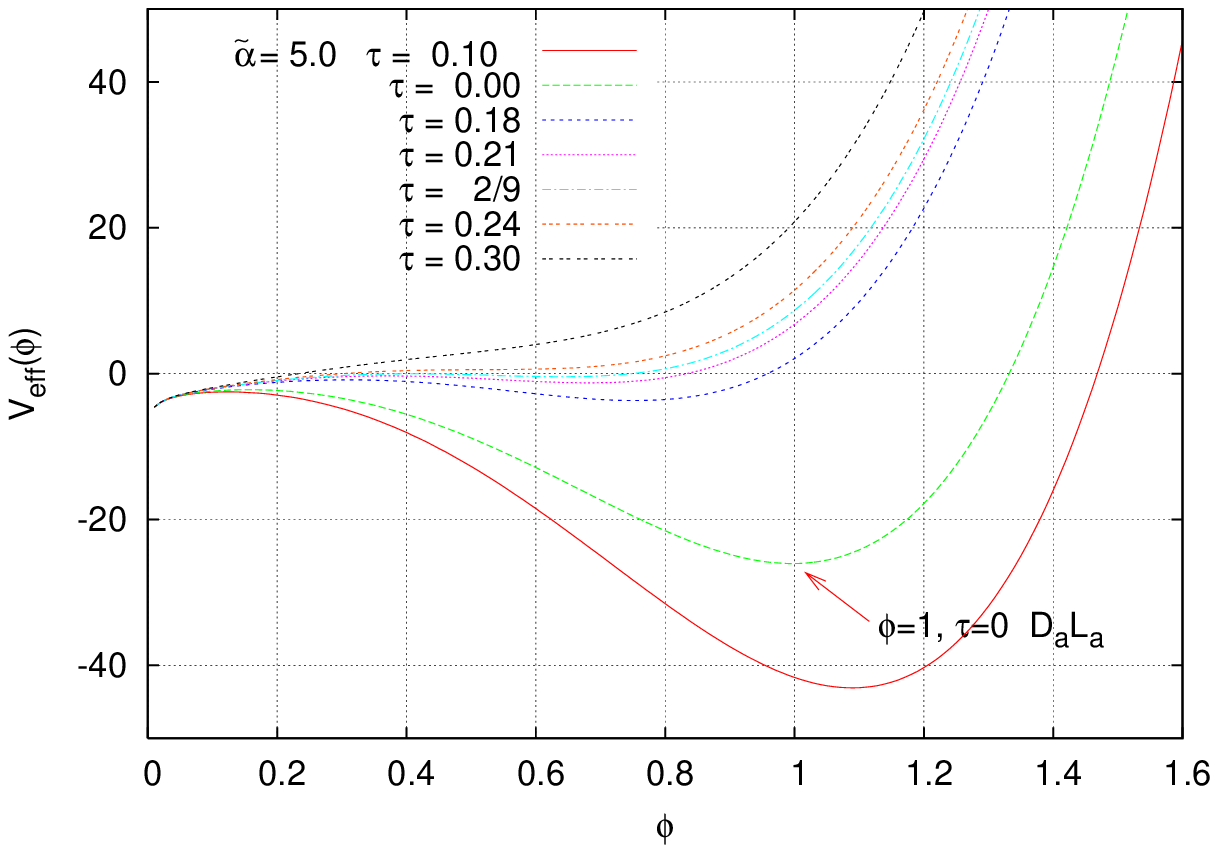}
\includegraphics[width=7.0cm,angle=0]{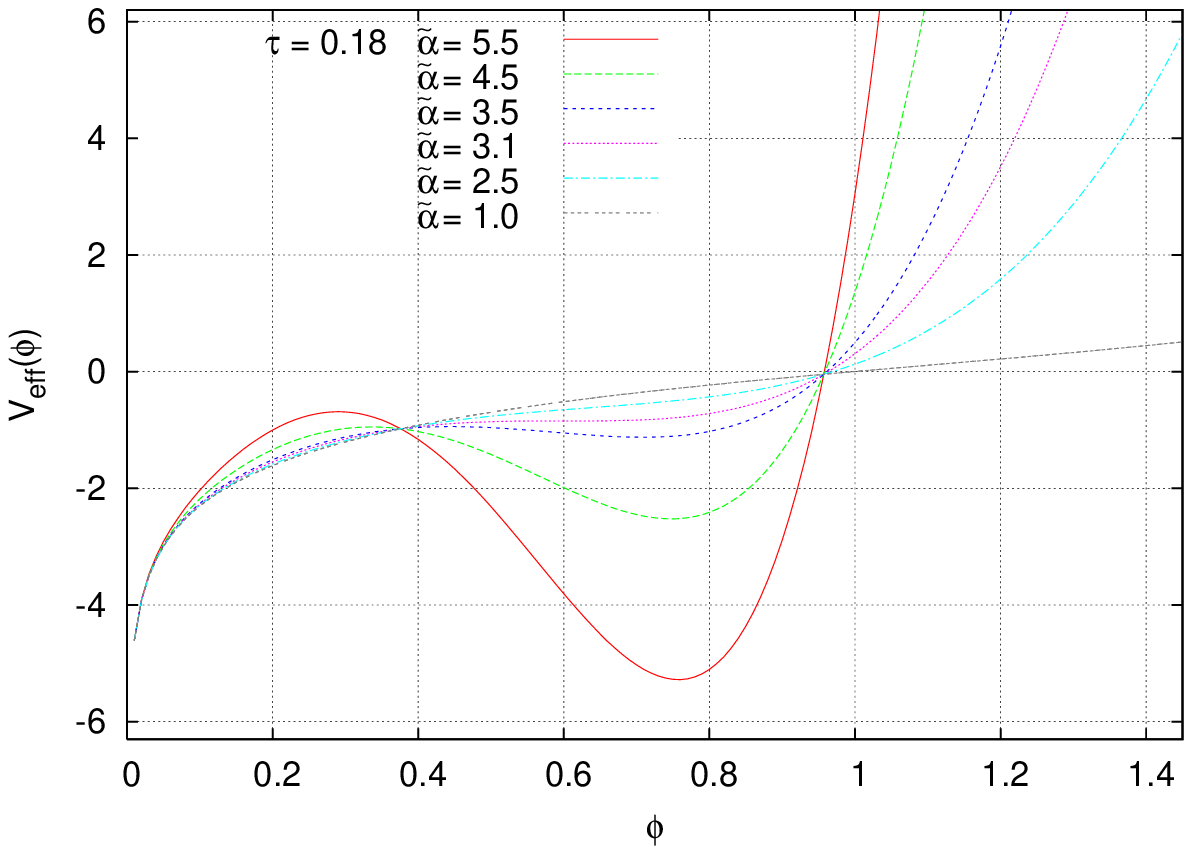}
\caption{(a) Effective potential for different values of $\tau$ for $\tilde{\alpha}=5$. (b) Effective potential for $\tau= 0.18$ for different values of $\tilde \alpha$}
\end{center}
\end{figure}

$V_{\rm eff}$ is plotted in figure 2. 
In figure 2(a) the potential is shown for a fixed
value of $\tilde\alpha =5$ for different values of $\tau$. In figure
2(b) we fix the value of $\tau= 0.18$ varying $\tilde{\alpha}$. From
both pictures we can see that there exists a region in which the fuzzy
sphere ceases to exist. This means that there exists a critical coexistence 
curve in the $(\tau,\tilde\alpha)$ plane where the model undergoes
a phase transition between the two phases.

These expressions predict that for 
sufficiently large $\tilde{\alpha}$ or low temperature and 
$\tau < \frac{1}{4}$ the fuzzy sphere is the ground state of the
model. The local minimum disappears for the simultaneous solution of
$V_{\rm eff}'(\phi)=0$ and $V_{\rm eff}''(\phi)=0$ so that
\begin{eqnarray}
4\phi^2_*-3\phi_*+2\tau=0
\end{eqnarray}
which gives the critical values for $\phi$ and $\tilde{\alpha}$.
\begin{eqnarray}\label{critline}
\phi_{*}=\frac{3}{8}\left(1+\sqrt{1-\frac{32\tau}{9}}\right) 
\quad \hbox{ and}\quad
g_{*}^2=\frac{1}{\tilde{\alpha}_*^4}
=\frac{{\phi}_*^2({\phi}_*-2\tau)}{8}.
\end{eqnarray}
These equations defines the phase diagram $(\tau,\tilde{\alpha})$ for the
present model. They prediction that at zero temperature 
i.e. $\tilde{\alpha}=\infty$ where $\phi_*=1/2$ and $\tau=1/4$.

The phase diagram is presented in Figure 3 
together with a comparison with numerical simulations. 
Numerical results are in excellent agreement with the 
prediction from effective potential for $\tau<\frac{2}{9}$ where 
the coexistence curve predicted in 
(\ref{critline}) clearly delimits the fuzzy sphere phase from the matrix
phase. The fuzzy sphere exists for $\tilde{\alpha}>\tilde{\alpha}_*$
and for $\tau<\tau_*$, where $\tau_*(\tilde\alpha)$ is obtained by inverting 
(\ref{critline}).

However, for $\frac{1}{4}>\tau >\frac{2}{9}$, simulations show that
{\it the coexistence curve asymptotes rapidly to the line
  $\tau=\frac{2}{9}$ for $\tilde\alpha$ greater than the special value
  $\tilde\alpha_*=12\left(\frac{4}{107+51\sqrt{17}}\right)^{1/4}\sim4.02$}.  
As we cross the critical value of $\tilde\alpha$, a rather exotic 
phase transition occurs where the geometry disappears as the temperature 
is increased. In the high
temperature phase, which we call a matrix phase, the order parameter
$\phi$ goes to zero as $N^{-1}$. In this phase the fluctuations are
insensitive to the value of $\tilde\alpha$, they are in fact
fluctuations around commuting matrices.  Since, in the large $N$
limit, $\tilde\alpha$ is unimportant we can rescale the matrices to
eliminate $\tilde\alpha$ from the quadratic term, defining
$X_a=\frac{\tilde\alpha}{\sqrt{N}} D_a$, we obtain
\begin{equation}
\label{rescaled-X-model}
S[X]=N Tr(-\frac{1}{4}[X_a,X_b]^2+\frac{2 i\tilde\alpha}{3\sqrt{N}}\epsilon_{abc}X_aX_bX_c+\frac{\tilde\alpha^2\tau}{N}X_a^2)
\end{equation}
and see that in this rescaled model both $\tau$ and 
$\tilde\alpha$ drop out of the model in the large $N$ limit.

\begin{figure}[ht]
\begin{center}\label{phasediagram}
\includegraphics[width=10.5cm,angle=0]{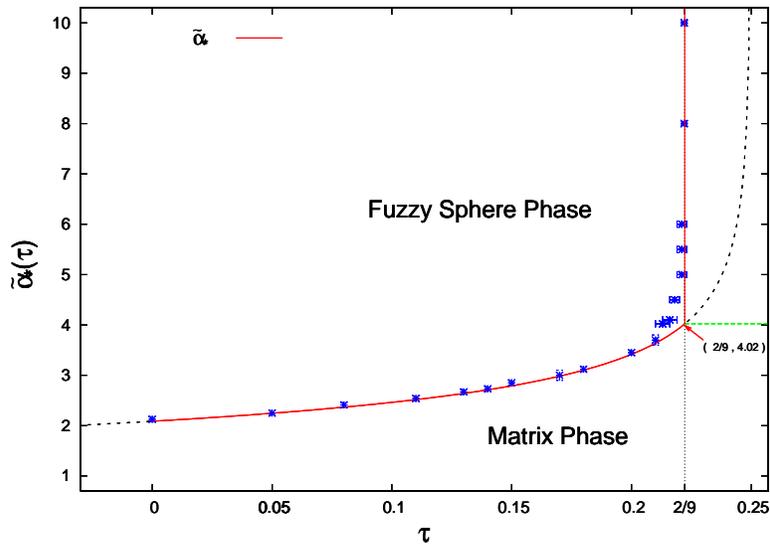}
\caption{Phase diagram $(\tau,\tilde{\alpha})$. The solid line corresponds to 
the theoretical prediction eq.(\ref{critline}) for $\tau<\tau_* =\frac{2}{9}$. The numerical points are obtained from simulations 
with matrix sizes $N=19,N=24,N=35$ and $N=64$. We see some rounding of 
the coexistence curve near $(\tilde\alpha_*=4.02,\;\tau=\frac{2}{9})$, before 
it asymptotes to the line $\tau=\frac{2}{9}$.}
\end{center}
\end{figure}

Next we discuss the numerical results which include 
the eigenvalue distributions for the configurations.

\subsection{Specific heat and phase transitions.}

There are several useful general identities that we can derive between
expectation values of observables, the simplest of these is 
the relation between different components of the action.
Since our action is a polynomial in the matrices $X_a$ it can be expressed
as $S=S_4+S_3+S_2$ where $S_k[\lambda X]=\lambda^k S[X]$.
Then if we scale the fields $X_a\rightarrow \lambda X_a$ in both the 
action and the measure, the partition function $Z$, is invariant and 
under an infinitesimal rescaling we obtain the constraint 
\begin{eqnarray}\label{ward}
4<S_4>+3<S_3>+2<S_2>=3(N^2-1),
\end{eqnarray}
which allows us to eliminate $<S_4>$. Defining ${\cal S}_k=\frac{<S_k>}{N^2}$ 
we have for large $N$ obtain 
\begin{eqnarray}\label{S-S3-S2}
{\cal S}=\frac{3}{4}+\frac{1}{4}{\cal S}_3+\frac{1}{2}{\cal S}_2.
\end{eqnarray}
For our model, with 
\begin{equation}\label{Xa}
X_a=\frac{\tilde\alpha}{\sqrt{N}} D_a, 
\end{equation}
we have 
\begin{eqnarray}
S_4&=&-\frac{N}{4}\Tr([X_{a},X_{b}]^2), \label{YM}\\ S_3& =& N\frac{2}{3}i\alpha \epsilon_{abc}\Tr(X_{a}X_{b}X_c), \label{Myers}\\ S_2&=&N\alpha^2\tau \Tr(X^2_{a}). \label{Xsq}
\end{eqnarray}

Where $\alpha=\frac{\tilde\alpha}{\sqrt{N}}$. In the fuzzy sphere phase we know fluctuations are around the configuration
$D_a=\phi L_a$ and we then have the predictions  (see \cite{longpaper} for 
details) the partition function $Z={\rm e}^{-F}$ with the free energy $F$ 
given by 
\begin{equation}
F(\tilde\alpha,\tau)=\frac{V_{\rm eff}}{N^2}+3\ln\tilde\alpha .
\end{equation}
where $V_{\rm eff}$ is evaluated with $\phi$ given by (\ref{phisol}).
From this we can then identify 
\begin{equation}
 \frac{<S_2>}{N^2}=\frac{\tau\tilde\alpha^4 \phi^2}{4}, \qquad 
\frac{<S_3>}{N^2}=-\frac{\tilde\alpha^4\phi^3}{6}\qquad \frac{<S_4>}{N^2}=\frac{3}{4}-\frac{\tilde\alpha^4\tau\phi^2}{8}+\frac{\tilde\alpha^4\phi^3}{8}
\end{equation}

In particular for the average of the action ${\cal S}$ and the specific heat
$C_v=\frac{1}{N^2}<(S-<S>)^2>$  we have 
\begin{equation}\label{cvth}
{\cal S}=\frac{3}{4}+\frac{\tilde{\alpha}^4\tau\phi^2}{8}-\frac{\tilde{\alpha}^4\phi^3}{24}\qquad \hbox{and }\qquad C_v=\frac{3}{4}+\frac{\tilde{\alpha}^5}{32}\phi(\phi-2\tau)\frac{d\phi}{d\tilde{\alpha}}.
\end{equation}

\section{The Matrix Phase}

In terms of $X_a$ of (\ref{Xa}) the model (\ref{mu-action}) becomes
\begin{equation}\label{actionXalphatilde}
S[X]=N \Tr \left[-\frac{1}{4}[X_a,X_b]^2+\frac{i}{3}\frac{\tilde\alpha}{\sqrt{N}}{\epsilon}_{abc}[X_aX_b]X_c+\tau \frac{\tilde\alpha^2}{N} X_a^2\right].
\end{equation}
We see that at high temperature where $\tilde\alpha\rightarrow0$,
provided fluctuations in $X_a$ do not grow too rapidly with either $\tilde\alpha$ or $N$ then in the large $N$ limit the model reduces to the 
pure Yang-Mills term
\begin{equation}\label{YMaction}
S[X]=-\frac{N}{4}\Tr [X_a,X_b]^2. 
\end{equation} 
Our numerical results support this and we find that 
$<\frac{Tr X_a^2}{N}>$ 
and
$\frac{2}{3}\tilde\alpha\epsilon_{abc}<Tr(X_a,X_bX_c)> $
are independent of $N$, see Figures 4.

We see immediately from (\ref{S-S3-S2}) and the reduction to (\ref{YMaction})
that in the high temperature, matrix phase we should expect both 
${\cal S}=\frac{3}{4}=C_v$ irrespective of the value of $\tilde\alpha$ 
and $\tau$. 
This is in accordance with our numerical results shown in Figures 5 and 7
and the earlier results of \cite{longpaper}, but there is some 
discrepancy with Figure 6 which we believe is due to finite size effects
as discussed below.

It is argued in \cite{OConnorThMp,FilevOConnor} that fluctuations are about a
background of commuting matrices whose eigenvalues are uniformly
distributed in the interior of a solid ball in ${\mathbb{R}}^3$. This in turn
predicts that the eigenvalue distribution of a single matrix, say
$X_3$, is the parabolic distribution
\begin{eqnarray}\label{parabola}
\rho(x)=\frac{3}{4 R^3}(R^2-x^2).
\end{eqnarray}
In \cite{FilevOConnor}, based on a two-loop approximation, 
it was estimated that $R\sim1.8$. This is in reasonable agreement with
our numerical results (see Fig 13) which give $R=2.0$. 
In Figure 13 we can see that the numerical results
for the eigenvalue distribution of $X_3$ are very well fit by the
parabola (\ref{parabola}).

\begin{figure}[t]
\begin{center}\label{figM29-a}
\includegraphics[width=7.0cm,angle=0]{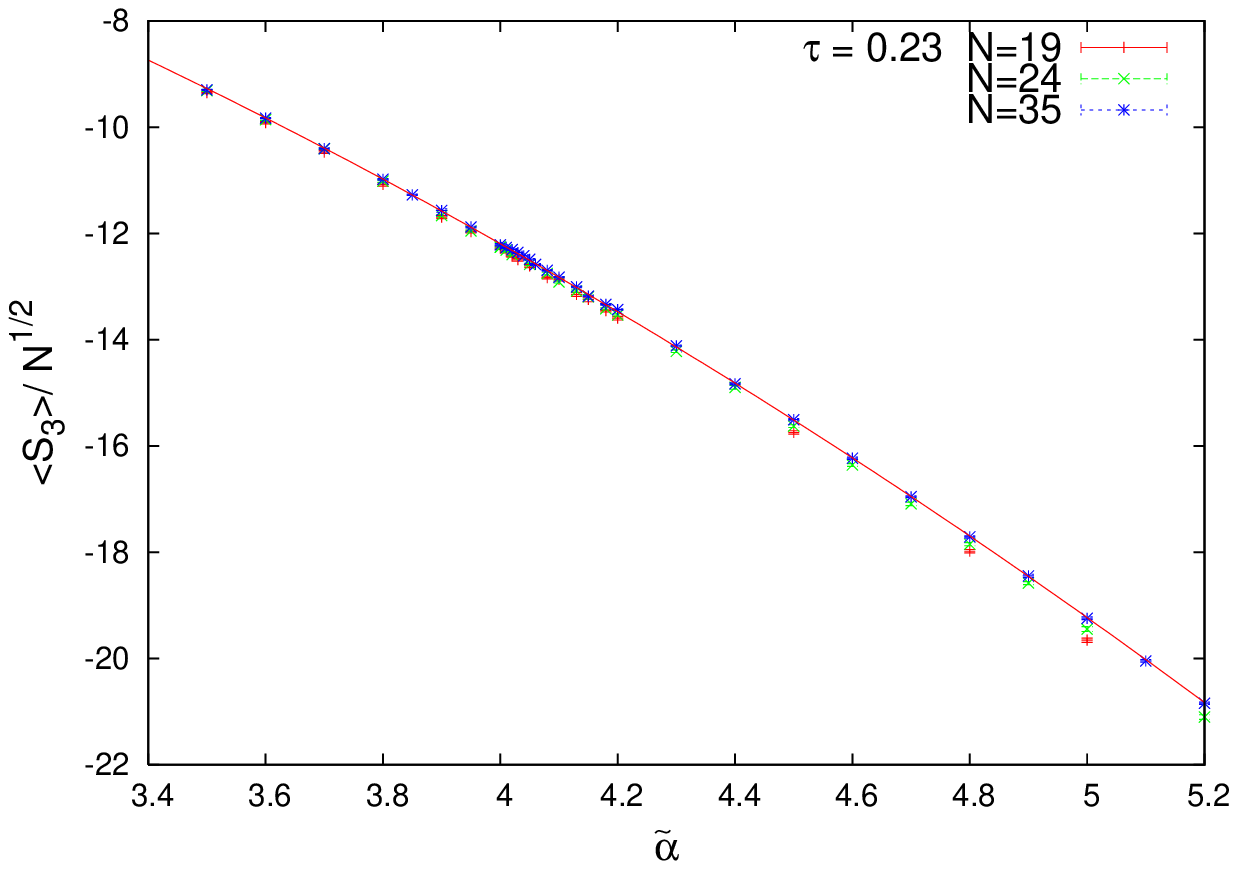}
\includegraphics[width=7.0cm,angle=0]{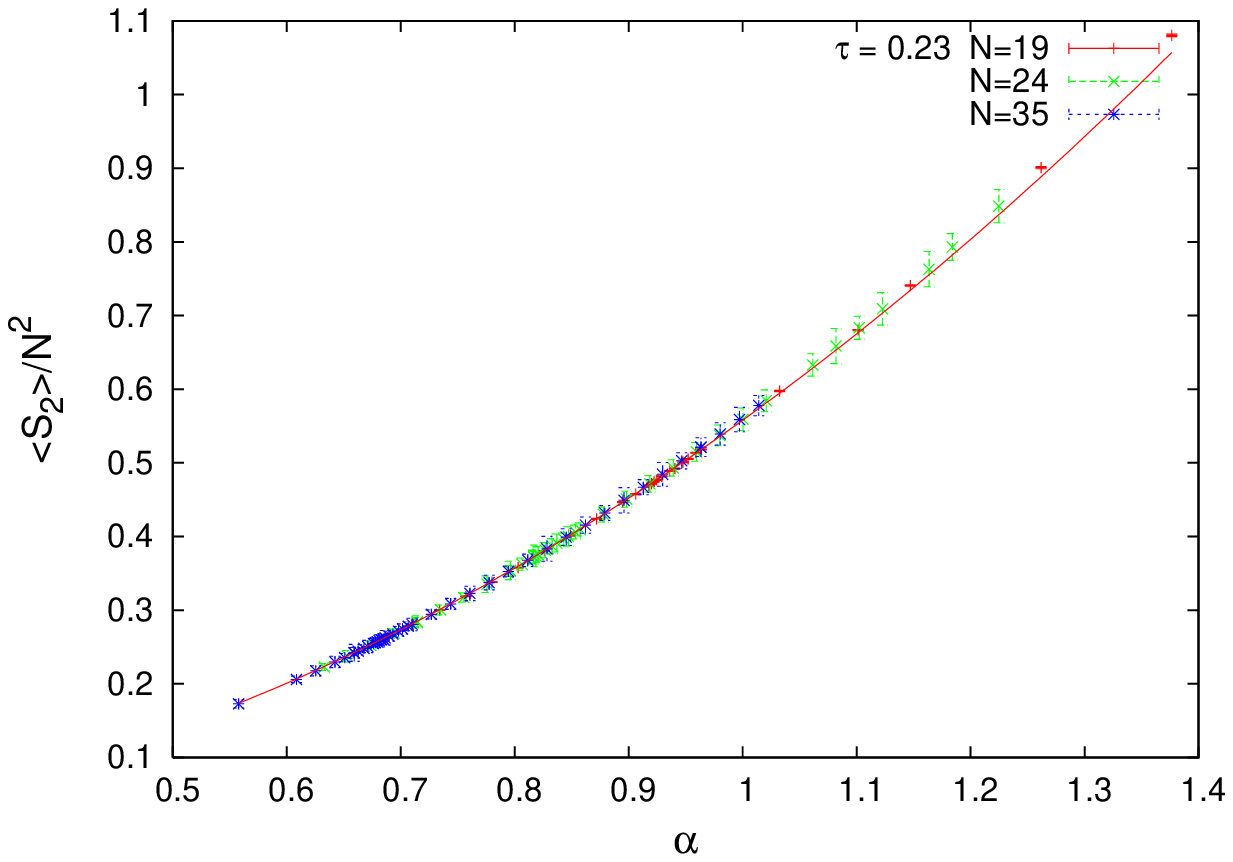}
\includegraphics[width=7.0cm,angle=0]{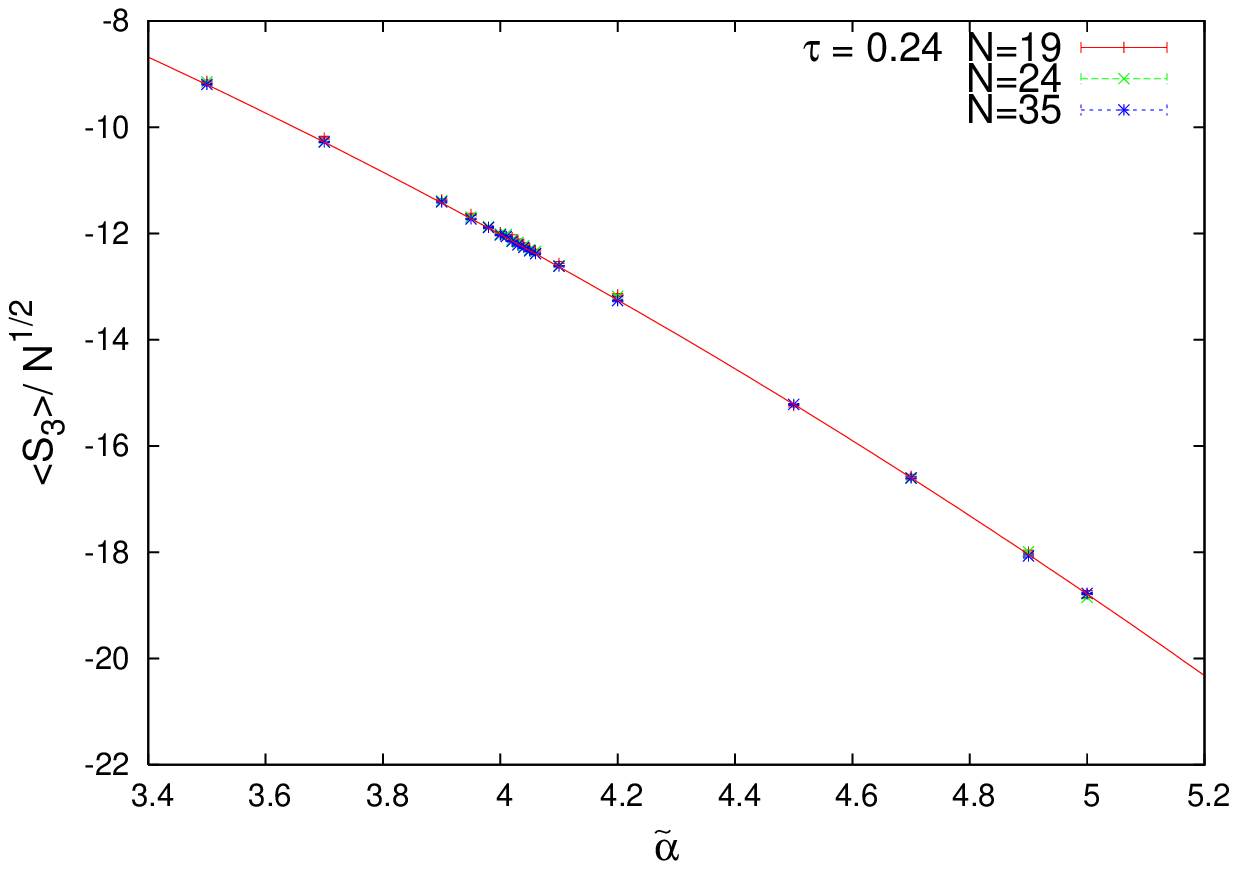}
\includegraphics[width=7.0cm,angle=0]{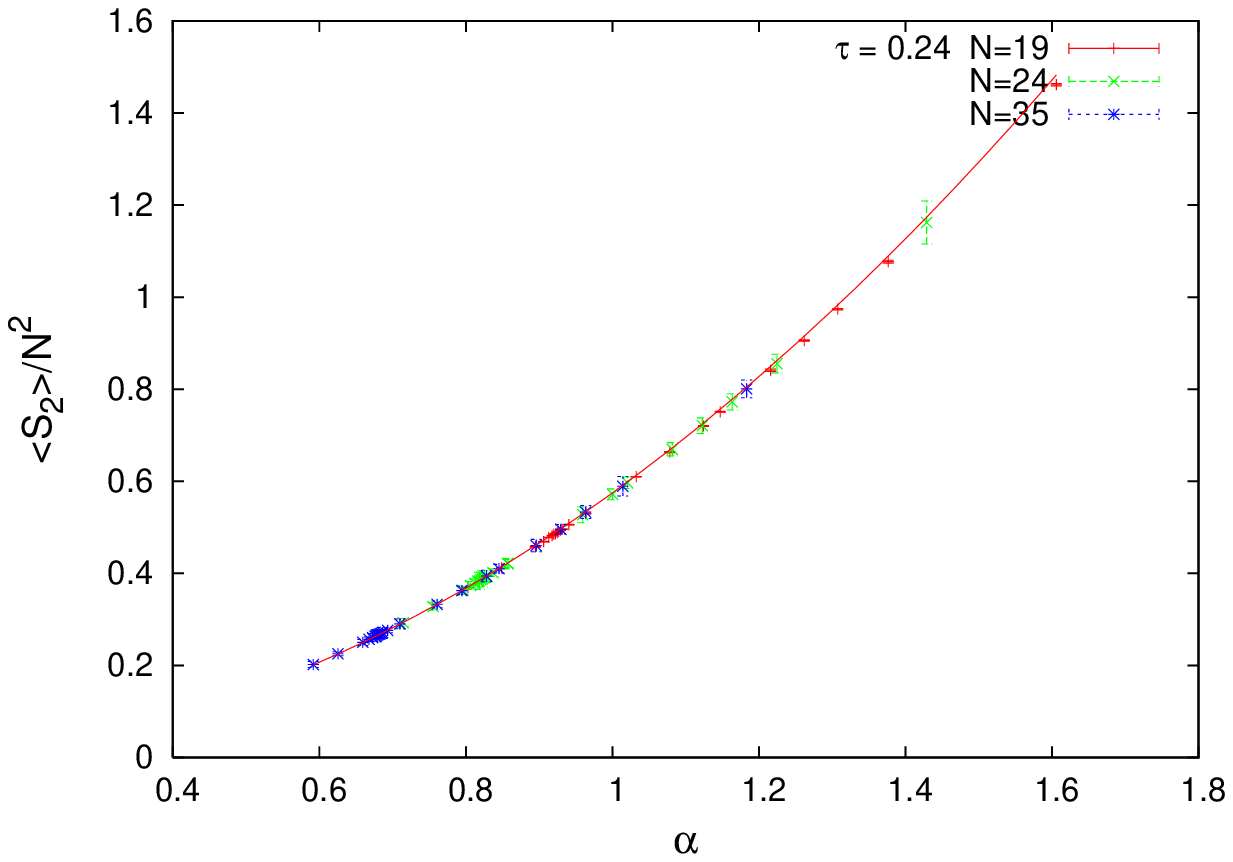}
\caption{$\frac{<S_3>}{N^{1/2}}$ as a function of $\tilde\alpha$ and $\frac{<S_2>}{N^2}$ as a function of $\alpha$ for $\tau=0.23,0.24$. We observe collapsed data in both plots in the range showed.}
\end{center}
\end{figure}

In the same figure we present the distribution for the commutator
$i[X_1,X_2]$ showing a symmetric distribution with support lying in
the interval $[-1.5,1.5]$. This indicates, that the fluctuations
around this background are non-commuting variables.

If we accept the parabolic distribution (\ref{parabola}) and radius $R=2$ 
we can predict 
\begin{eqnarray}
<\frac{\Tr}{N}( X^2_a)> = 3\int^R_{-R}dx_1\rho(x_1)x^2_1 = \frac{3R^2}{5}=\frac{12}{5}.
\end{eqnarray}
Therefore we have
\begin{eqnarray}\label{trXsq}
{\cal S}_2&=& \alpha^2\tau \frac{<\Tr X^2_a >}{N} = \tau\frac{\tilde\alpha^2}{N}\frac{12}{5},
\end{eqnarray}
which goes to zero for fixed $\tilde\alpha$ in the large $N$ limit. Our numerical results are in very good agreement with this result for $\mathcal{S}_2$, as we can see in figure 4 (right). The value of the radius extracted from here is $R=2.01$, in accordance with the value obtained from the eigenvalues of $X_3$. However, as $\tilde\alpha$ increases we can observe small deviations. The value $R=2$ corresponds to the limit of infinite temperature and large $N$. 

From our numerical simulations we see that ${\cal S}_3$ goes to zero even more rapidly than ${\cal S}_2$. In figure 4 we see that the curves for different $N$ collapse when we 
\begin{eqnarray}
\frac{<S_3>}{N^{1/2}}=\left<\frac{2}{3}i\tilde{\alpha} \epsilon_{abc}\Tr X_{a}X_{b}X_c\right>
\end{eqnarray}
as a function of $\tilde\alpha$. For different values of
$\frac{2}{9}<\tau <\frac{1}{4}$ we find $\frac{S_3}{\tilde\alpha}$
behaves linear in $\tilde\alpha$, however the approximation becomes
poorer as $\tau \to \frac{1}{4}$ and higher corrections subleading in
$N$ should be taken into account. The data for $\tau=0.23$ is well fit
by
\begin{eqnarray}
\frac{<S_3>}{N^{1/2}} = 0.142533 \;\tilde\alpha -0.799823 \;\tilde\alpha^2,\qquad \qquad\frac{<S_2>}{N^{2}}=0.557965\;\alpha^2.
\end{eqnarray}
For $\tau=0.24$ we find
\begin{eqnarray}
\frac{<S_3>}{N^{1/2}} = 0.0243563 \;\tilde\alpha -0.757264 \;\tilde\alpha^2,\qquad \qquad \frac{<S_2>}{N^{2}}=0.574869\;\alpha^2
\end{eqnarray}

From which we deduce that ${\cal S}_3\rightarrow 0$ as $N^{-\frac{3}{2}}$.

Taking this into account we can write the expectation value of the action $\frac{<S>}{N^2}$ given by eq. (\ref{S-S3-S2})
\begin{eqnarray}\label{finitesizecorrectionstoS}
{\cal S}&=& \frac{3}{4}+\frac{1}{4N}(0.142533 \;\tilde\alpha -0.799823 \;\tilde\alpha^2)+\tau\frac{\tilde\alpha^2}{N}\frac{6}{5} \nonumber \\
     &=& \frac{3}{4}+\frac{1}{4N^{1/2}}(0.142533 \;\alpha) -\frac{1}{4}(0.799823 \;\alpha^2)+\tau \alpha^2\frac{6}{5};\qquad \mbox{for} \;\;\tau=0.23,  \\
{\cal S}&=& \frac{3}{4}+\frac{1}{4N}(0.024356 \;\tilde\alpha -0.757264 \;\tilde\alpha^2)+\tau\frac{\tilde\alpha^2}{N}\frac{6}{5} \nonumber \\
     &=& \frac{3}{4}+\frac{1}{4N^{1/2}}(0.024356 \;\alpha) -\frac{1}{4}(0.757264 \;\alpha^2)+\tau \alpha^2\frac{6}{5};\qquad \mbox{for} \;\;\tau=0.24, \\
  {\cal S} &=& \frac{3}{4}+\frac{1}{4N^{1/2}}(-0.159707 \;\alpha) -\frac{1}{4}(0.701843 \;\alpha^2)+\tau \alpha^2\frac{6}{5};\qquad \mbox{for} \;\;\tau=0.25. 
\end{eqnarray}
The expectation values for the action $\mathcal{S}$ are shown in
figure 8 for these last two particular values of $\tau$.

\section{Numerical Results}

We now describe the results of the numerical results for the
expectation value of the action $<S>$ and the specific heat
$C_v/N^2=<(S-<S>)^2>$ as function of $\tilde{\alpha}$ for a fixed
value of $\tau$.

\begin{figure}[t]
\begin{center}\label{figcvs}
\includegraphics[width=7.0cm,angle=0]{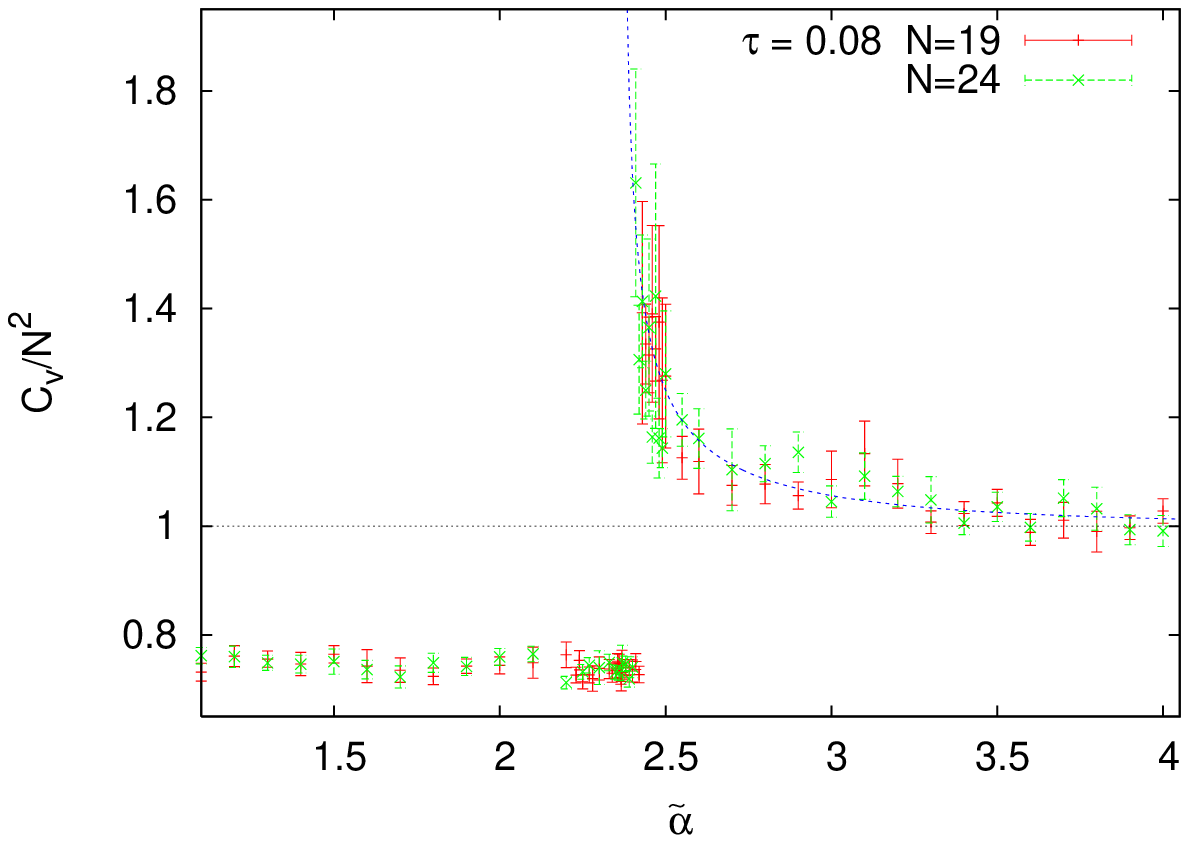}
\includegraphics[width=7.0cm,angle=0]{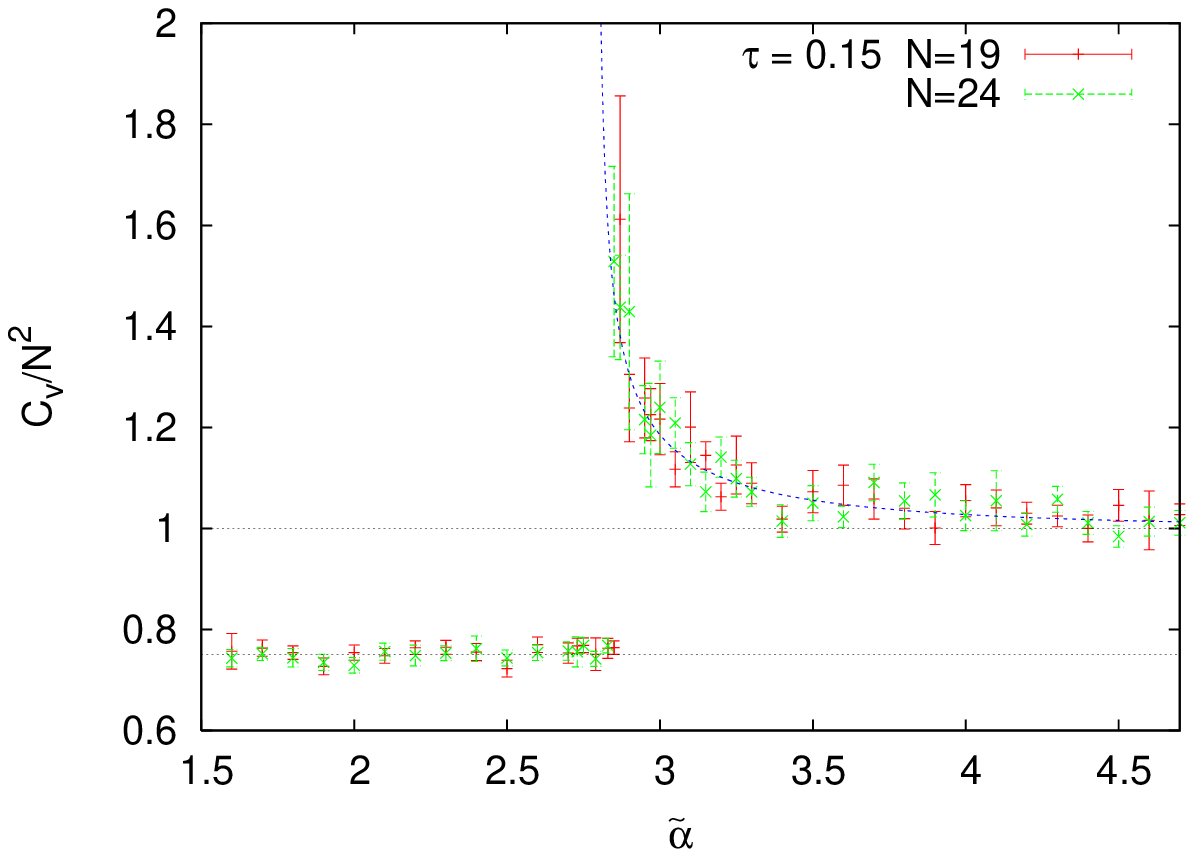}
\caption{The specific heat $C_v$ as a function of $\tilde\alpha$ 
for different $\tau$ values, $\tau<2/9$. 
The curve dashed line corresponds to expression (\ref{cvth}) 
for the specific heat.}
\end{center}
\end{figure}

\begin{figure}[t]
\begin{center}\label{figM29-b}
\includegraphics[width=7.0cm,angle=0]{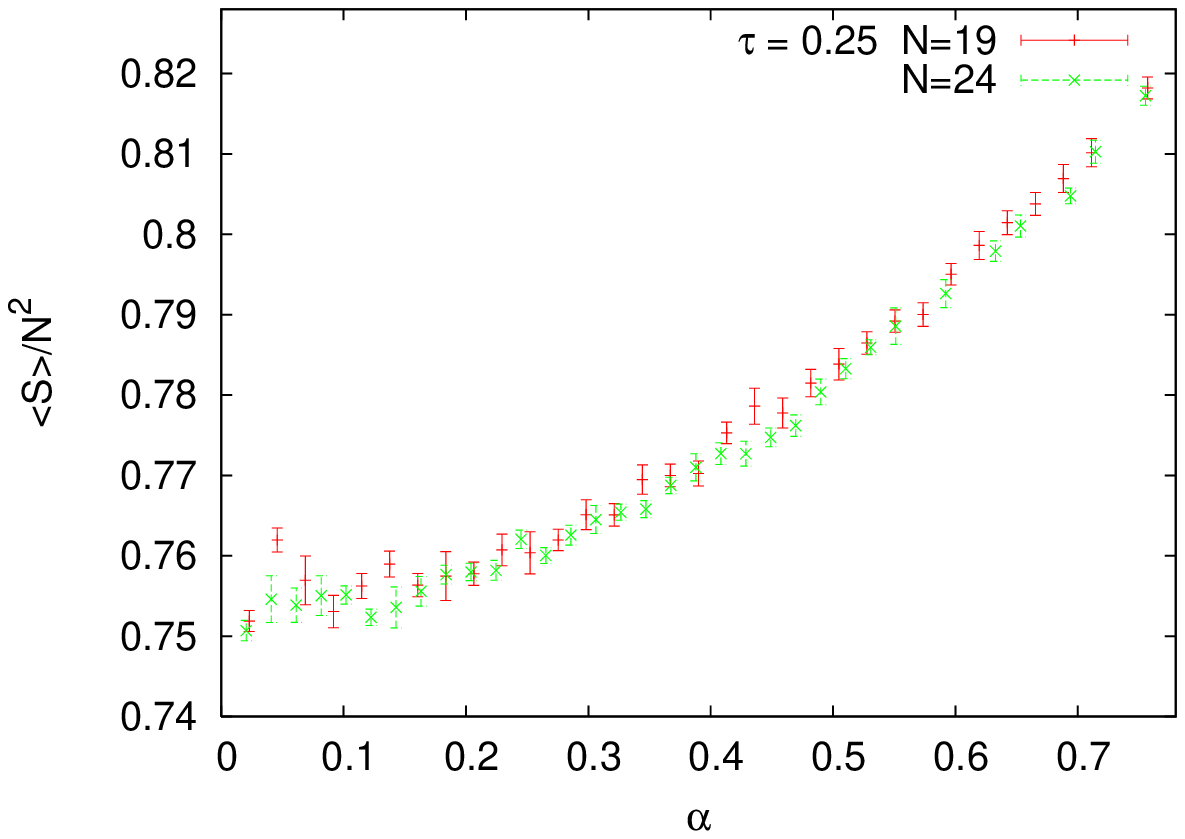}
\includegraphics[width=7.0cm,angle=0]{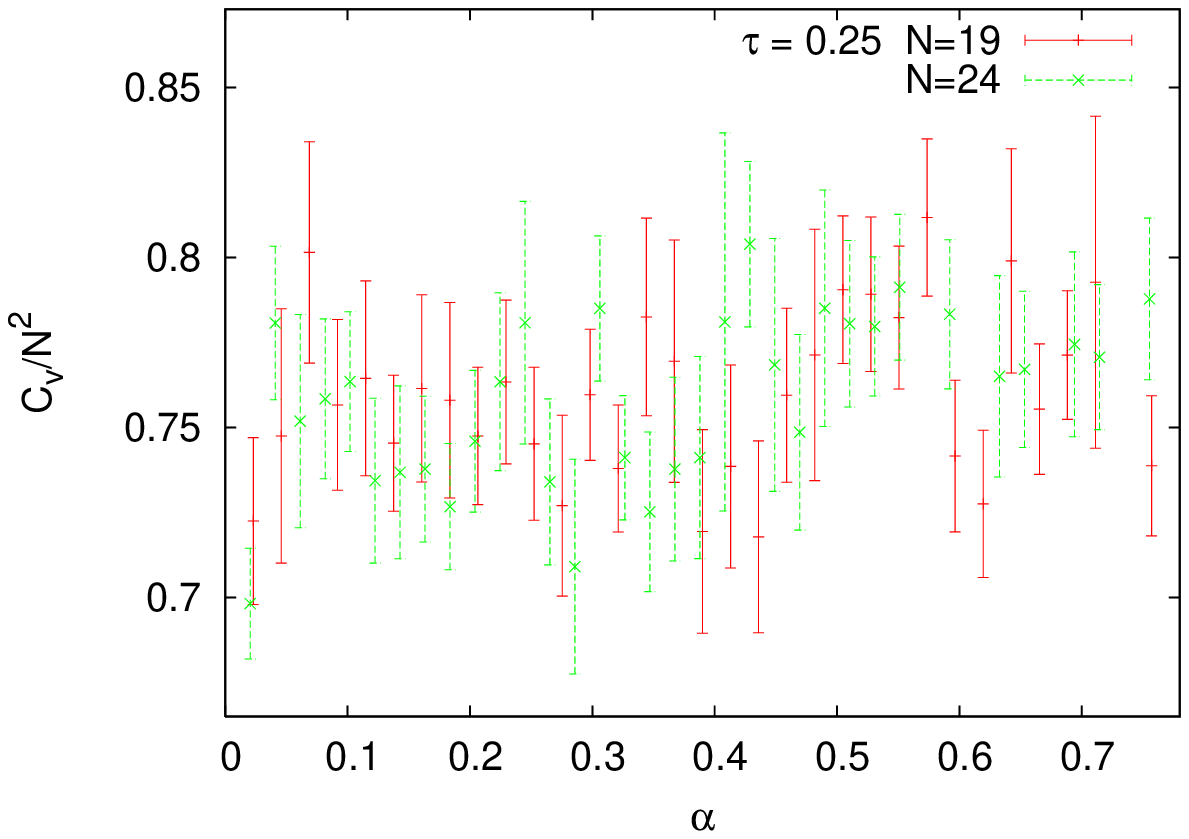}
\caption{Observables $<S>, C_v$ for $\tau=0.25$. The range of $\alpha$ shown corresponds to $\tilde\alpha<3.7$}
\end{center}
\end{figure}

\begin{figure}[t]
\begin{center}\label{figM29-c}
\includegraphics[width=7.0cm,angle=0]{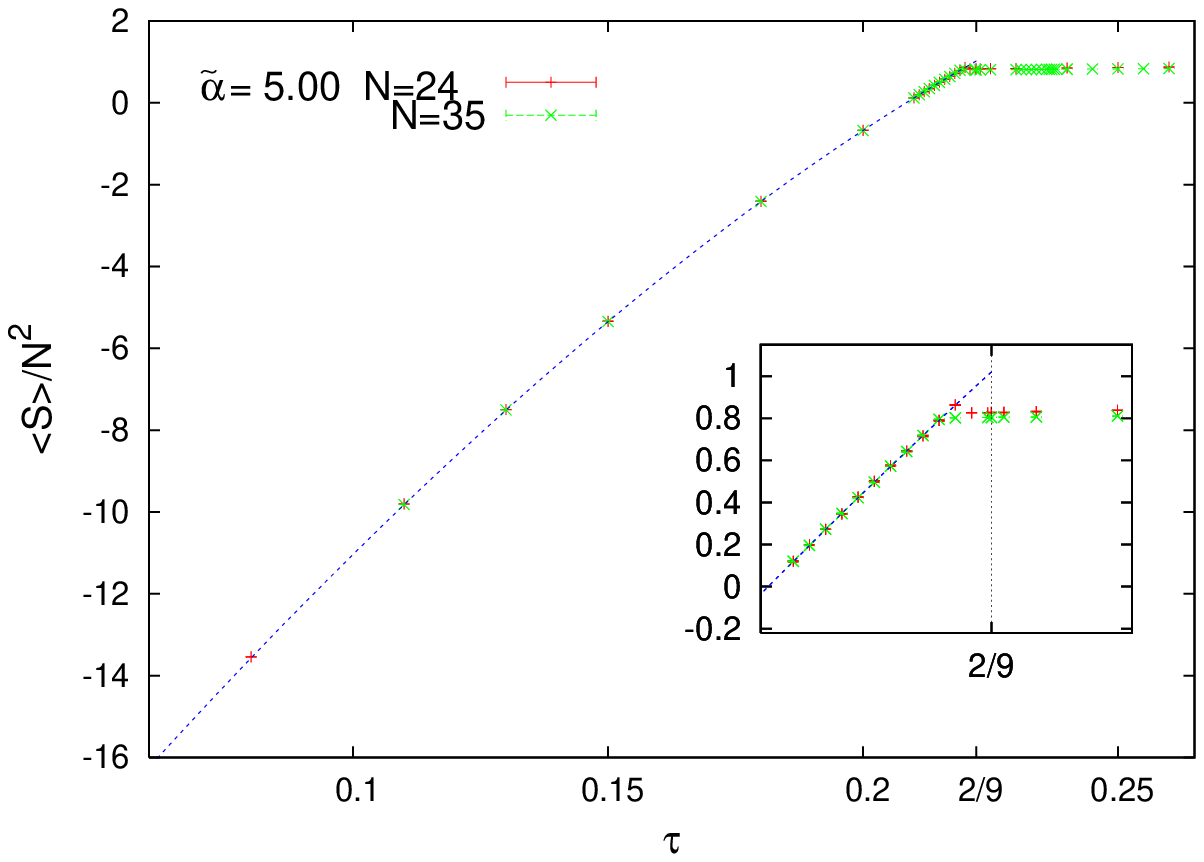}
\includegraphics[width=7.0cm,angle=0]{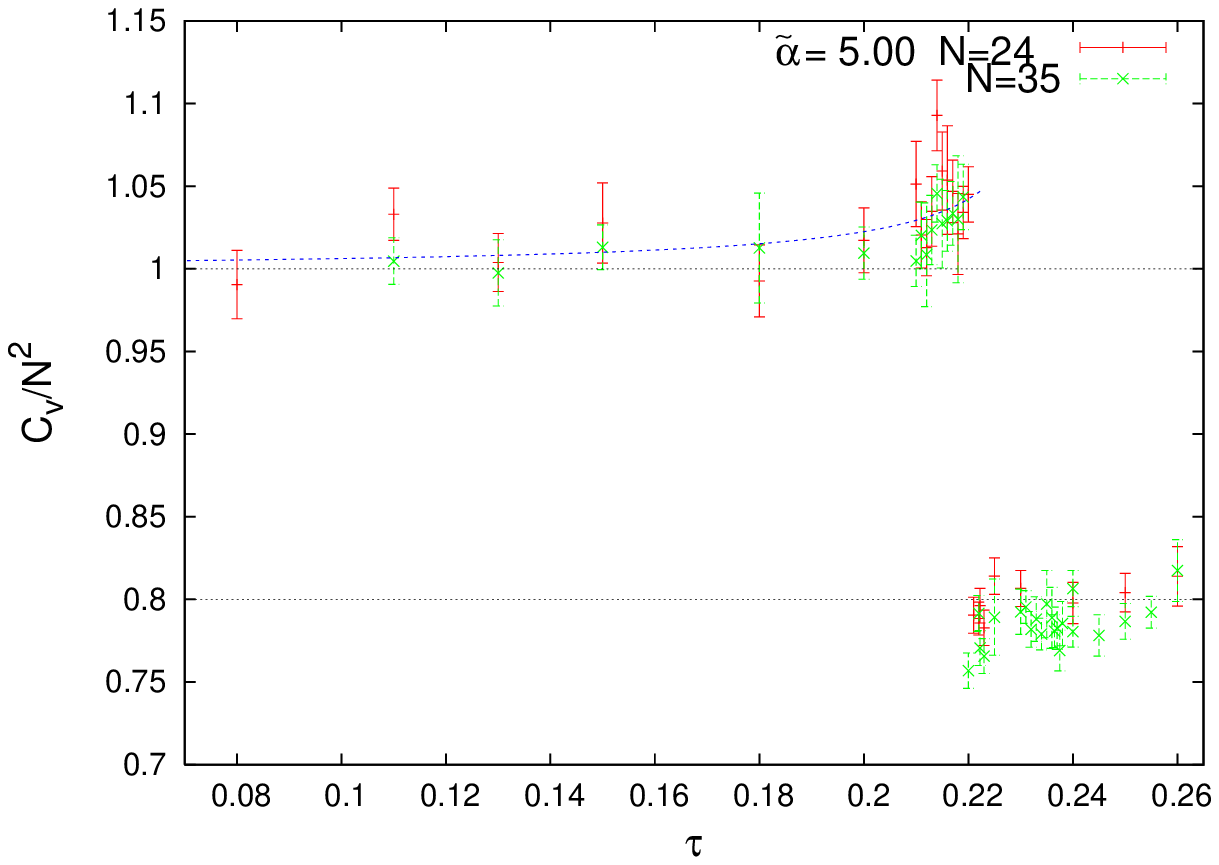}
\caption{Observables $<S>, C_v$ as function of $\tau$ for $\tilde\alpha=5$. 
The transition occurs at $\tau=0.219\pm 0.005$.}
\end{center}
\end{figure}

\begin{figure}[t]
\begin{center}\label{figM29-d}
\includegraphics[width=7.0cm,angle=0]{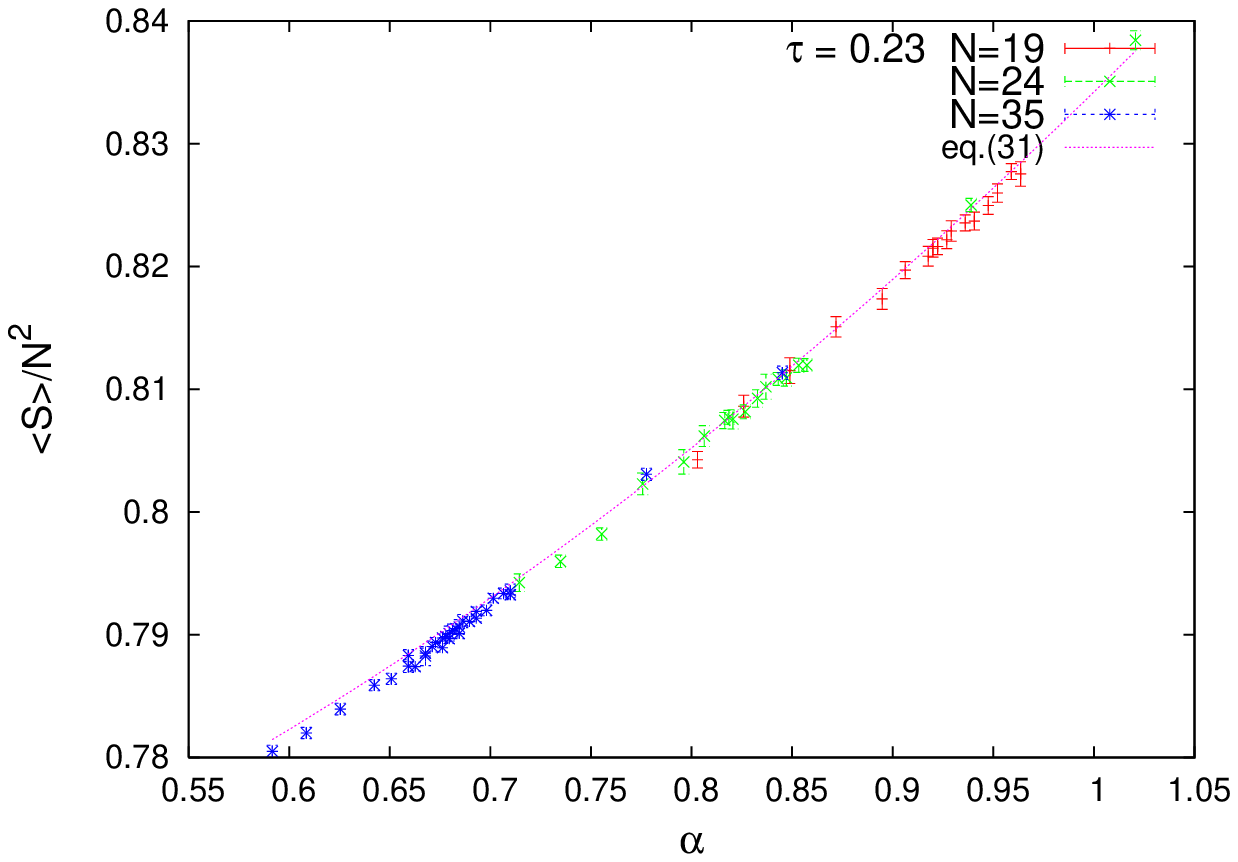}
\includegraphics[width=7.0cm,angle=0]{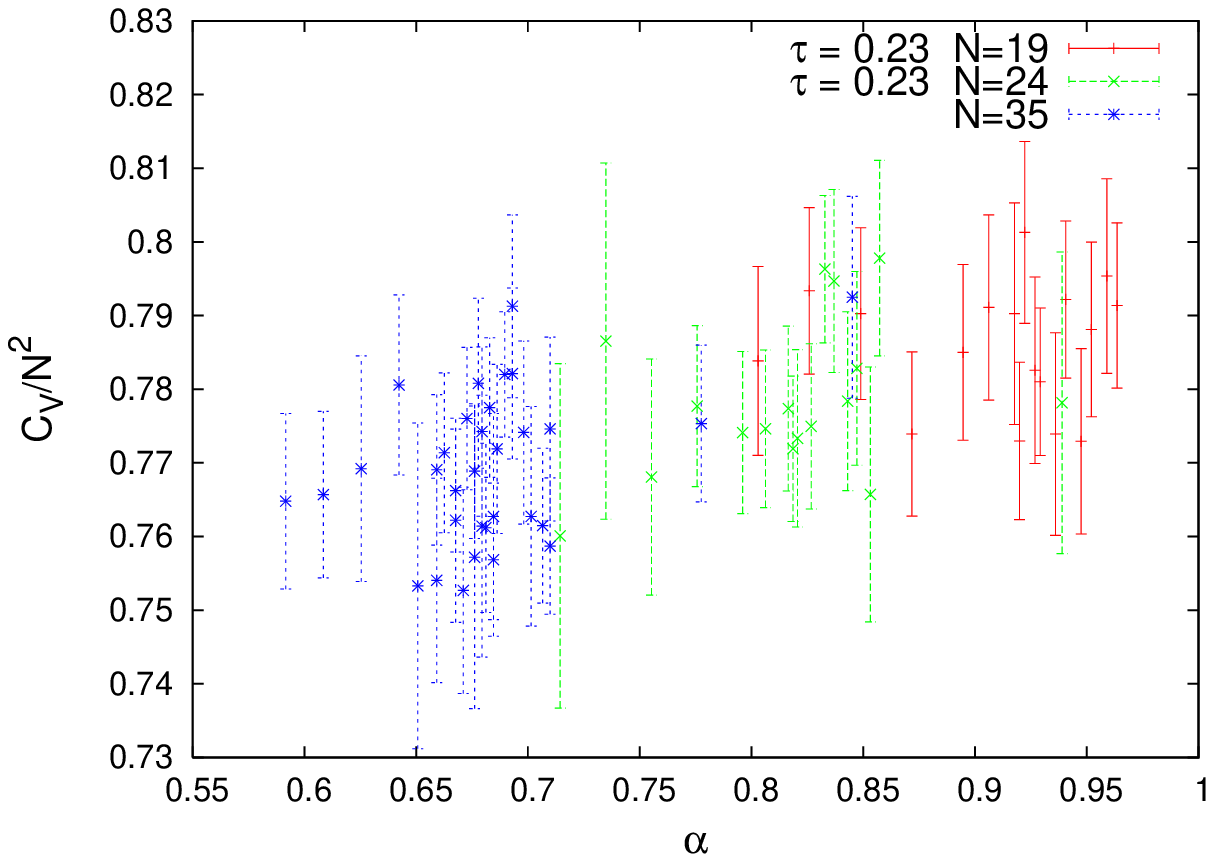}
\includegraphics[width=7.0cm,angle=0]{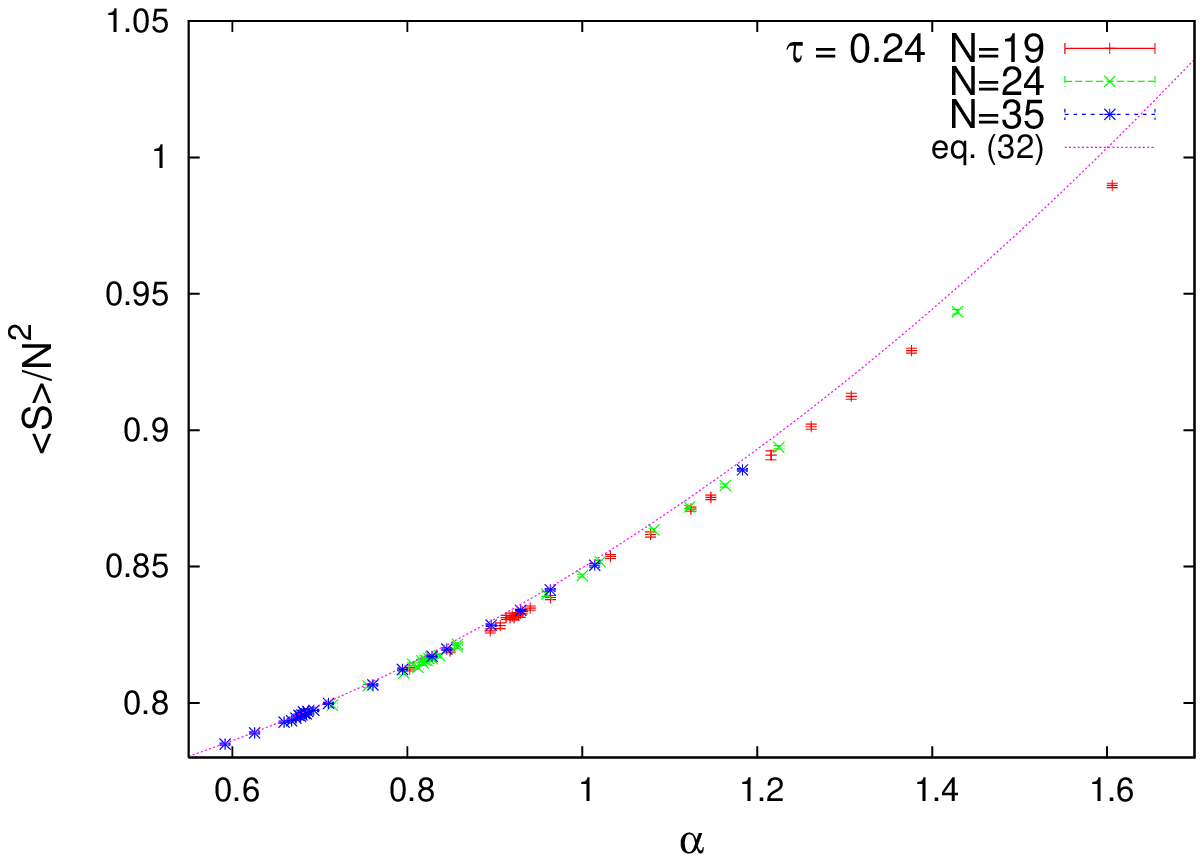}
\includegraphics[width=7.0cm,angle=0]{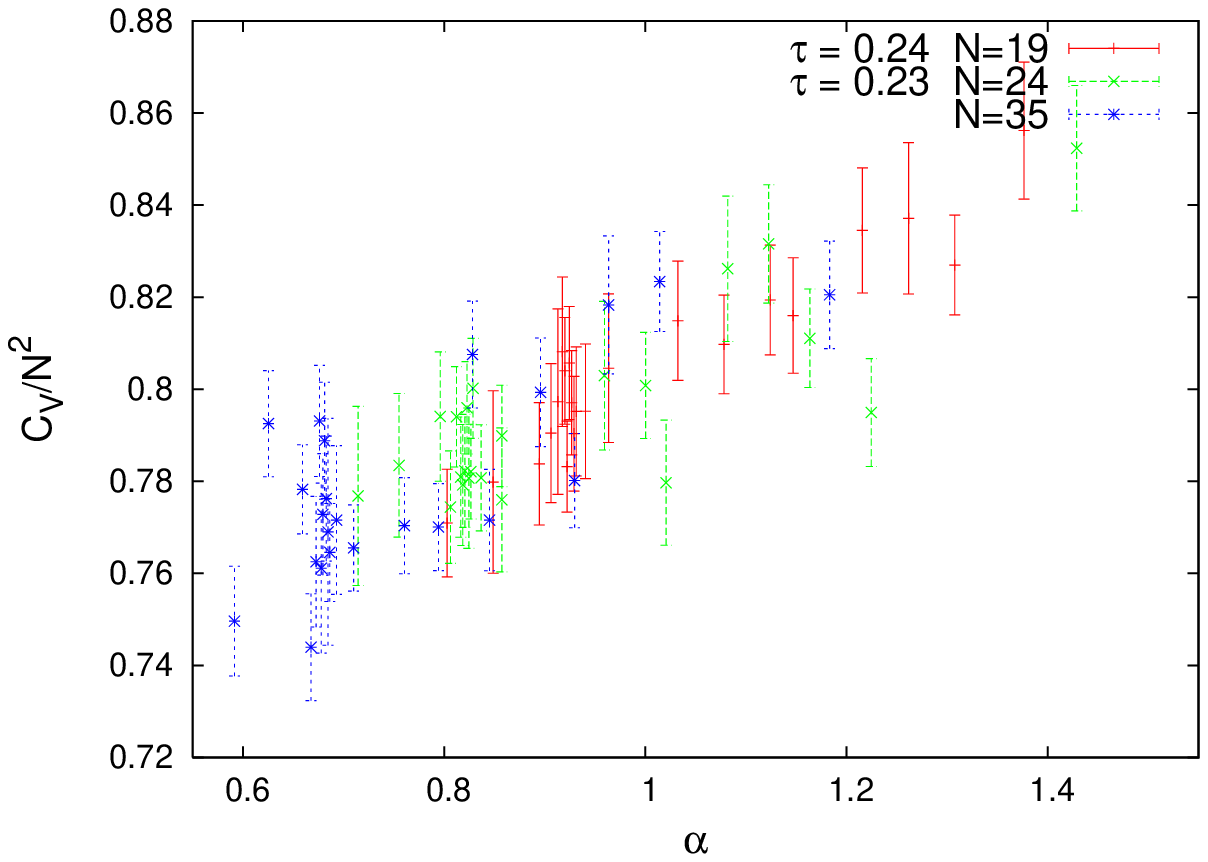}
\caption{Expectation value of the action $<S>$ and specific heat $C_v$ for $\tau=0.23,0.24$ as a function of $\alpha=\tilde\alpha/\sqrt{N}$. These values of $\tau$ lie in the matrix phase.}
\end{center}
\end{figure}

For all values of $\tau<2/9$ we observed that there is a 
discontinuity in the expectation value of the action
$<S>$ across the transition line in Figure 3. This jump in ${\cal S}$ 
is well predicted by the theoretical expression 
\begin{equation}
\Delta{\cal S}=-\tilde\alpha^{4}(-\frac{\tau\phi^2_*}{8}+\frac{\phi^3_*}{24})
\end{equation}
with $\phi_*$ given by equation (\ref{phisol}) evaluated on the transition.
The specific heat is non-analytic along the transition, but the nature
of its non-analyticity differs on the critical line
$\tau=\frac{2}{9}$ and the critical curve
$(\tau<\frac{2}{9},\tilde\alpha_*(\tau))$. In the former case the
specific heat has the standard jump of a 1st order transition while on
the latter it diverges as the transition is approached from the low
temperature or fuzzy sphere phase side of the transition but there is no
observable increase in the specific heat as the transition is
approached from the high temperature side.

To identify the transition we choose that value of $\tilde\alpha$ that
gives the specific heat and identify this as $\tilde{\alpha}_*$.  In
figure 5 we show the curve of the specific heat $C_v$ for fixed values
of $\tau=0.08$ and $\tau=0.15$.  The critical value of the coupling
$\tilde{\alpha}_*$ measured from simulation is $2.40 \pm 0.05$ for
$\tau=0.08$ whereas we found a value of $\tilde{\alpha}_*=2.85 \pm
0.05$ when $\tau=0.15$.  These are to be compared to the prediction
(\ref{critline}) from which we get the values $\tilde{\alpha}_*=2.366$
and $\tilde{\alpha}_*=2.790$ respectively. As we can see the numerical
results are in good agreement with the our theoretical predictions.

Thought the transition is characterised by a latent heat, 
as it was pointed out in \cite{prlROY,longpaper}, it
is unusual in that there are also critical fluctuations in the 
specific heat as the transition is approached from the low temperature
phase for $\tau<\frac{2}{9}$. These lead to a divergent specific heat with specific heat exponent 
$\alpha=1/2$.  Our simulations are consistent with 
the expected divergence in the low temperature phase.

In figure 6, where we show ${\cal S}$ for $\tau=1/4$ and $\tilde\alpha<3.7$.
We see the data collapses when plotted against $\alpha$.  We also show that
the specific heat is fluctuating around $C_v=\frac{3}{4}$.
These are consistent with a uniform distribution of commuting 
$X_a$ whose eigenvalues are distributed within a ball of radius $R=2.0$. 

In figure 7, we show ${\cal S}$ and the specific heat as a function of
$\tau$ for $\tilde\alpha=5.0$ and we observe that the transition occurs
at $\tau=0.219\pm 0.005$ a value consistent with $\tau=\frac{2}{9}$.
We have chosen $\tilde\alpha=5.0$ 
to represent the typical behaviour for 
$\tilde\alpha>\tilde\alpha_*(\tau=2/9)=4.02$, so that 
the curve crosses the critical line $\tau_*=2/9$. 
Our theoretical prediction for the critical value of the specific heat when 
the transition is approached from the low temperature phase, eq.
(\ref{cvth}) predicts $C_v(\tau=2/9,\tilde\alpha=5)=1.02$, 
which is in agreement with our simulations.
Figure 7 also shows that ${\cal S}$ has a bend but there is no apparent, jump.
It also appears that the bend occurs at a slightly larger value than 
$\tau=\frac{2}{9}$. The precision of our numerical results show that the bend 
occurs at $\tau=0.222\pm 0.002$, showing small deviations when $\tilde\alpha$ is close (and larger) to $\tilde\alpha_*(\tau=2/9)=4.02$ where the transitions take place at slightly larger values than $\tau=\frac{2}{9}$ (see figure 3), however this might be due to finite size effects.
The specific heat shows there are no strong fluctuations, 
so the classical theory should be a good approximation
and in fact predicts the critical point $\tau_*=2/9$ see eq.
(\ref{critline}). The finite size effects 
in the specific heat are consistent (\ref{finitesizecorrectionstoS})
and account for the deviation from the limiting large $N$ value of $C_v=0.75$.

\subsection{Eigenvalue distributions}

By measuring the eigenvalue distribution of the matrices $X_a$ and the
commutators $i[X_a,X_b]$ we can investigate in more detail the
features of the configurations at quantum level. The model has global
$SO(3)$ invariance, therefore each of the three matrices has the same 
eigenvalue distribution and also each of the commutators share a 
common distribution.  We also measure the eigenvalues for the
$2N\times 2N$ matrix
\begin{equation}
{\mathbb C}=\sigma_a X_a
\end{equation}
where $\sigma_a$ are the three Pauli matrices. This matrix is
particularly useful because it encodes information from the three
matrices $X_a$ simultaneously and its spectrum can be easily computed
for specific configurations. For instance, when $X_a=\alpha L_a$,
where $L_a$ are IRRs of $SU(2)$, with $L_aL_a=\frac{N^2-1}{4}$, we have
\begin{eqnarray}
(\sigma_aL_a)^2&=&L^2_a-\sigma_aL_a \;\;\;\mbox{then}\qquad
(\sigma_aL_a+\frac{1}{2})^2=\frac{N^2}{4}\mathbf{1}
\end{eqnarray}
and we find that with ${\mathbb C}\to U^{-1}{\mathbb C}_{diag}U$ 
with $U\in U(2N)$ and noting that ${\mathbb{C}}$ is 
traceless and is the $SU(2)$ tensor product 
$2\otimes N= (N+1)\oplus(N-1)$ we see that the eigenvalue multiplicities 
are $N-1$ for $-\frac{N}{2}-\frac{1}{2}$ and $N+1$ 
for $\frac{N}{2}-\frac{1}{2}$ therefore
\begin{equation}\label{espC-X}
{\mathbb{C}}_{diag}=\alpha\left(\frac{(N-1)}{2}\mathbf{1}_{N+1}\oplus\frac{(-N-1)}{2}\mathbf{1}_{N-1}\right).
\end{equation}
More generally for a reducible representation $J_a$ of (\ref{classpotmu})
we have
\begin{equation}\label{espC-reducible}
{\mathbb{C}}_{diag}=\alpha\oplus_{i}\left(\frac{(n_i-1)}{2}\mathbf{1}_{n_i+1}\oplus\frac{(-n_i-1)}{2}\mathbf{1}_{n_i-1}+{\bf 0}\right).
\end{equation}
where ${\bf 0}$ corresponds to the one dimensional representations in $J_a$.
We see that ${\mathbb{C}}$ is sensitive to the representation 
content of the matrices. The spectrum of ${\mathbb{C}}$ will have distinct 
eigenvalues for the different $SU(2)$ IRRs present in $X_a$.

We also measure the eigenvalues of the Dirac operator
\begin{eqnarray}
\mathbb{D}:=\sigma_a[X_a,\;\cdot]\label{Dirac},
\end{eqnarray}
where $[X_a,\cdot]$ means $X_a$ acting as commutator:
$[X_a,M]=X_aM-MX_a$ for any matrix $M\in Mat_N$.

The spectrum of $\mathbb{D}=\sigma_a[X_a,\cdot]$ can be easily
computed when $X_a=\alpha L_a$. For this particular case we have
\begin{eqnarray}
\mathbb{D} =\sigma_a[X_a,\;\cdot]= \alpha\;\sigma \cdot\mathcal{L}=\alpha\left[{\cal J}_a^2-{\cal L}_a^2-\frac{3}{4}\right]=\alpha\left[j(j+1)-l(l+1)-\frac{3}{4}\right],
\end{eqnarray}
with $j=l\pm \frac{1}{2}$ with $l$ taking the values $l=0,1,..,N-1$
and we obtain the spectrum
\begin{eqnarray}
spec\{\mathbb{D}\}&=&\alpha\left\{\begin{array}{rcl}
            l; &j=l+\frac{1}{2}, & g(l)=2(l+1) \\
           -(l+1); &j=l+\frac{1}{2}, &g(l)=2l \\
            0; & l=0,& g(l)=2\end{array}\right.
\end{eqnarray}
$g(l)$ is the degeneracy. 
The Dirac operator $\mathbb{D}$ has therefore the spectrum 
\begin{eqnarray}\label{espDirac}
spec\{\mathbb{D}\}=\alpha \{-N,-(N-1),\dots,-3,-2,0,1,2,3,\dots,(N-1)\}. 
\end{eqnarray}
This spectrum is well reproduced in our numerical simulations for the
parameter range of the phase diagram Figure 3 corresponding to the 
fuzzy sphere, see Figure 10.

\subsection{Numerical results for eigenvalue distributions}

\subsubsection{The $\tau=0$  case}
We begin with our results for the case $\tau=0$. This situation
corresponds to the model studied in
\cite{nishi:numeS2,prlROY,longpaper} and the system crosses the phase
boundary of Figure 3 at $\tilde{\alpha}=2.08$ as predicted from the
effective potential or $\tilde{\alpha}=2.01\pm 0.01$ as measured from
simulations.

In the low temperature phase (large $\tilde{\alpha}$), 
the dominant configurations are IRRs of $SU(2)$, $D_a\sim L_a$. 
Figure 9 shows the eigenvalue distribution for $N=24$ and 
$\tilde{\alpha}=5.00$ and $\tau=0$. 
The simulation indeed shows that the spectrum of both $D_3$ and $i[D_1,D_2]$ 
are discrete and equally spaced consistent with an IRR of $SU(2)$.

In figure 10, the eigenvalues for the matrix $C=\sigma_aD_a$ and the
Dirac operator ${\cal D}=\sigma_a[D_a,\cdot]$ are shown\footnote{Note:
  We have defined $\mathbb{C}=\alpha {\cal C}$ and 
  $\mathbb{D}=\alpha {\cal D}$. Also, the operator ${\cal D}$ 
  differs from the standard Dirac operator \cite{bal-lect} on the 
  sphere, this would now be ${\cal D}+\phi$.} 
and are consistent with the spectrum (\ref{espC-X})
for matrix ${\mathbb{C}}$, and (\ref{espDirac}) for $\mathbb{D}$.

\begin{figure}[ht]
\begin{center}\label{fig:XandComm-mu0}
\includegraphics[width=7.0cm,angle=0]{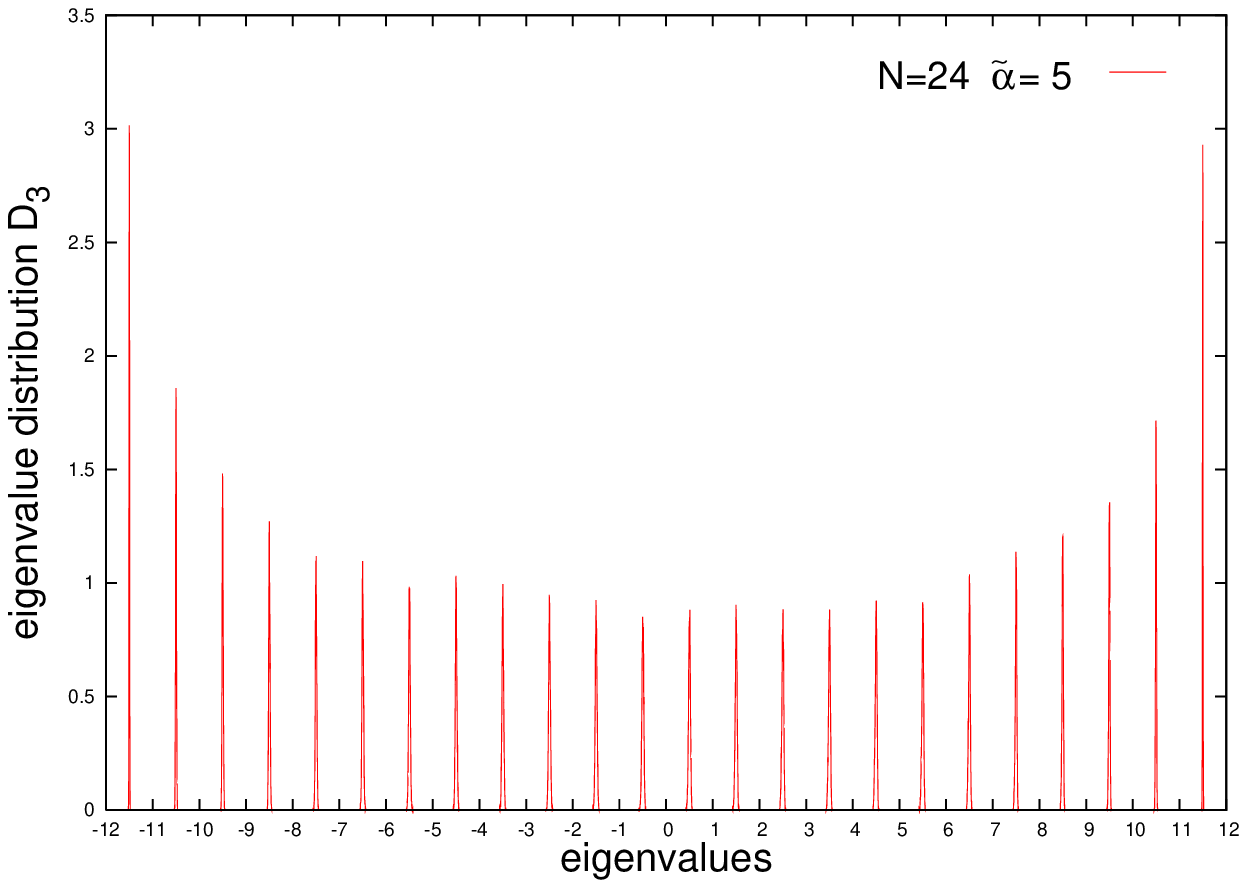}
\includegraphics[width=7.0cm,angle=0]{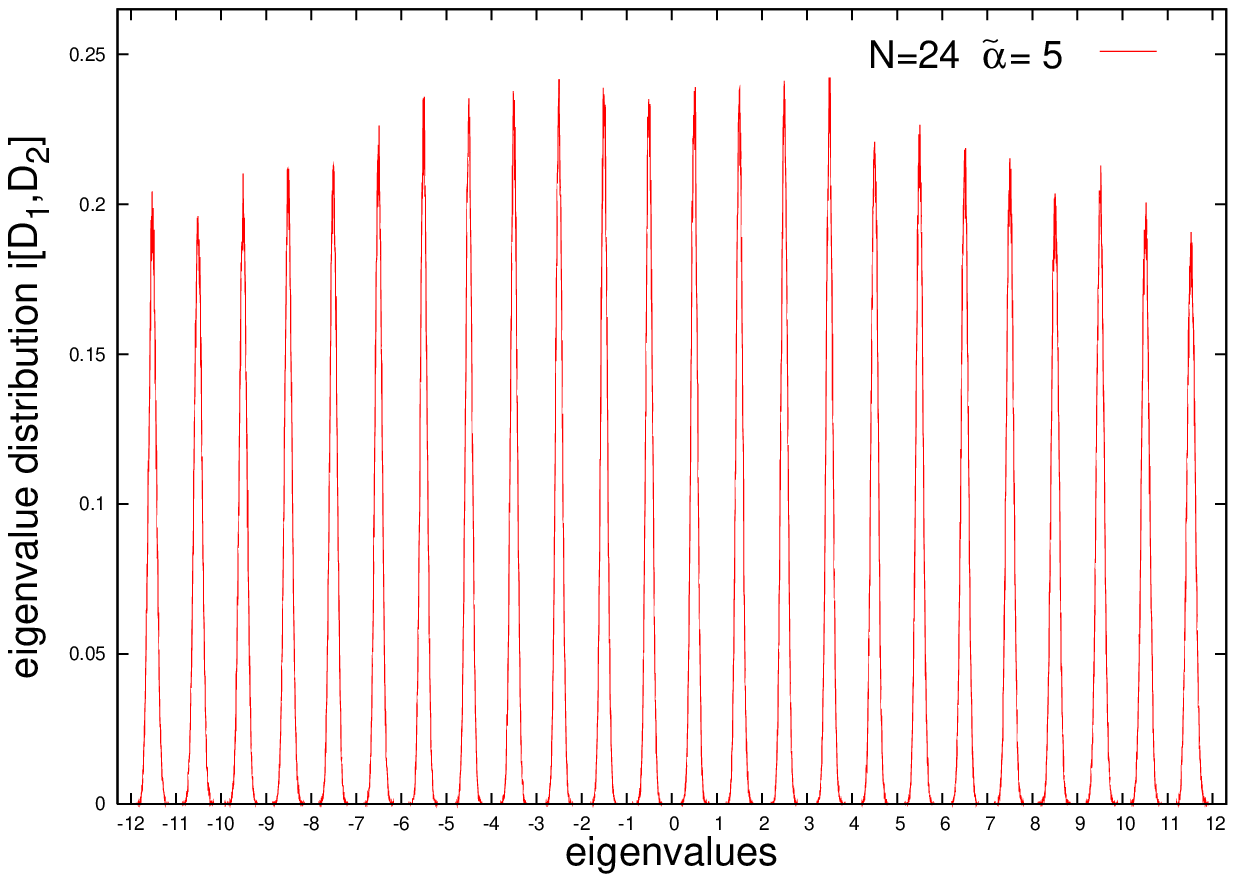}
\caption{Eigenvalues for $D_3$ and $i[D_1,D_2]$. We observe a discrete spectrum. $N=24$, $\tilde{\alpha}=5.00$, $\tau=0.0$.}
\end{center}
\end{figure}

\begin{figure}[ht]
\begin{center}\label{fig:CandDirac-mu0}
\includegraphics[width=7.0cm,angle=0]{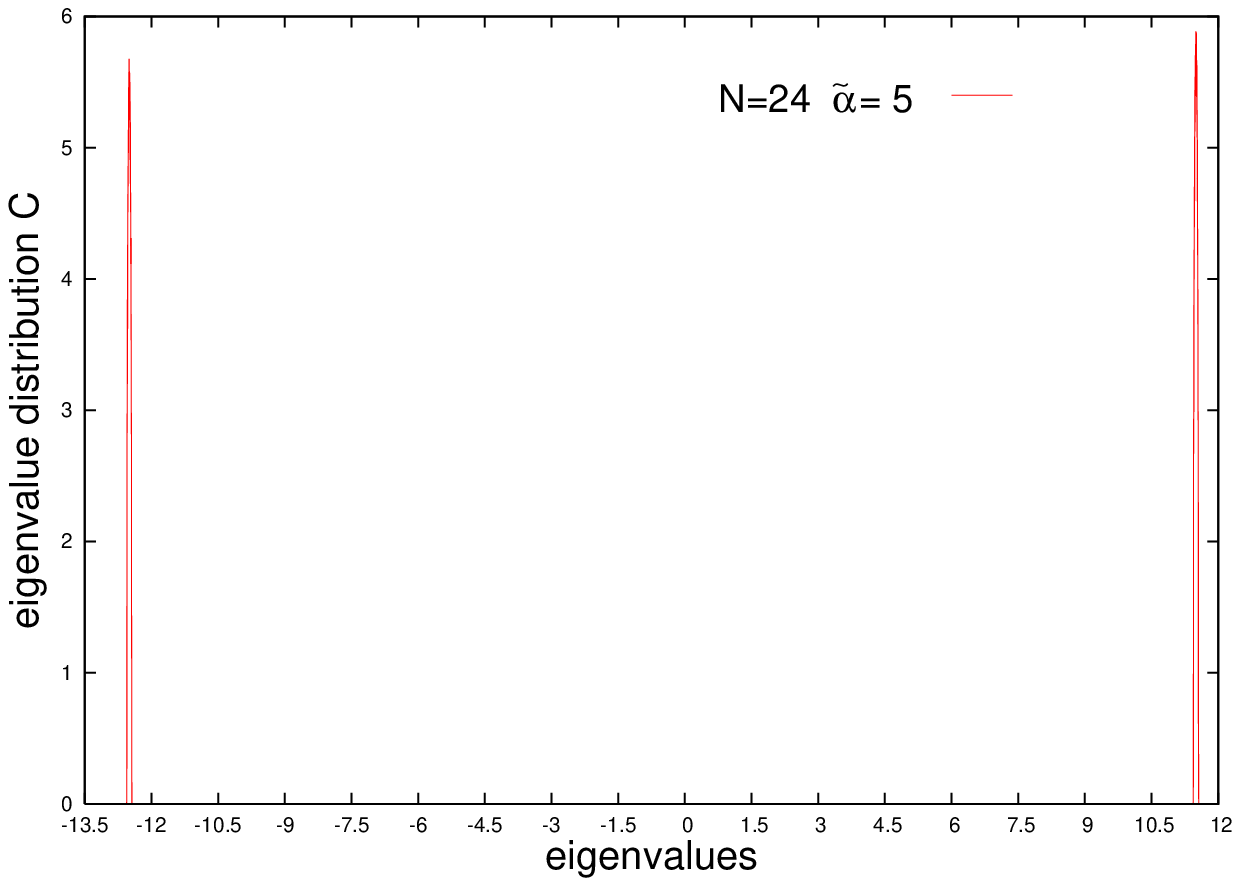}
\includegraphics[width=7.0cm,angle=0]{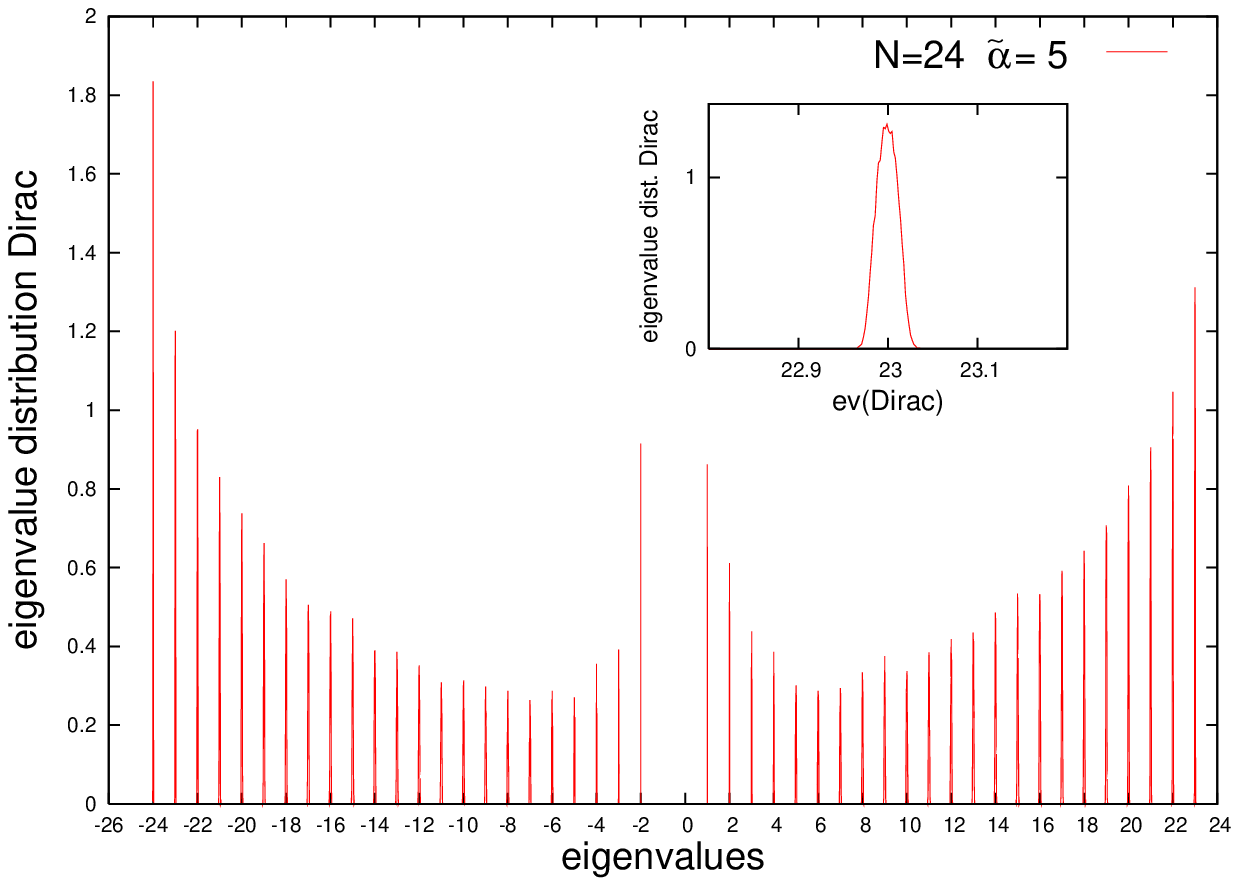}
\caption{Eigenvalues for matrix $C=\sigma_aD_a$ and Dirac operator ${\cal D}=\sigma_a[D_a,]$. $N=24$, $\tilde{\alpha}=5.00$, $\tau=0.0$. The eigenvalues for the matrix $C$ are nicely distributed in two peaks around $C=-\frac{25}{2}=-12.5$ and $C=+\frac{23}{2}=11.5$ as expected. While the operator ${\cal D}$ 
reproducing the spectrum (\ref{espDirac}).} 
\end{center}
\end{figure}

In the matrix phase, we observe a continuous spectrum for the matrices
$X_a$; we also observe that the eigenvalue distribution for $X_a$ has
an increasing number of oscillations and becomes a smooth convex
spectrum for large $N$ and as might be expected from
(\ref{actionXalphatilde}) the spectrum of $X_a$ is largely independent
of $\tilde\alpha$. The eigenvalue distribution of each $X_a$ is
symmetric around zero and its support is localised within the interval
$[-2,2]$ and the parabola law (\ref{parabola}) fits excellently with
$R=2.0$.  The distribution for the commutator is also symmetric around
zero.

The matrix $\mathbb{C}=\sigma_aX_a$ shows a distribution 
with two maxima around $\pm 2$. The distribution, however, 
is not symmetric with the peak on the right hand slightly higher than the 
one on the left. There is a small effect of non-zero $\tilde\alpha$ which 
is larger than any effect on $X_a$ or $i[X_a,X_b]$. 
The distribution for the operator $\mathbb{D}$ has three peaks for small $N$,
with the central peak around zero disappearing as $N$ is increased.
The distribution also shows a slight distortion with the right hand peak larger 
than the left. Numerical results are shown in figure 11 and 12 for $N=24,35$ with $\tilde{\alpha}=1.00$, in figure 12 we also show the case $N=24$ with $\tilde{\alpha}=0.60$ in order to compare.

\begin{figure}[ht]
\begin{center}
\includegraphics[width=7.0cm,angle=0]{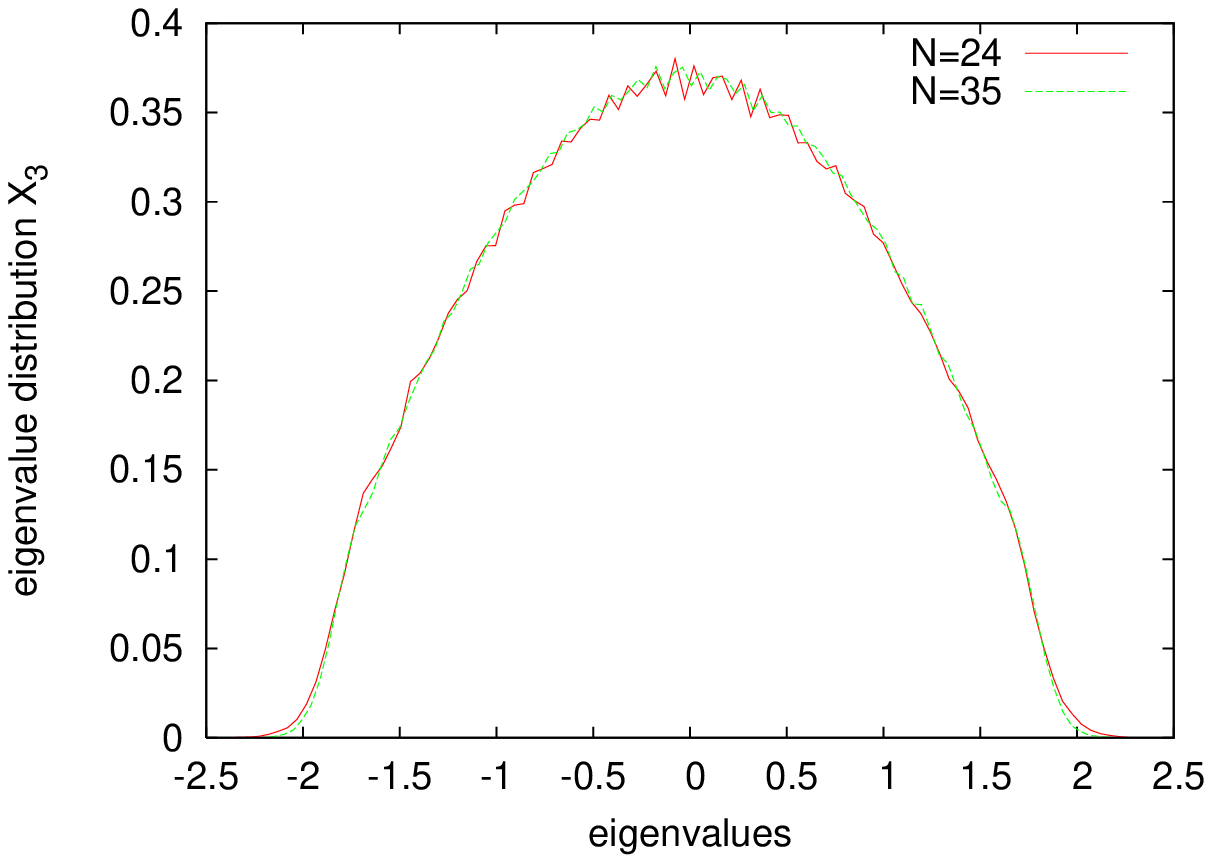}
\includegraphics[width=7.0cm,angle=0]{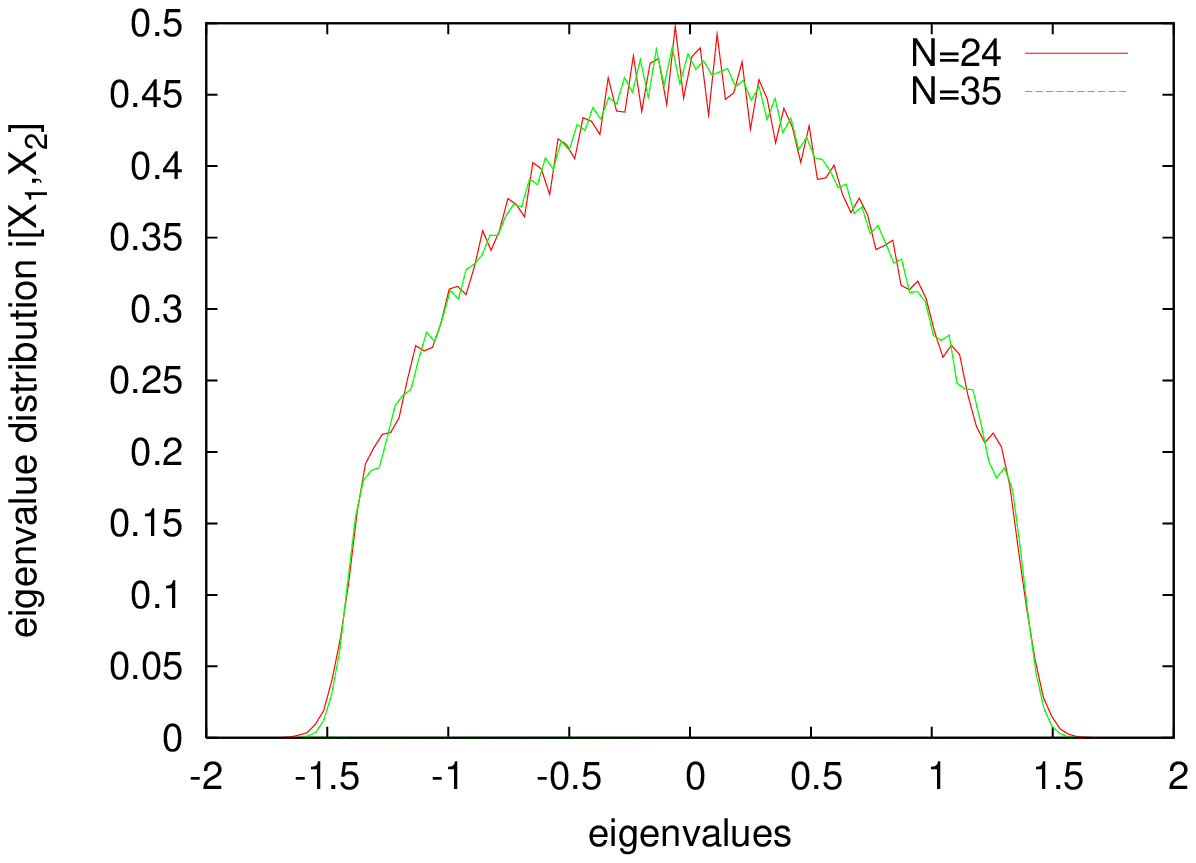}
\caption{Eigenvalues for $X_3$ and $i[X_1,X_2]$ in matrix phase. $N=24$, $\tilde{\alpha}=1.00$, $\tau=0.0$.}
\end{center}
\end{figure}

\begin{figure}[ht]
\begin{center}
\includegraphics[width=7.0cm,angle=0]{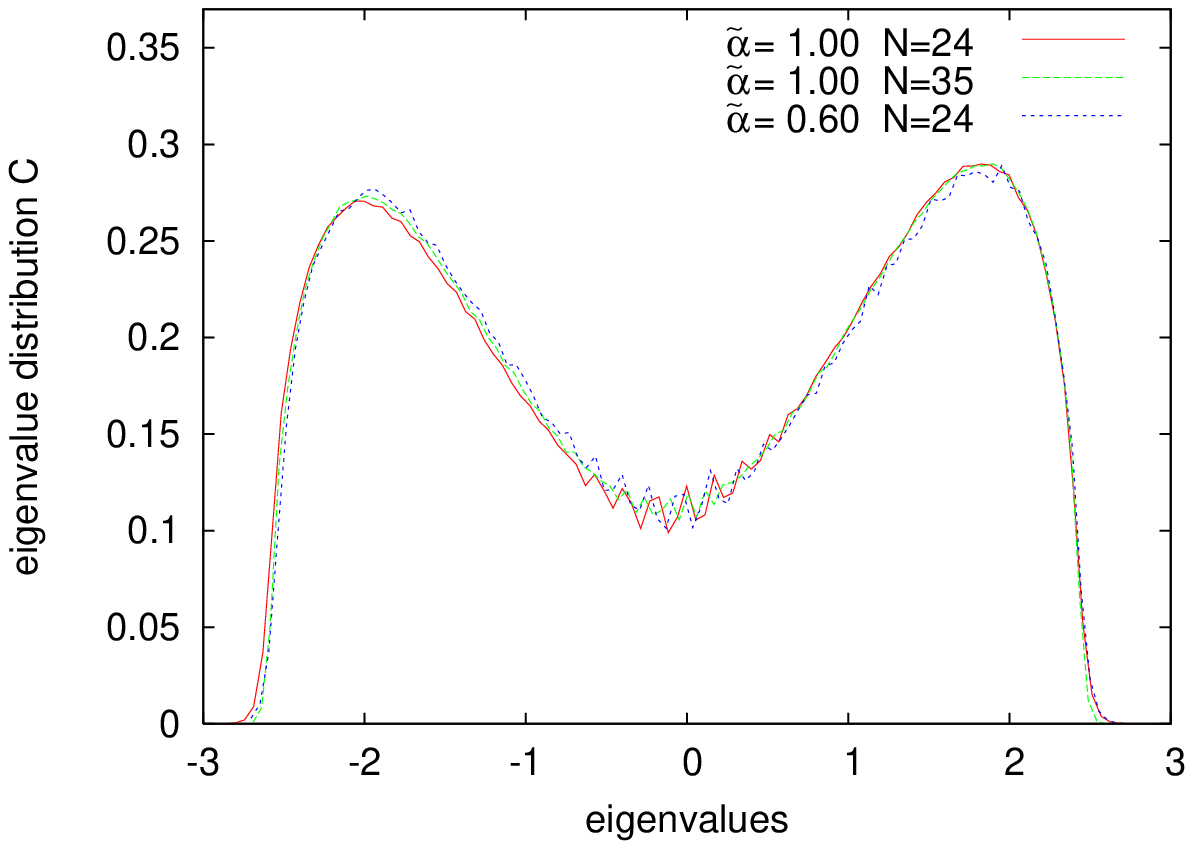}
\includegraphics[width=7.0cm,angle=0]{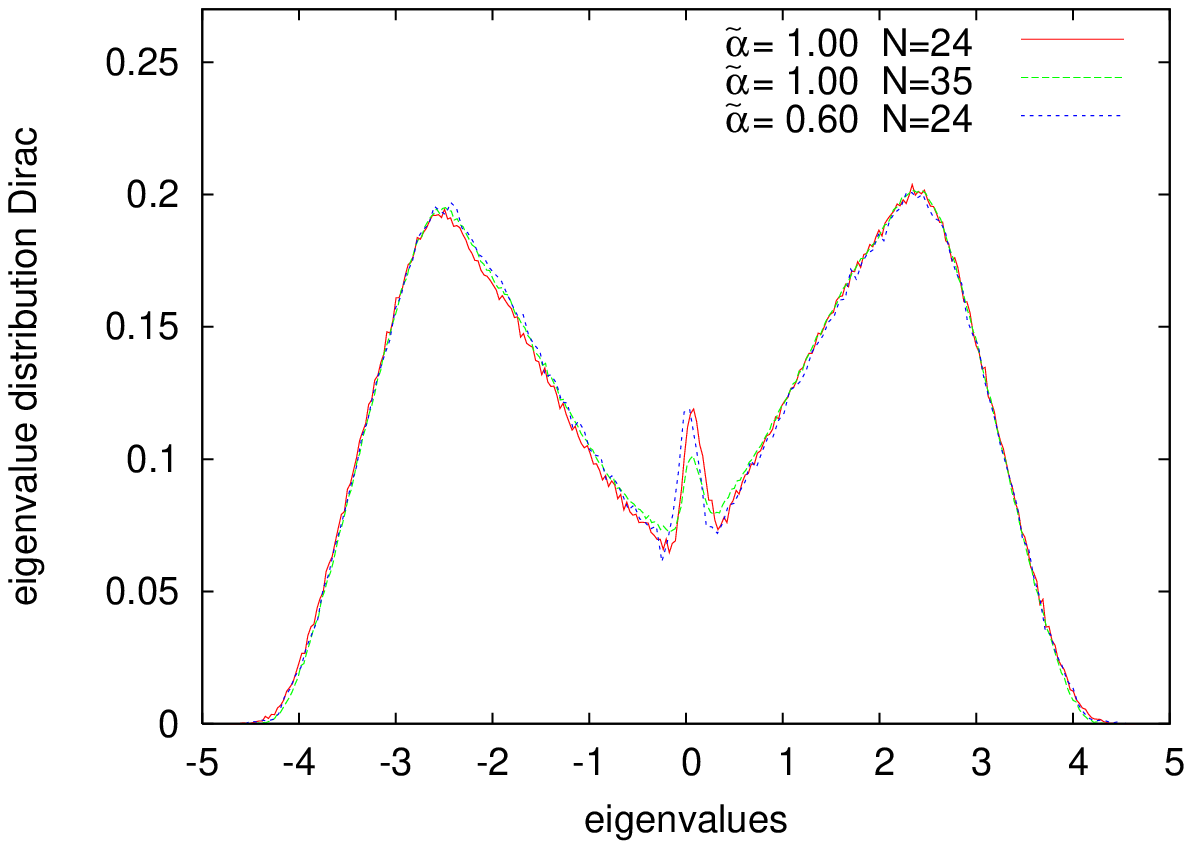}
\caption{Eigenvalues for matrix $\mathbb{C}=\sigma_aX_a$ and Dirac operator $\mathbb{D}=\sigma_a[X_a,]$ in the matrix phase $N=24$, $\tilde{\alpha}=0.60, 1.00$, and $N=35$, $\tilde{\alpha}=1.00$ for $\tau=0.0$. The eigenvalue distributions for the matrix $\mathbb{C}$ and operator $\mathbb{D}$ are asymmetric} 
\end{center}
\end{figure}

We highlight the special case of both $\tau=\alpha=0$ which we refer to
as the pure Yang-Mills matrix model and show the distributions of
$X_a$, $i[X_a,X_b]$, in figure 13 and those of ${\mathbb{C}}$ and
${\mathbb{D}}$ in figure 14. These are all symmetric and consistent
with our interpretation of the matrix phase as fluctuations around
commuting matrices whose joint eigenvalue distribution is a solid ball
of radius $R=2.0$.

\begin{figure}[ht]
\begin{center}
\includegraphics[width=7.0cm,angle=0]{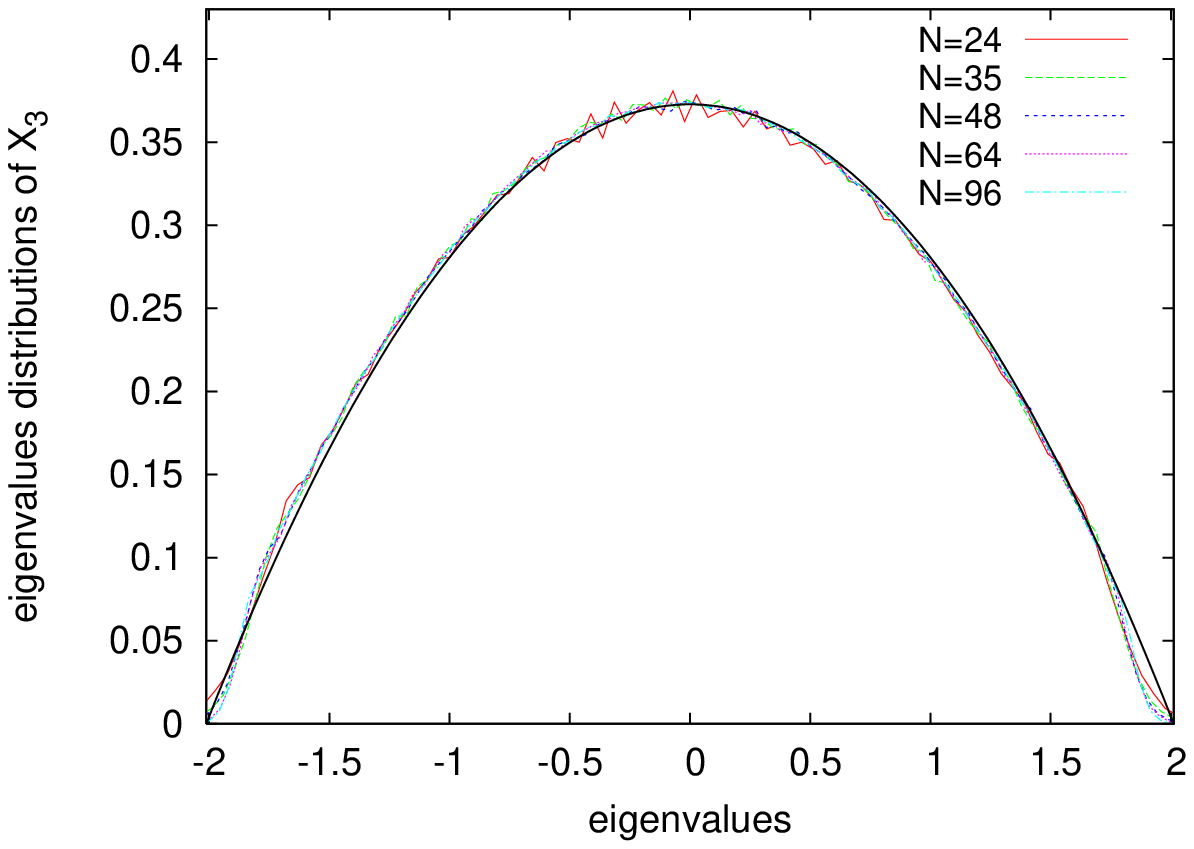}
\includegraphics[width=7.0cm,angle=0]{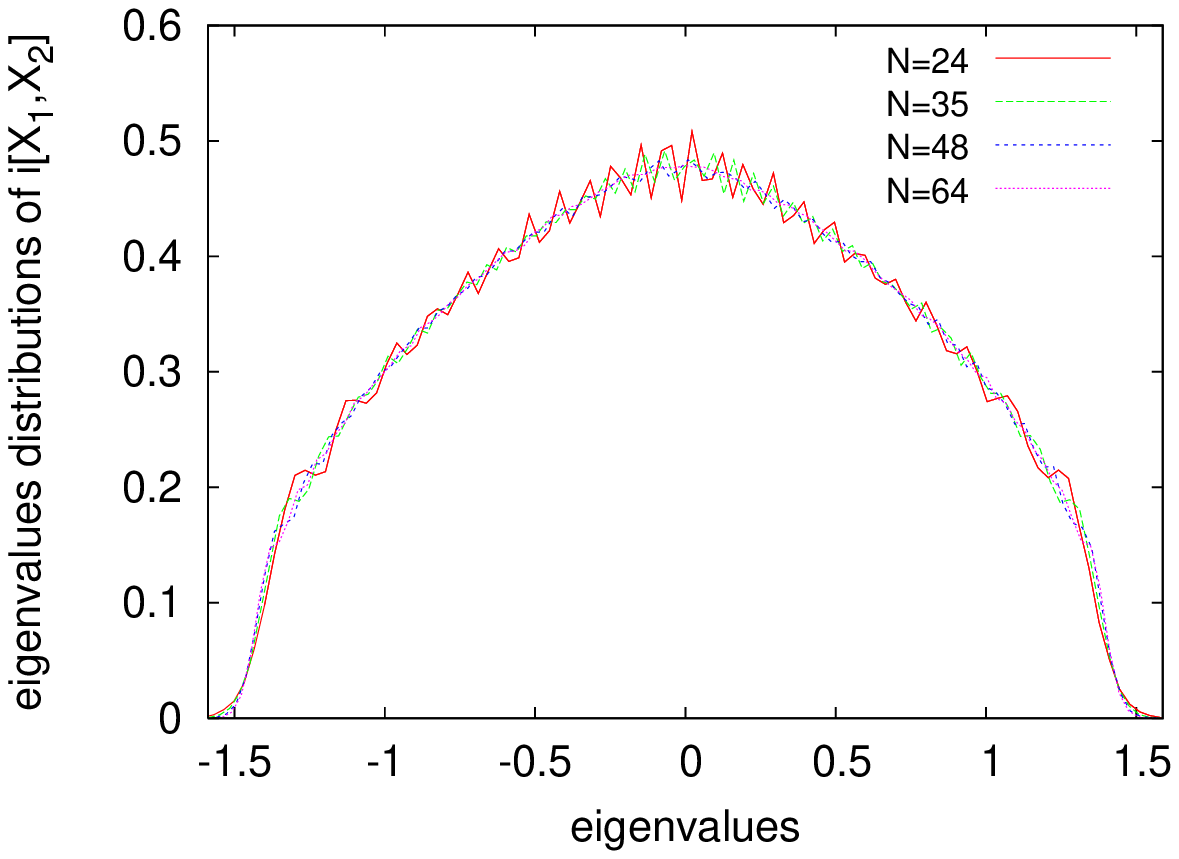}
\caption{Eigenvalues for $X_3$ (left) and $i[X_1,X_2]$ (right) for pure Yang-Mills matrix model for different $N$. The eigenvalue 
distribution of $X_3$ is fit by the parabolic distribution (\ref{parabola}) with $R=2.0$ (solid line) and is consistent with a background of commuting matrices whose 
eigenvalues are uniformly distributed 
inside a solid 3-ball of radius $R$.}
\end{center}
\end{figure}

The eigenvalue distribution for $\mathbb{C}=\sigma_aX_a$ has two peaks at
approximately $\pm 1.9$. In the large $N$ limit the Dirac operator
$\mathbb{D}$, has two peaks peaks located at approximately $\pm2.4$
and support of its spectrum lies in the interval $[-4,4]$.

\begin{figure}[ht]
\begin{center}
\includegraphics[width=7.0cm,angle=0]{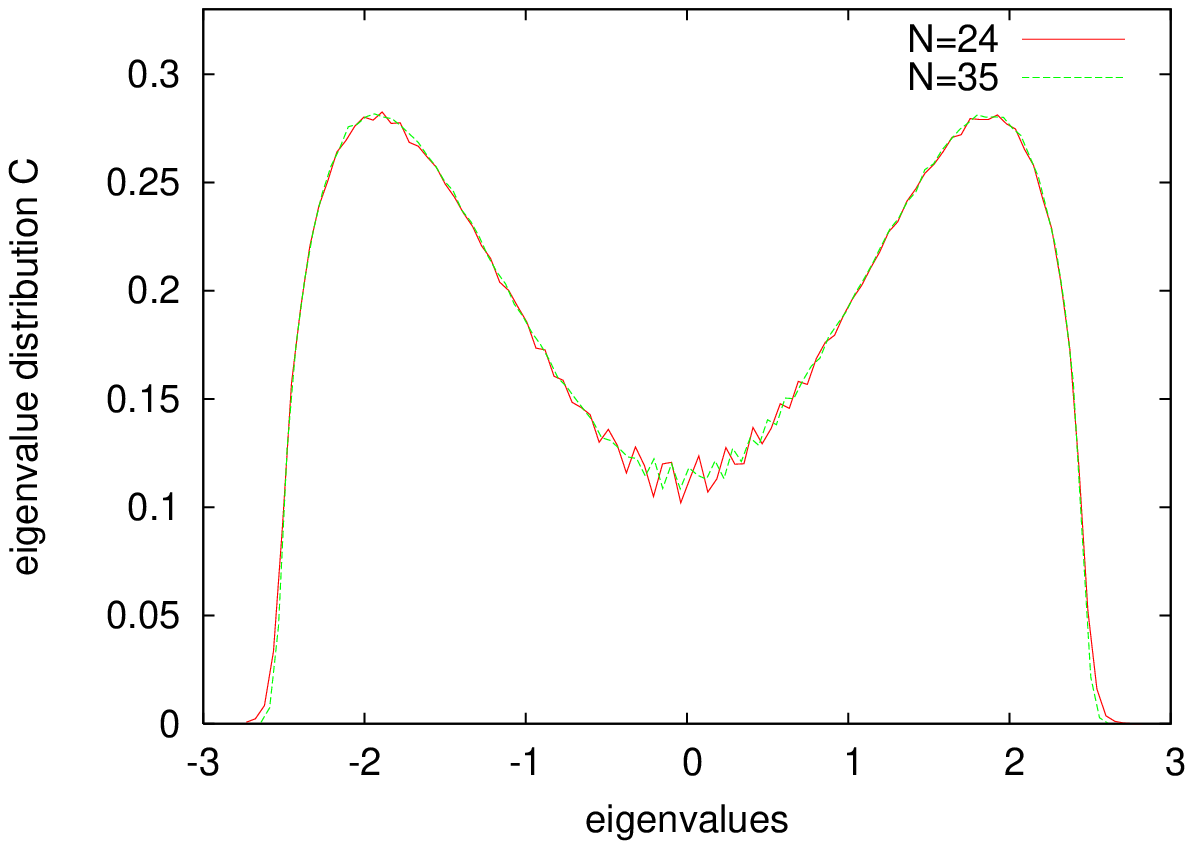}
\includegraphics[width=7.0cm,angle=0]{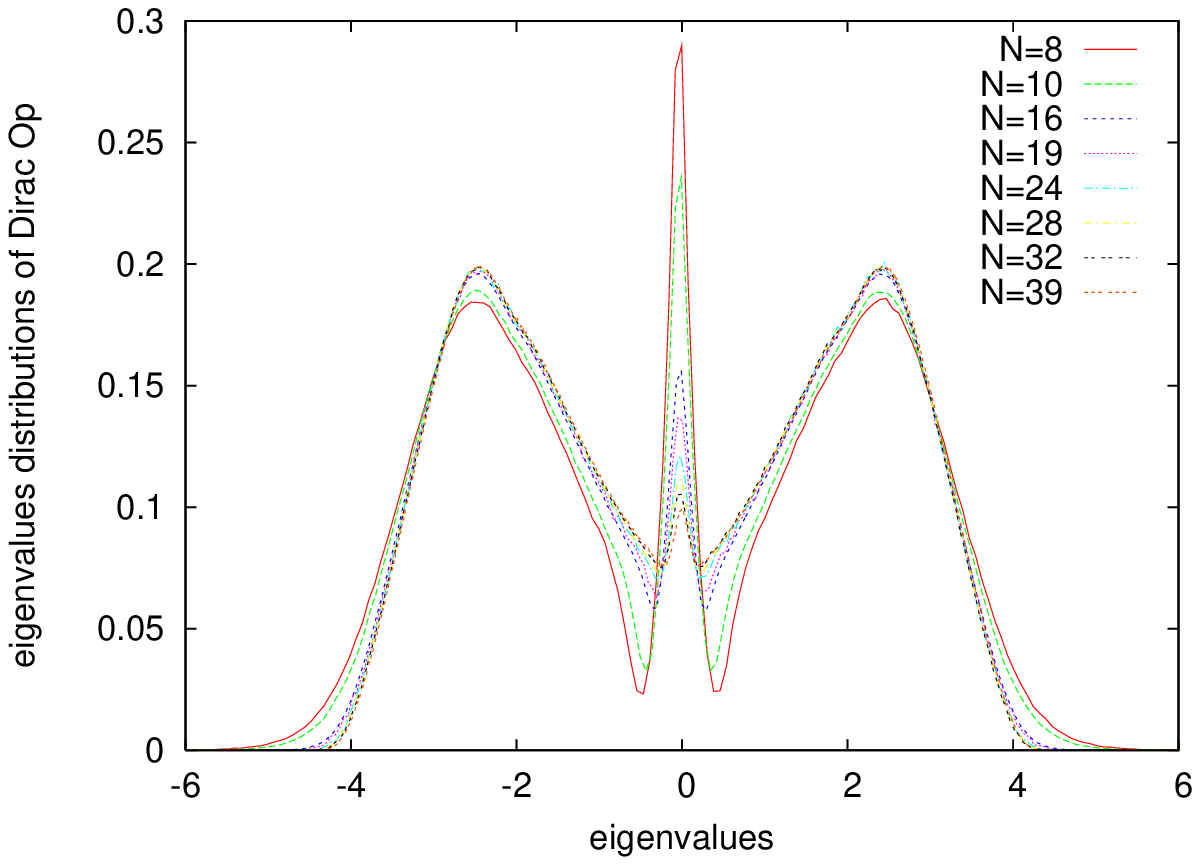}
\caption{Eigenvalues for matrix $\mathbb{C}=\sigma_aX_a$ and the Dirac operator $\mathbb{D}=\sigma_a[X_a,]$ for the pure Yang-Mills matrix model. Both distributions are symmetric.} 
\end{center}
\end{figure}

\subsubsection{The $\tau\neq0$ case }

We again focus on the eigenvalue distributions of the observables; $D_a$, $i[D_a,D_b]$, $C=\sigma_aD_a$ and the operator Dirac ${\cal D}$ in the fuzzy sphere
phase and $X_a$, $i[X_a,X_b]$, $\mathbb{C}$ and $\mathbb{D}$ in 
the matrix phase.  By a close inspection of the spectrum of these matrices 
we confirm that the numerical results are in good accord with 
$\phi$ as predicted by equation (\ref{phisol}) in the fuzzy sphere phase as we will now describe. 

\paragraph{Fuzzy sphere phase.}
The eigenvalue distributions in the region of the parameter space in
which the fuzzy sphere solution $D_a=\phi L_a$ exists is delimited by
the critical line (\ref{critline}) and defined for
$\tilde{\alpha}>\tilde{\alpha}_*$ whenever\footnote{The model is
  unstable for $\tau<0$, yet since tunneling is suppressed in the large
  $N$ limit and the fuzzy sphere phase is in fact stable.}
$\tau<2/9$. The effect of $\tau$ is to increase the critical point
$\tilde{\alpha}_*$ according to (\ref{critline}), in other words, the
critical temperature at which the transition occurs, is lowered. For a
specific set of parameter $N,\tilde{\alpha}$ we can measure the value
for $\phi$ for a fixed value $\tau$.  Knowing that in this phase the
ground state is an IRR of $SU(2)$, i.e. $D \sim \phi
\;diag(-s,-(s-1),\cdots,+(s-1),+s)$ (with $s$, the spin labeling the
IRR: $s=\frac{N-1}{2}$) we can extract the value of $\phi$ measured in
simulations since
$\phi=\frac{1}{s}\;\mbox{eigenvalues($D$)}(=\frac{\sqrt{N}}{\tilde{\alpha}s}\;\mbox{eigenvalues($X$)})$.
For instance, in Figure 15(a) for $N=37$, $\tilde{\alpha}=5$,
$\tau=0.18$ using the largest eigenvalue, the simulation gives
$\phi=0.7539$ while the prediction (\ref{phisol}) gives $\phi=0.75$.

Selecting the first positive and negative eigenvalues of $D_a$, which
should correspond to $ev(L_3)=\pm1$, we obtain
$\mbox{ev}(D_a)_{1}=0.73$ and $\mbox{ev}(D_a)_{-1}=-0.77$, (see figure
15(b)).  Taking the average modulus we can get an estimate for $\phi$
of $\phi=0.75$ while the predicted value from eq.(\ref{phisol}) is
$\phi=0.7537$.  We can now replot $D_a/\phi$ and the operator
$\mathbb{D}/\phi$ in figure 16; we can observe that the configurations
are indeed around $D_a/\phi \sim L_a$.  Our numerical results are in
excellent agreement with the analytical predictions.

\begin{figure}[ht]
\begin{center}\label{fig:FS-mu-a}
\includegraphics[width=7.0cm,angle=0]{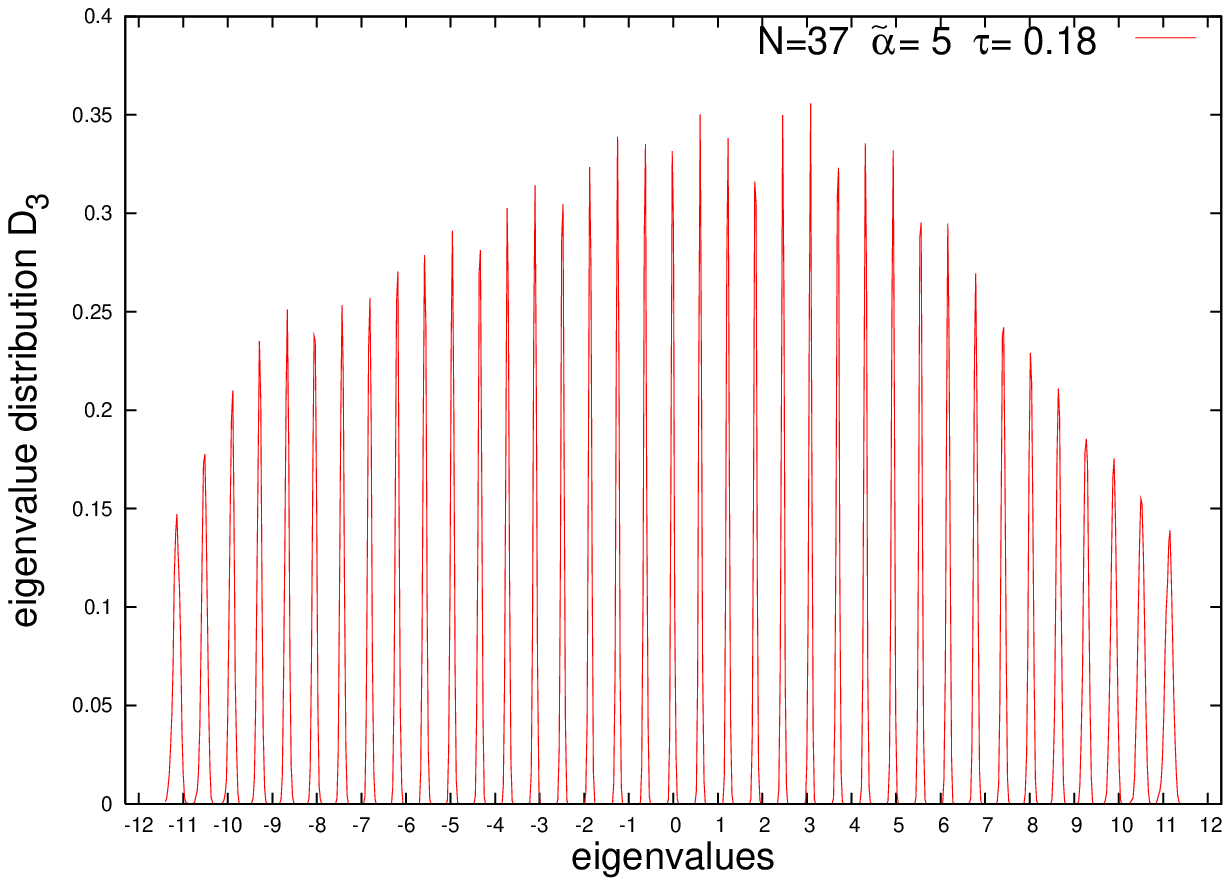}
\includegraphics[width=7.0cm,angle=0]{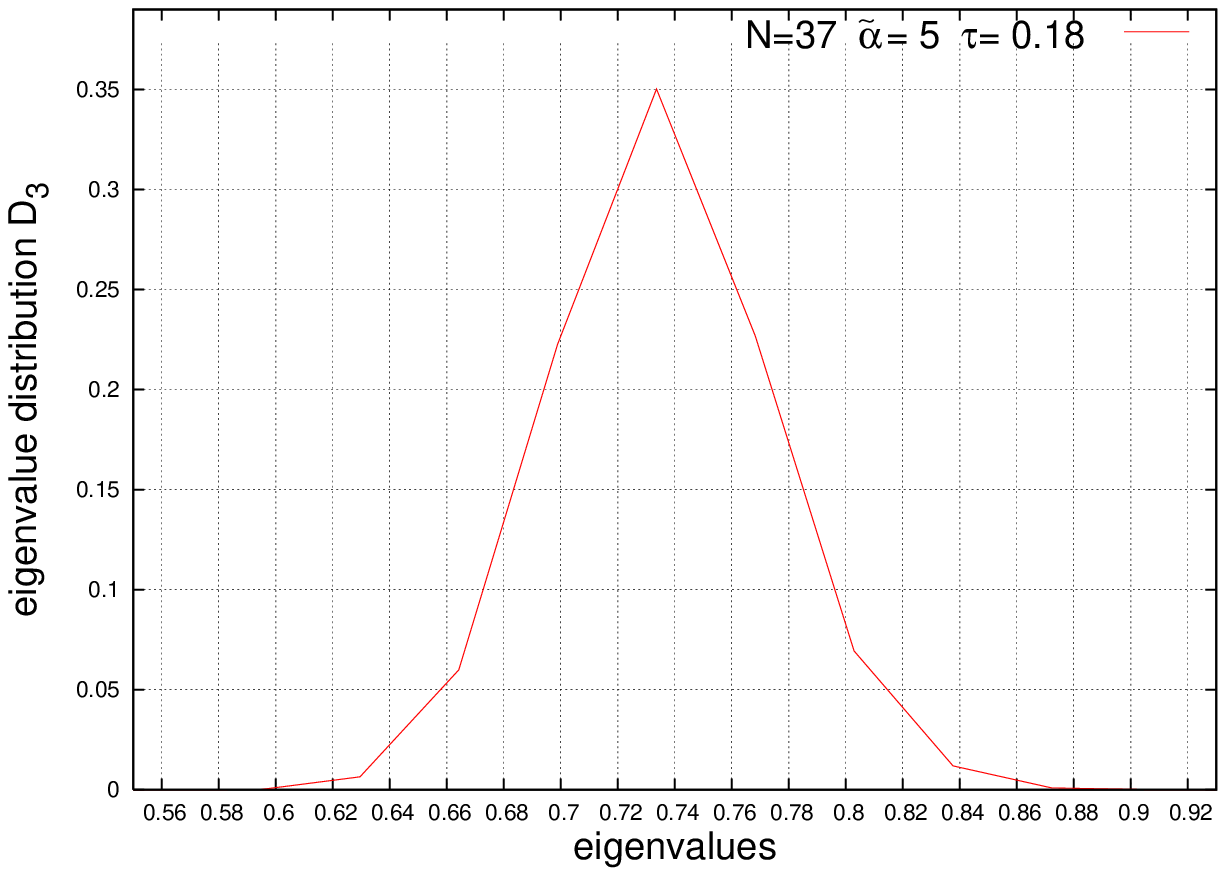}
\caption{(a) Eigenvalues for matrix $D_a$ deep inside the fuzzy sphere region $\tilde{\alpha}=5$ and $\tau=0.18$. (b) On the right a zoom for the first peak on the right side of zero eigenvalue of $D_a$ from which the value for $\phi$ is measured.}
\end{center}
\end{figure}

\begin{figure}[ht]
\begin{center}\label{fig:FS-mu-b}
\includegraphics[width=7.0cm,angle=0]{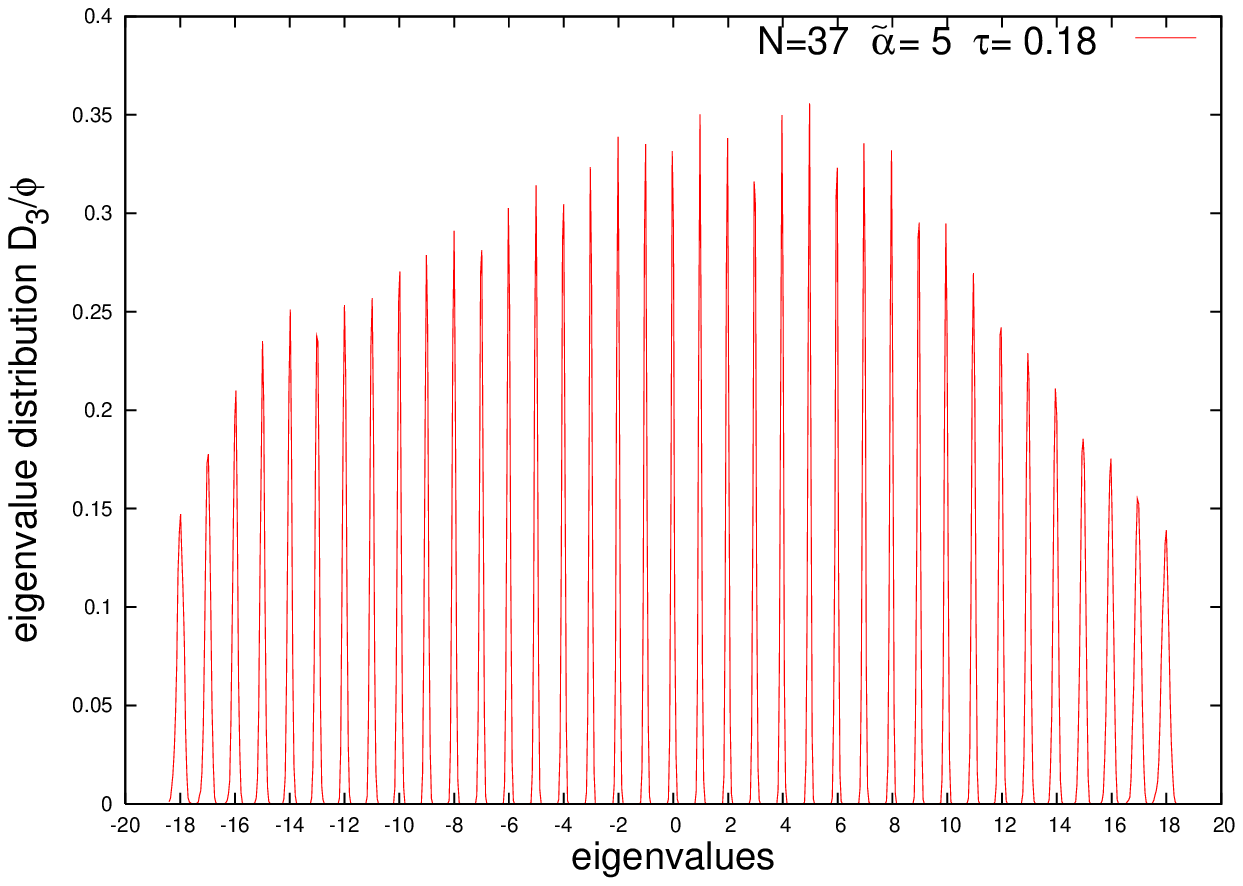}
\includegraphics[width=7.0cm,angle=0]{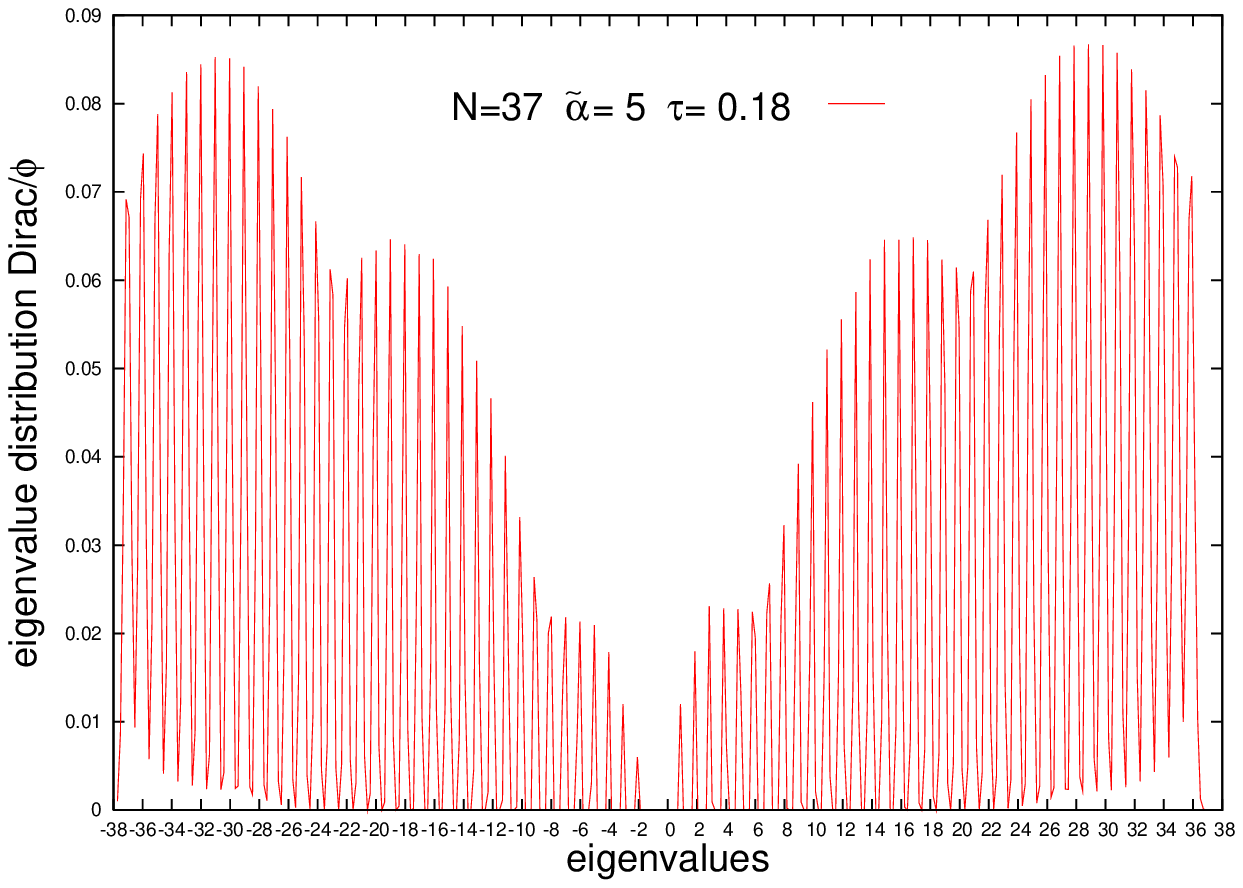}
\caption{Eigenvalues for matrix $D_a$ and the operator $\mathcal{D}=\sigma_a[D_a,\cdot]$ deep inside the fuzzy sphere region $\tilde{\alpha}=5$ and $\tau=0.18$. The eigenvalues are divided by the measured value of $\phi=0.75$ in order to show that $D_a/\phi$ are fluctuating around $L_a$, irrep. $SU(2)$ for $N=37$.} 
\end{center}
\end{figure}

Alternatively, we can extract $\phi$ from the eigenvalue distribution of matrix $C=\sigma_aD_a$. 
In the fuzzy sphere we have $D_a=\phi L_a$. We know that the eigenvalues of $C$ are given by
\begin{equation}\label{espC}
C_{\pm}=\alpha\phi\left(\pm\frac{N}{2}-\frac{1}{2}\right).
\end{equation}
Therefore by choosing for instance $C_+$ the value for $\phi$ can be extracted from the numerical results for the corresponding set of parameters. We obtain similar result by choosing $C_-$. See figure 17.

\begin{figure}[ht]
\begin{center}\label{fig:FS-mu-c}
\includegraphics[width=7.0cm,angle=0]{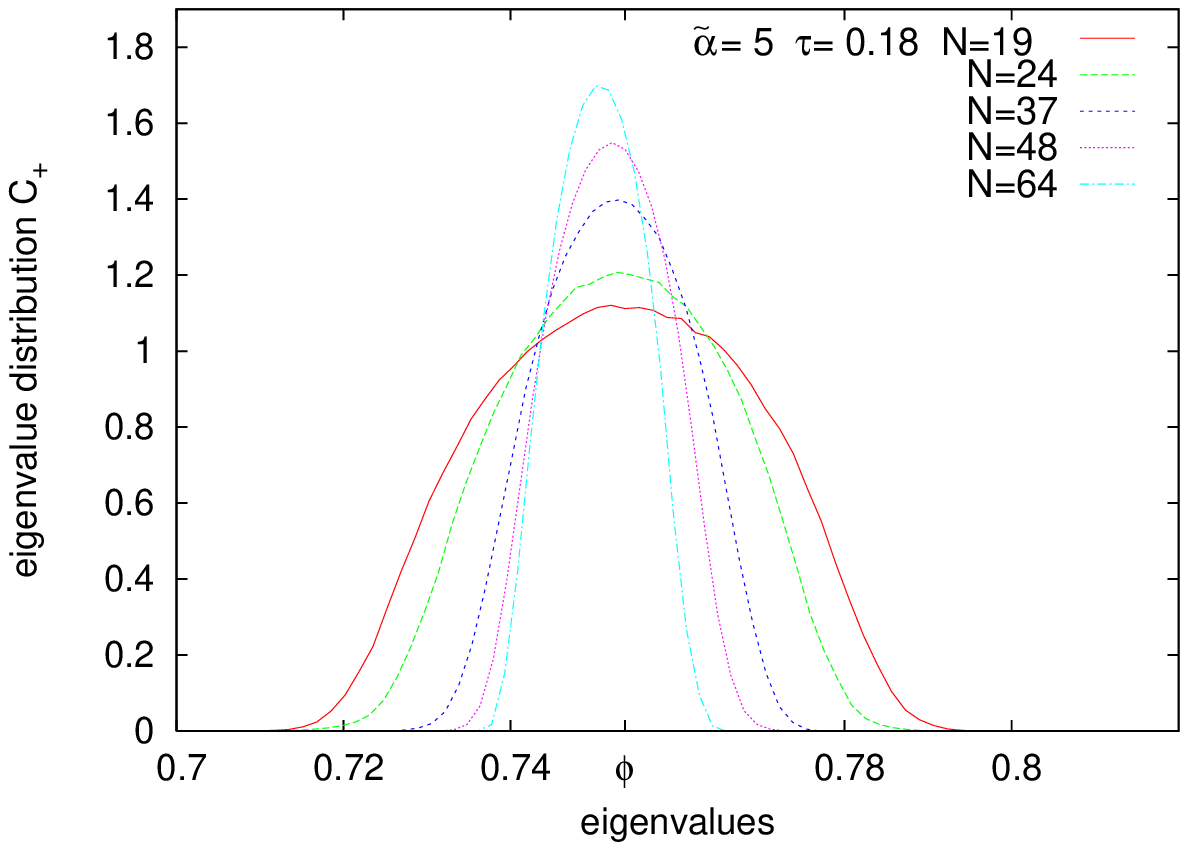}
\includegraphics[width=7.0cm,angle=0]{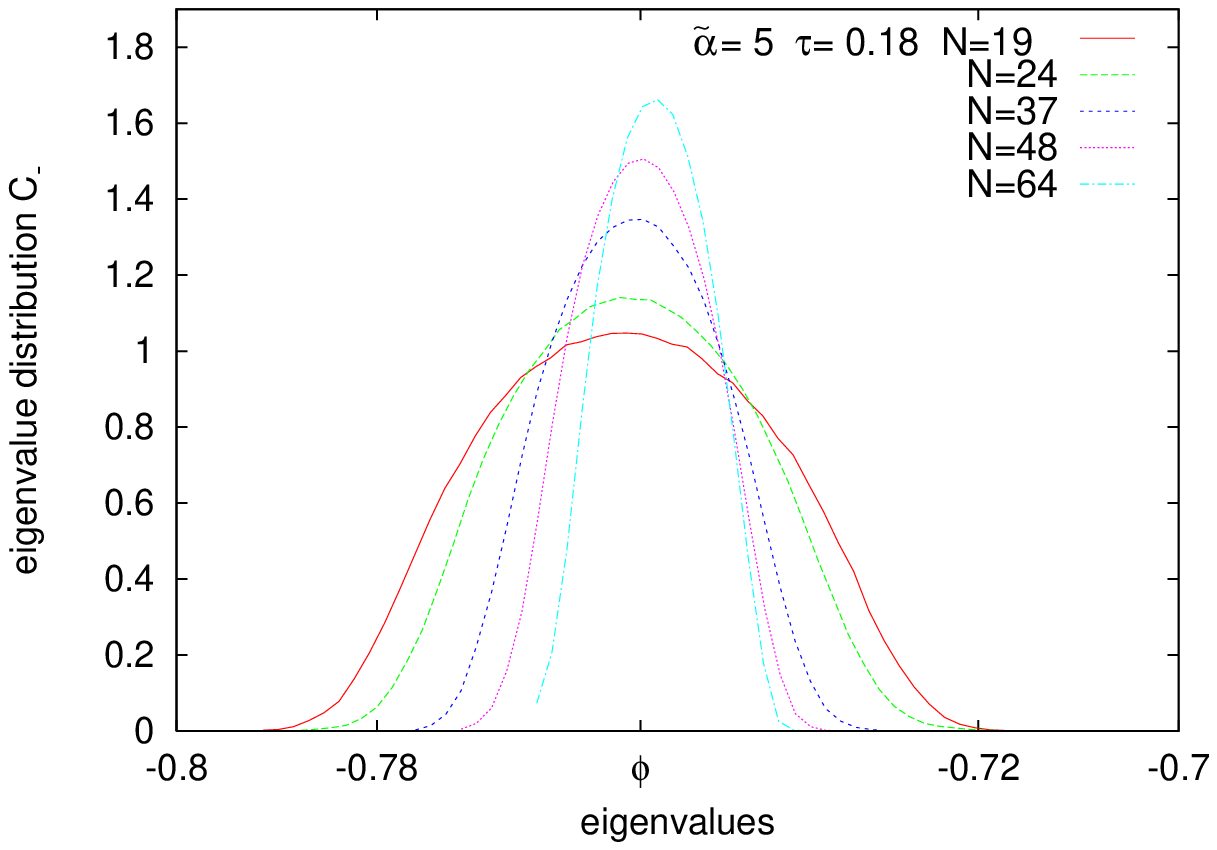}
\caption{Here we use the eigenvalue distribution of the matrix $C=\sigma_aD_a$ in order to extract the value of $\phi$ since $\frac{2\sqrt{N}C_+}{\tilde\alpha(N-1)}=\phi$. The parameters are given by $\tilde{\alpha}=5$ and $\tau=0.18$, for different values of $N$. We can see that the numerical value for $\phi$ is in very good agreement with the theoretical prediction $\phi=0.7537$ given by eq.(9). On the left the corresponding plot for $C_-$, where $\phi$ is given by $\phi=-\frac{2\sqrt{N}C_-}{\tilde\alpha(N+1)}$.} 
\end{center}
\end{figure}

\begin{figure}[ht]
\begin{center}\label{fig:MatrixPhase-X-Comm-mu-0.15}
\includegraphics[width=7.0cm,angle=0]{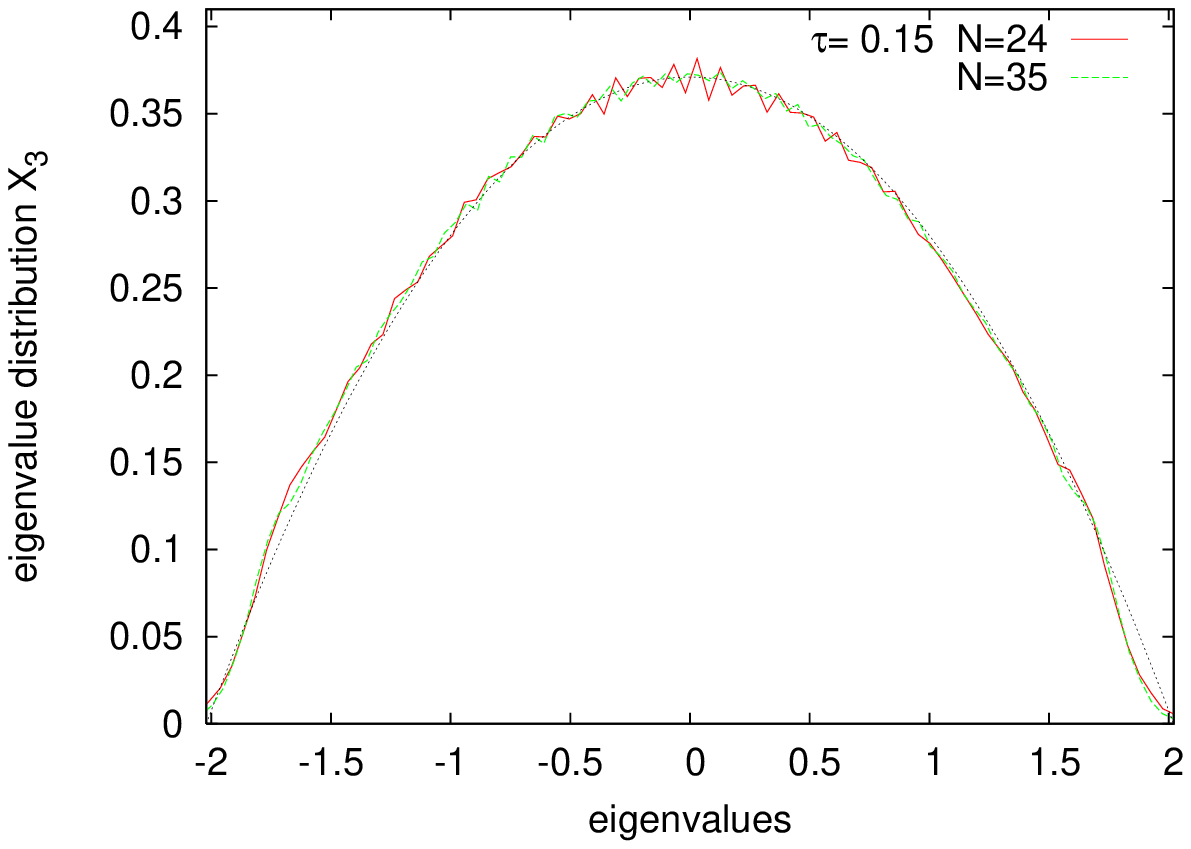}
\includegraphics[width=7.0cm,angle=0]{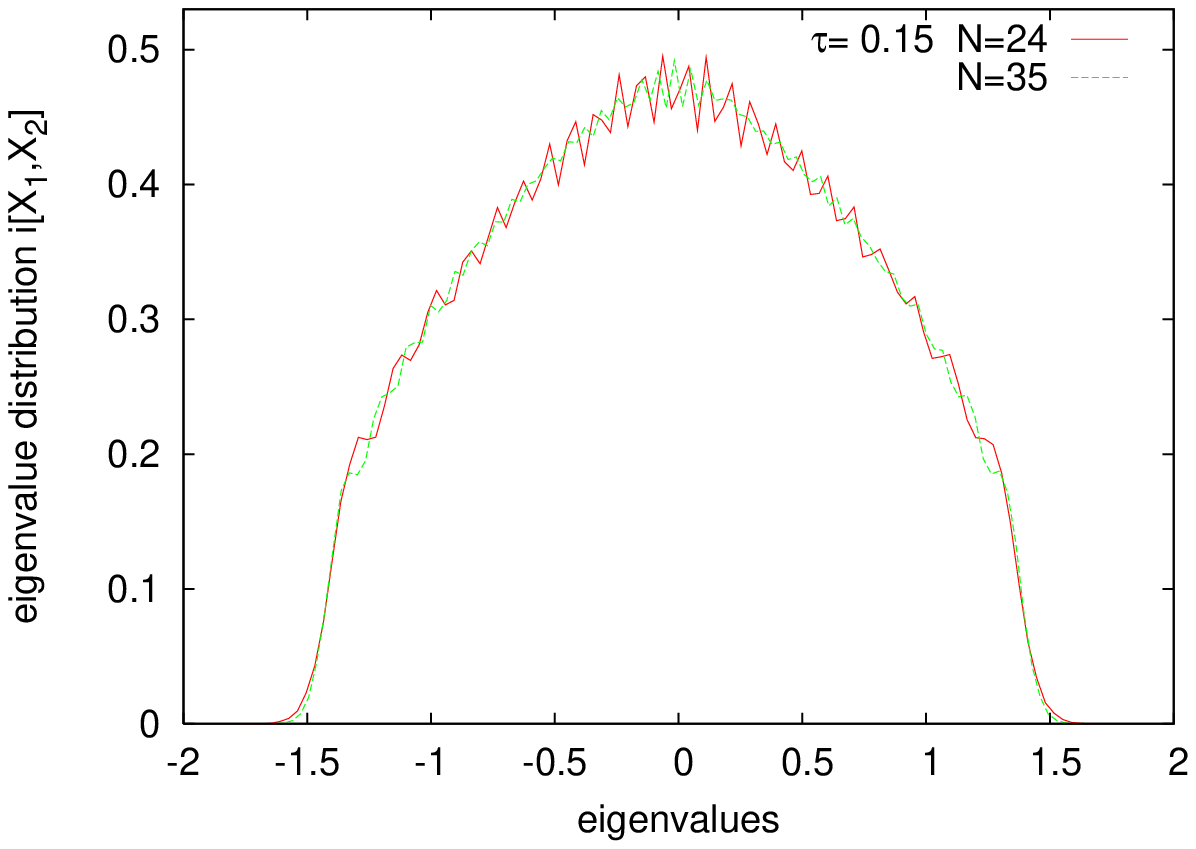}
\caption{Eigenvalues for $X_3$ and $i[X_1,X_2]$ for different $N$, $\tau=0.15$ in the matrix phase $\tilde{\alpha}=1$. Eigenvalues distribute uniformly inside a 3-solid ball.}
\end{center}
\end{figure}

\begin{figure}[ht]
\begin{center}\label{fig:MatrixPhase-C-Dirac-mu-0.15}
\includegraphics[width=7.0cm,angle=0]{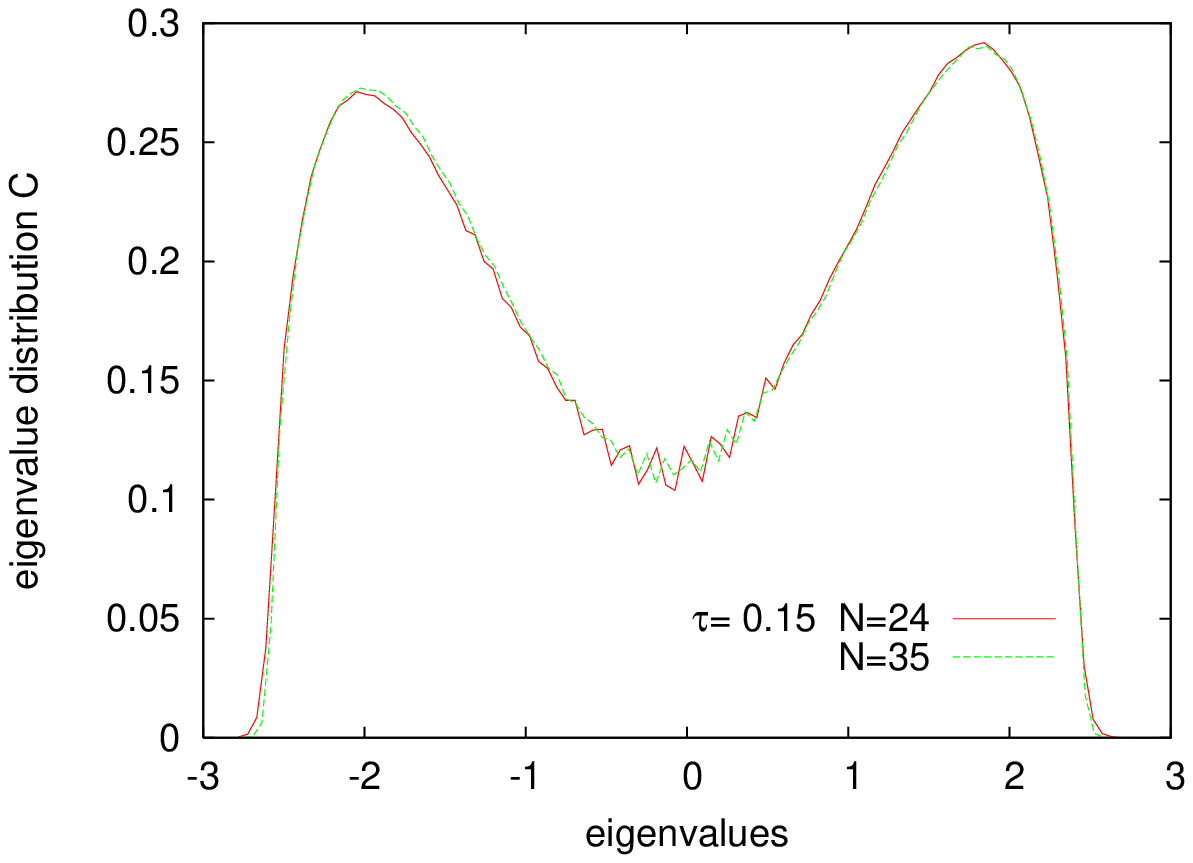}
\includegraphics[width=7.0cm,angle=0]{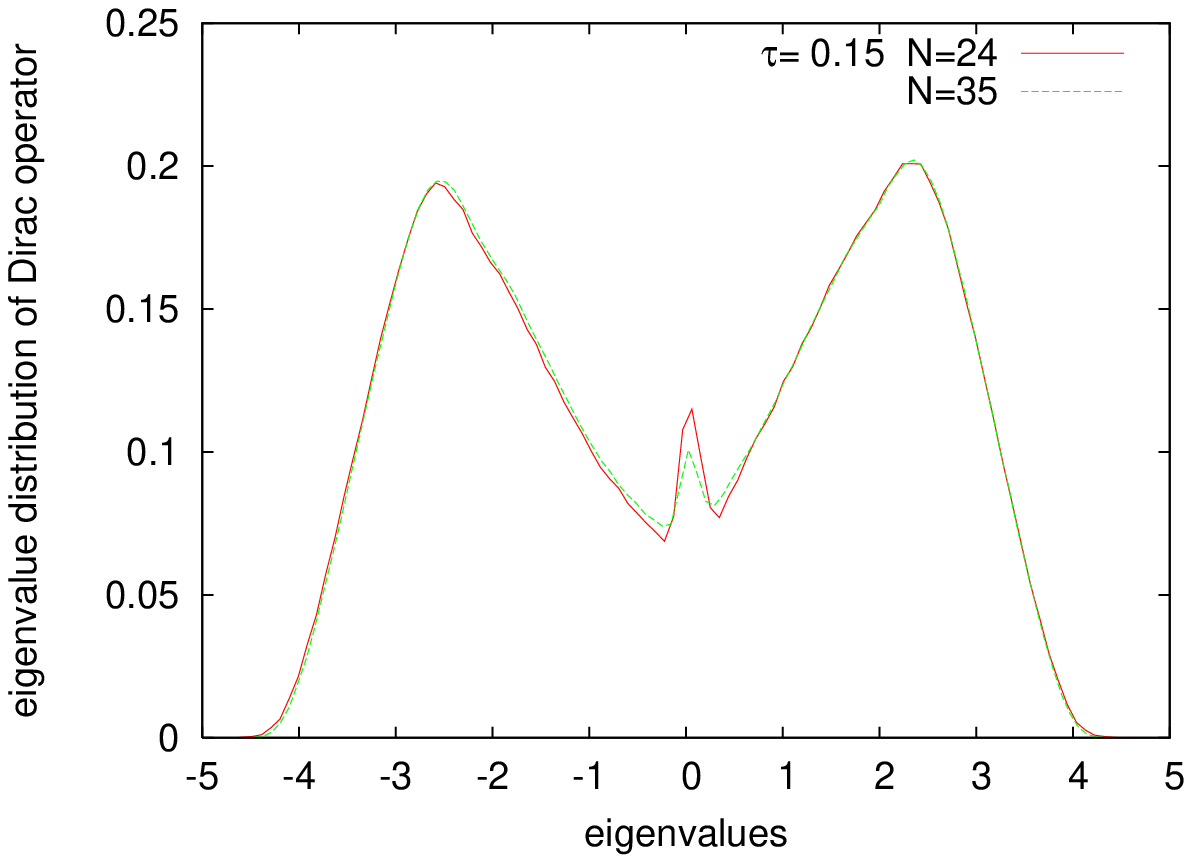}
\caption{(a) Eigenvalues for matrix $\mathbb{C}=\sigma_aX_a$. (b) $\mathbb{D}=\sigma_a[X_a,]$ in the matrix phase for $\tau=0.15$, $\tilde{\alpha}=1$} 
\end{center}
\end{figure}

\begin{figure}[ht]
\begin{center}\label{fig:effect-mu-MP}
\includegraphics[width=7.0cm,angle=0]{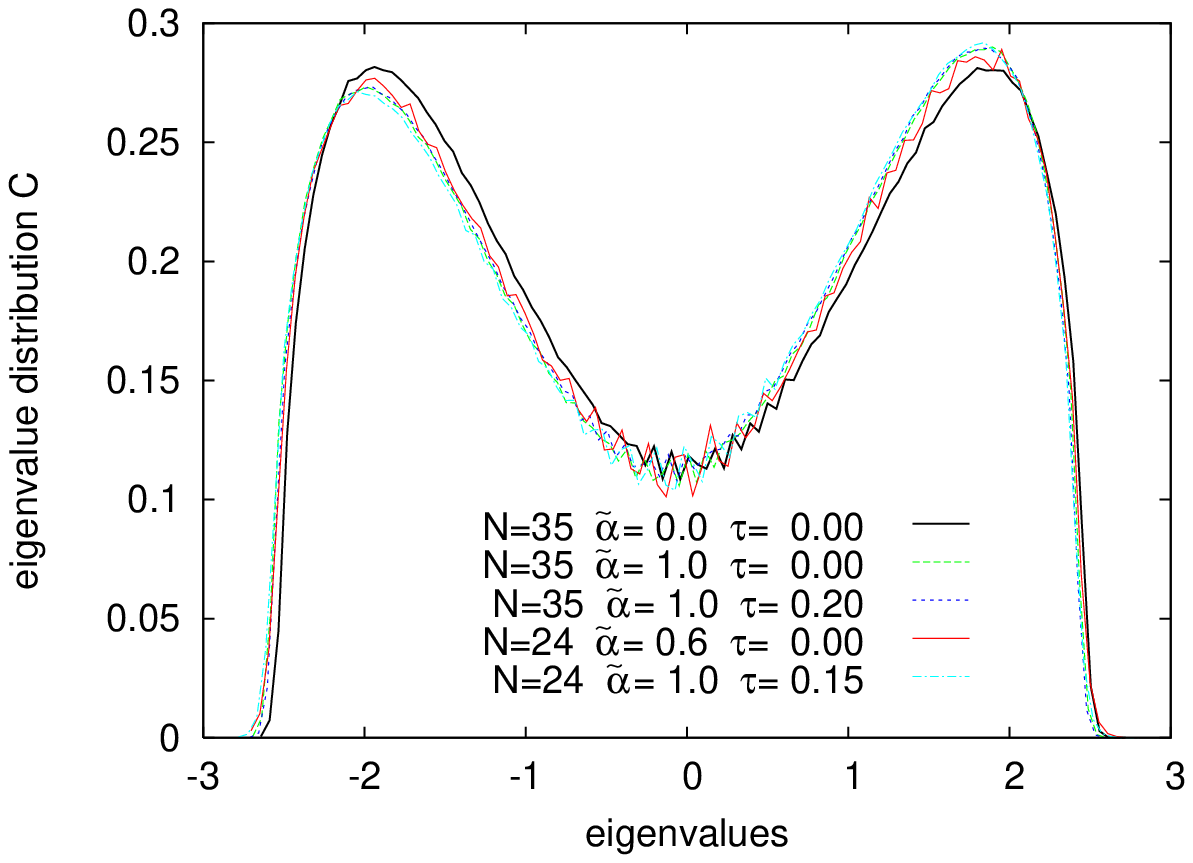}
\includegraphics[width=7.0cm,angle=0]{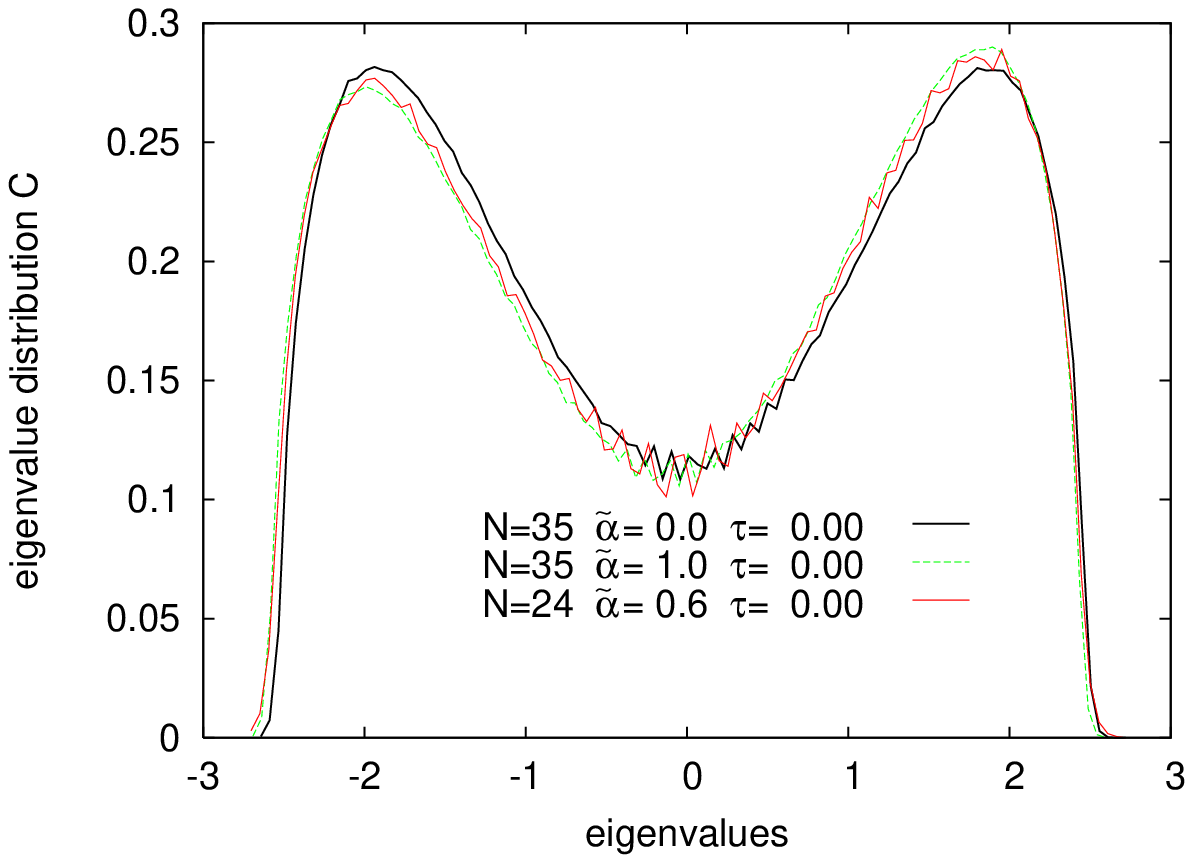}
\includegraphics[width=7.0cm,angle=0]{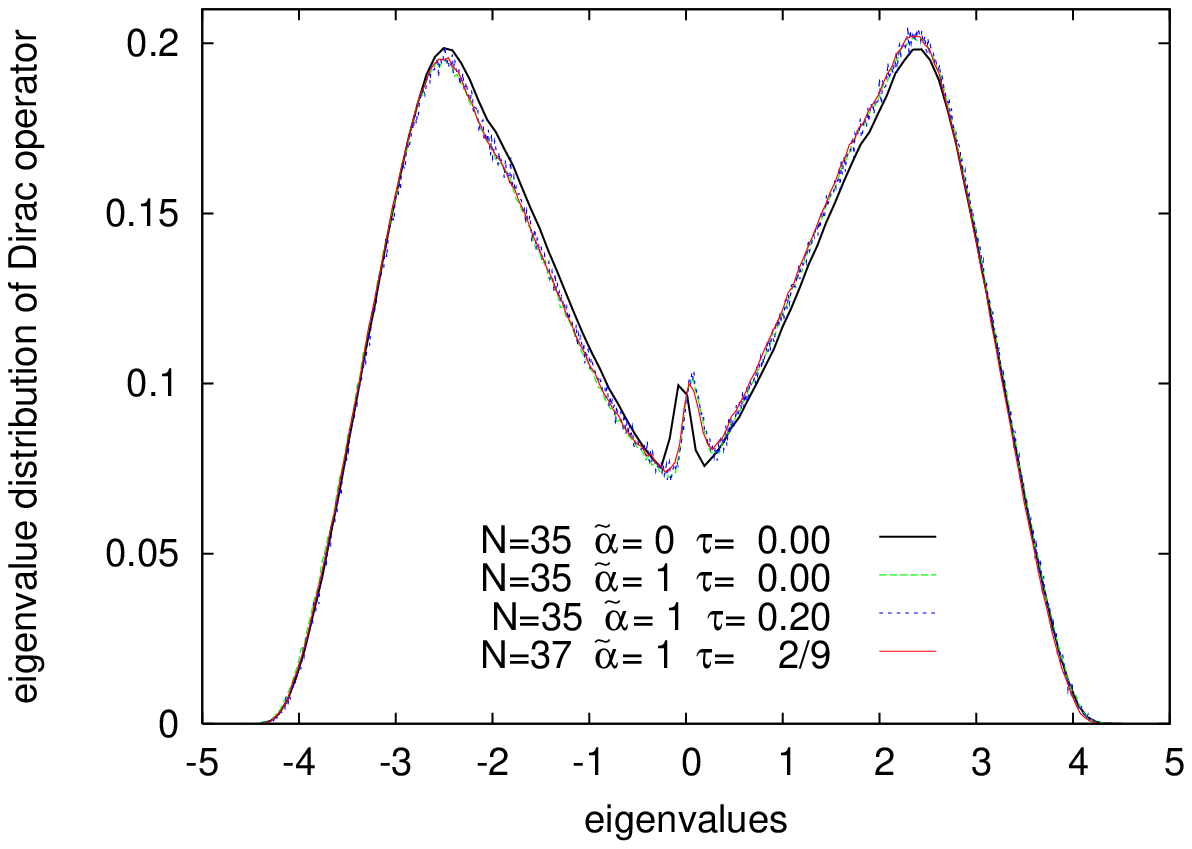}
\caption{Effect of Myers term on the eigenvalues for matrix $\mathbb{C}=\sigma_aX_a$ (left) and Dirac operator $\mathbb{D}=\sigma_a[D_a,]$ (bottom) in the matrix phase. We can see that the Myers term has a small effect of the distributions, they are no longer symmetric as compared with the pure Yang-Mills (in black thick line).} 
\end{center}
\end{figure}

\begin{figure}[ht]
\begin{center}\label{evo-FS-mu-d}
\includegraphics[width=7.0cm,angle=0]{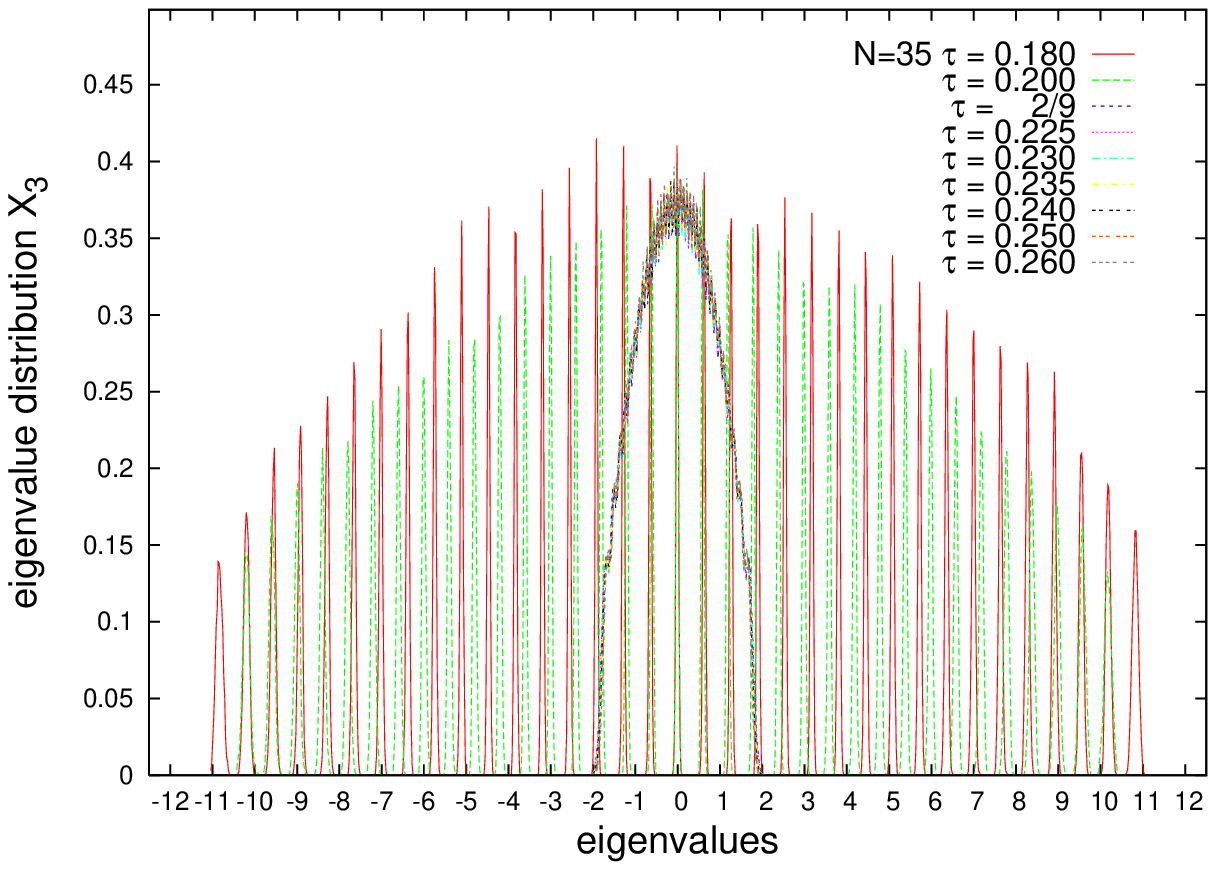}
\includegraphics[width=7.0cm,angle=0]{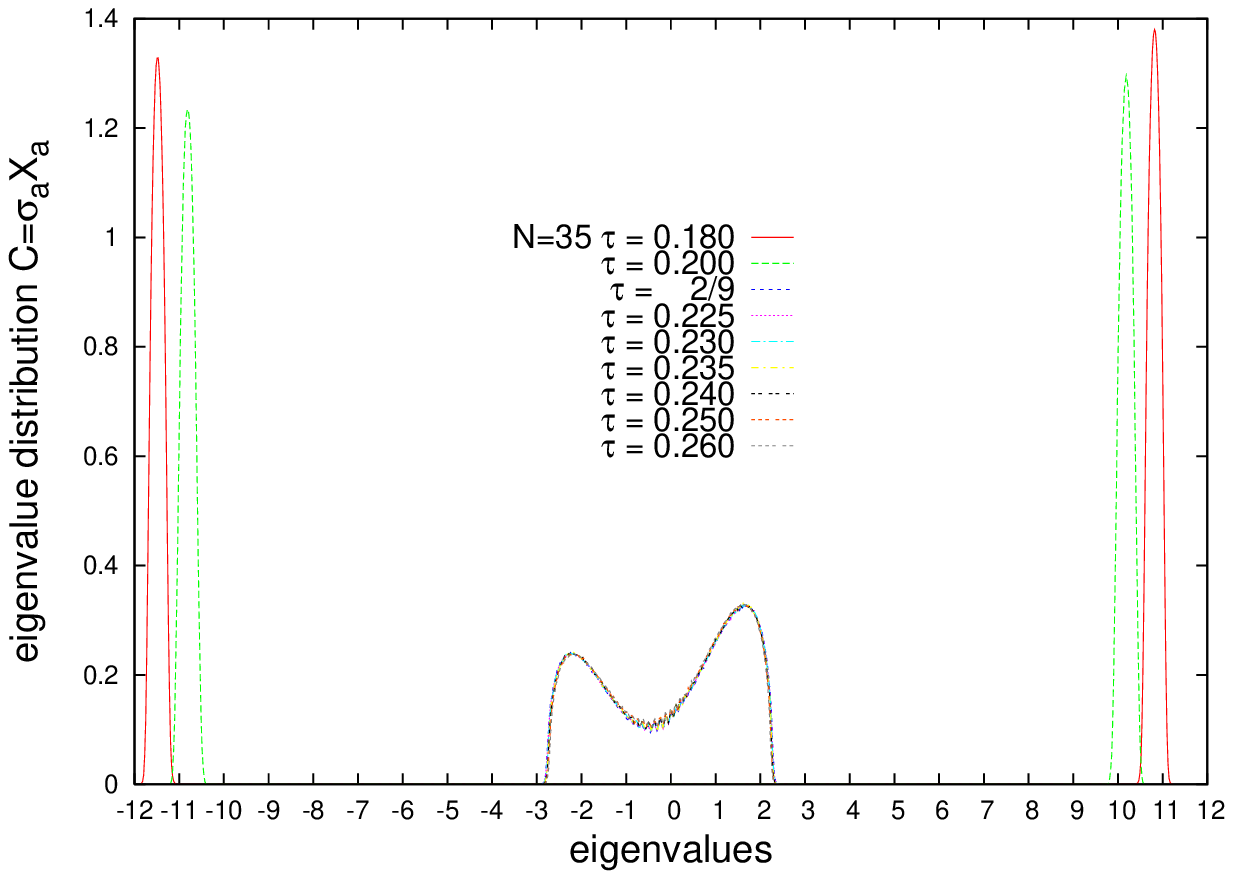}
\caption{Evolution for eigenvalues for matrix $X_3$ for fixed $\tilde{\alpha}=5$ for different values of $\tau$. At $\tau=2/9$ the fuzzy sphere disappears. On the right eigenvalues for the matrix $\mathbb{C}=\sigma_aX_a$.} 
\end{center}
\end{figure}

\begin{figure}[ht]
\begin{center}\label{fig:FS-C-tau4.02}
\includegraphics[width=7.0cm,angle=0]{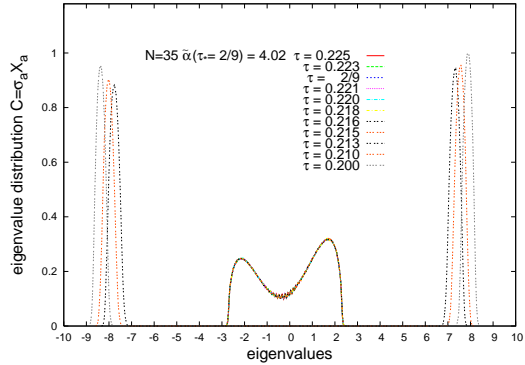}
\caption{The spectrum of ${\mathbb{C}}=\sigma_aX_a$
for varying $\tau$ at $\tilde{\alpha}(\tau=2/9)=4.02$. We see that the
transition occurs at $\tau=0.213\pm0.002$.}
\end{center}
\end{figure}

\begin{figure}[ht]
\begin{center}\label{fig:Phaseup2-9}
\includegraphics[width=7.0cm,angle=0]{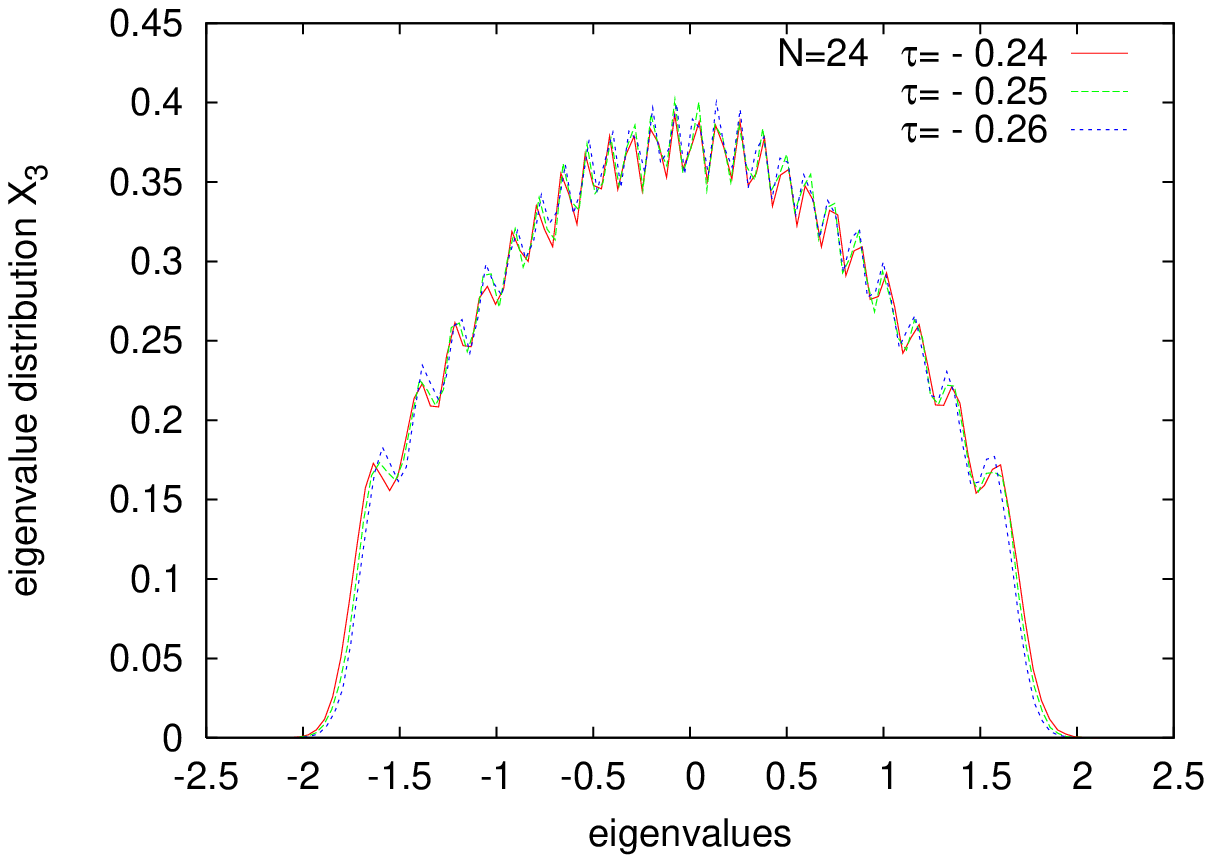}
\includegraphics[width=7.0cm,angle=0]{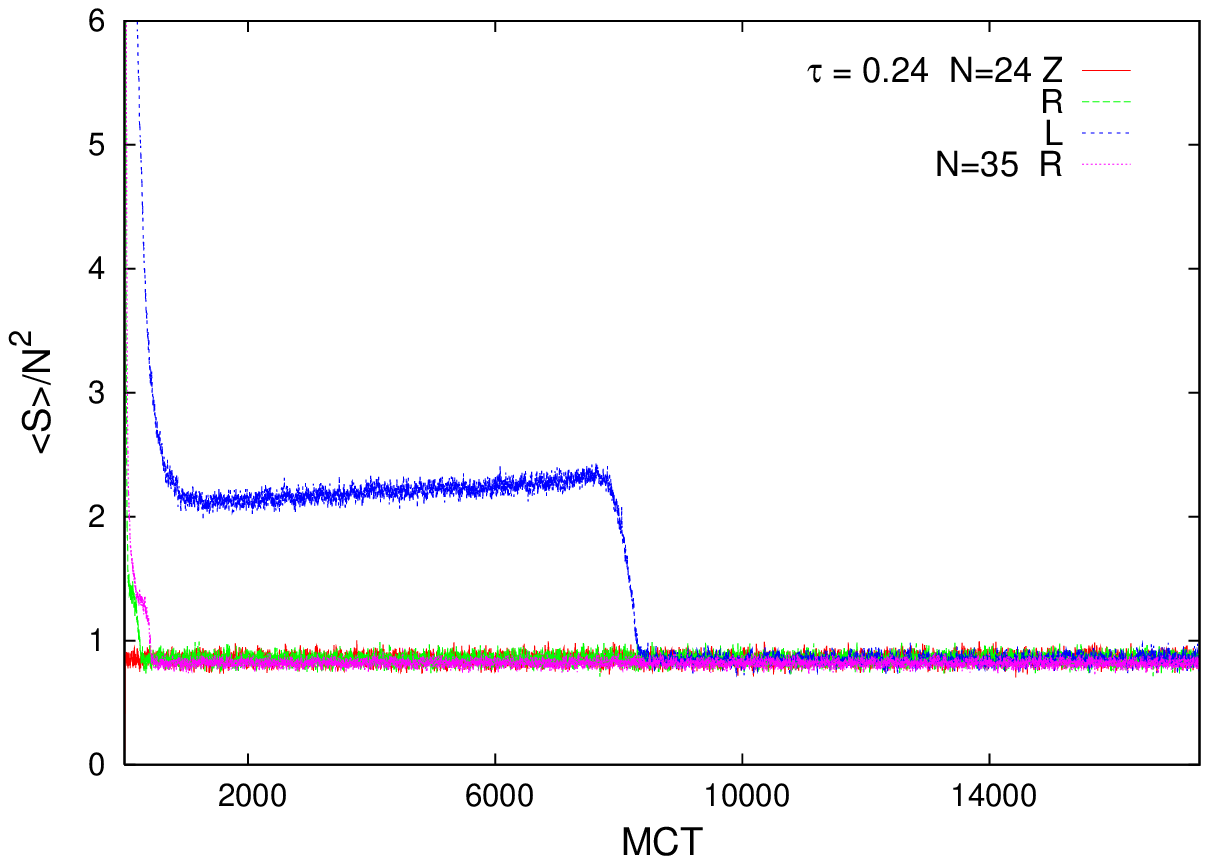}
\includegraphics[width=7.0cm,angle=0]{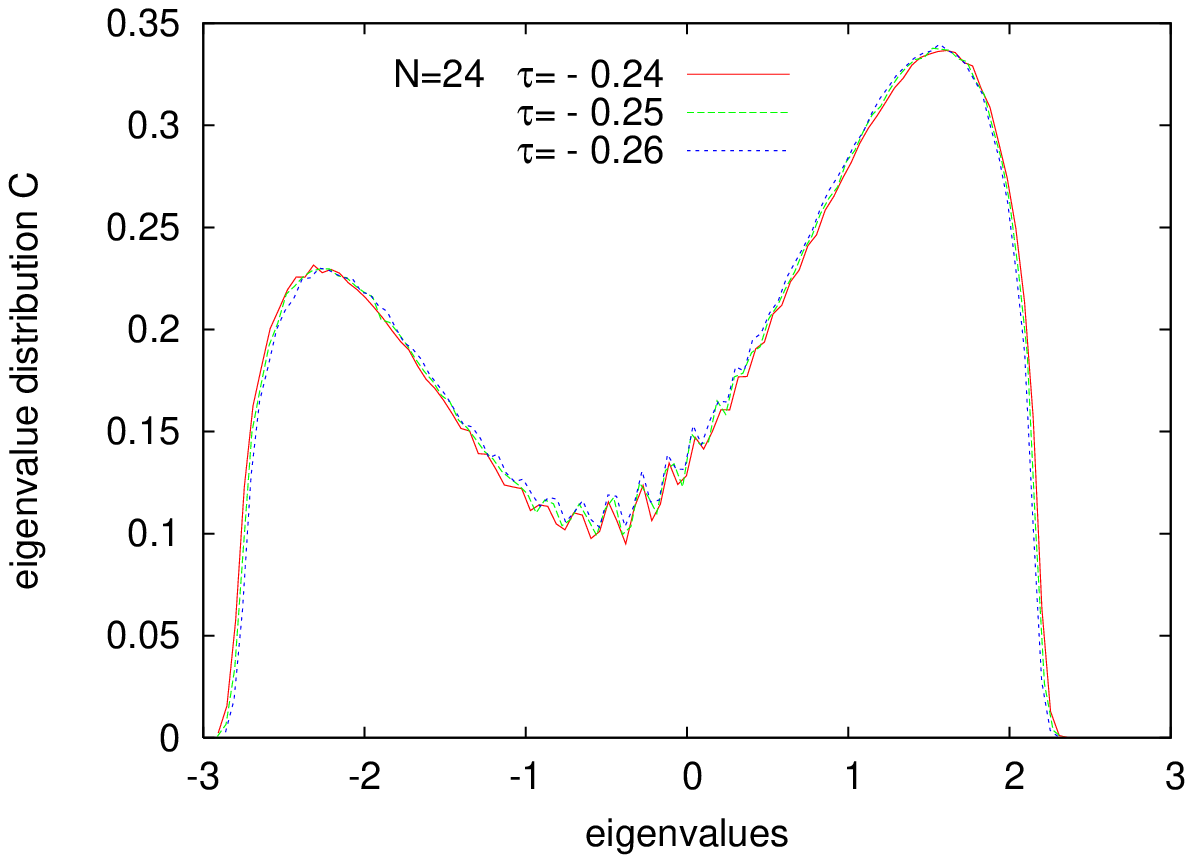}
\includegraphics[width=7.0cm,angle=0]{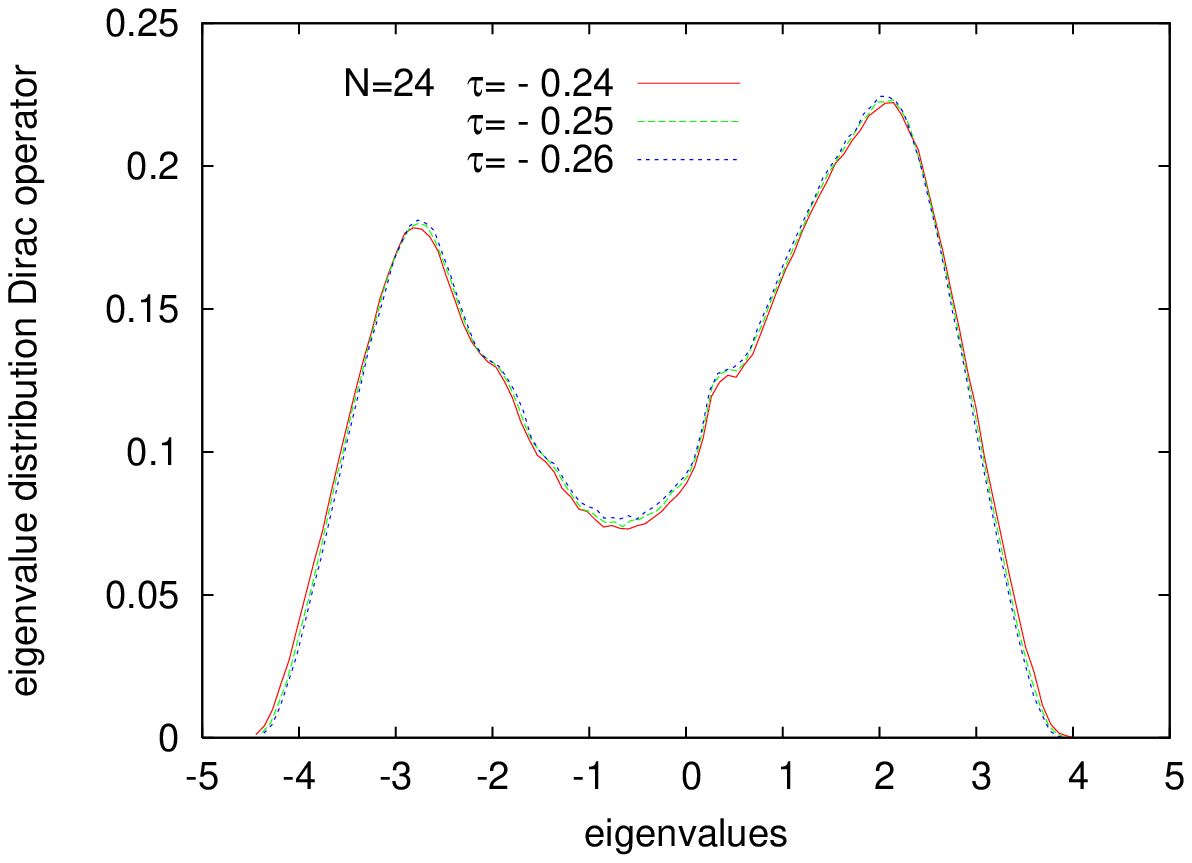}
\caption{The eigenvalue distributions for $X_3$, 
$\mathbb{C}$ and ${\mathbb{D}}$ for 
$\tilde{\alpha}=5.0$ and $\tau>2/9$. We also show the thermalisation 
history for $N=24$ and  $\tilde{\alpha}=5.0$ with $\tau=0.24$,  
for different initial conditions, $D_a=0$ (Z), random start (R) 
and fuzzy sphere start $D_a=L_a$ (L). }
\end{center}
\end{figure}

\begin{figure}[ht]
\begin{center}\label{fig:effect-Myers-MP}
\includegraphics[width=7.0cm,angle=0]{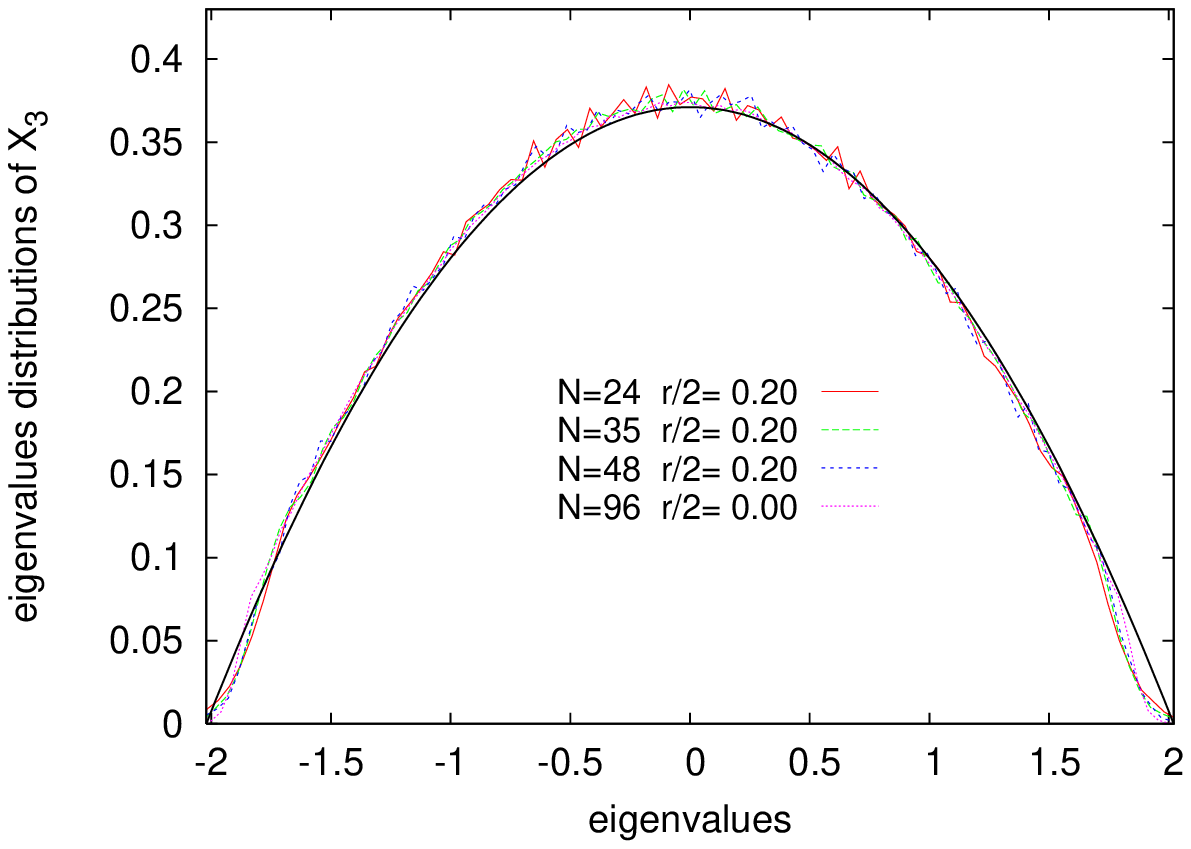}
\includegraphics[width=7.0cm,angle=0]{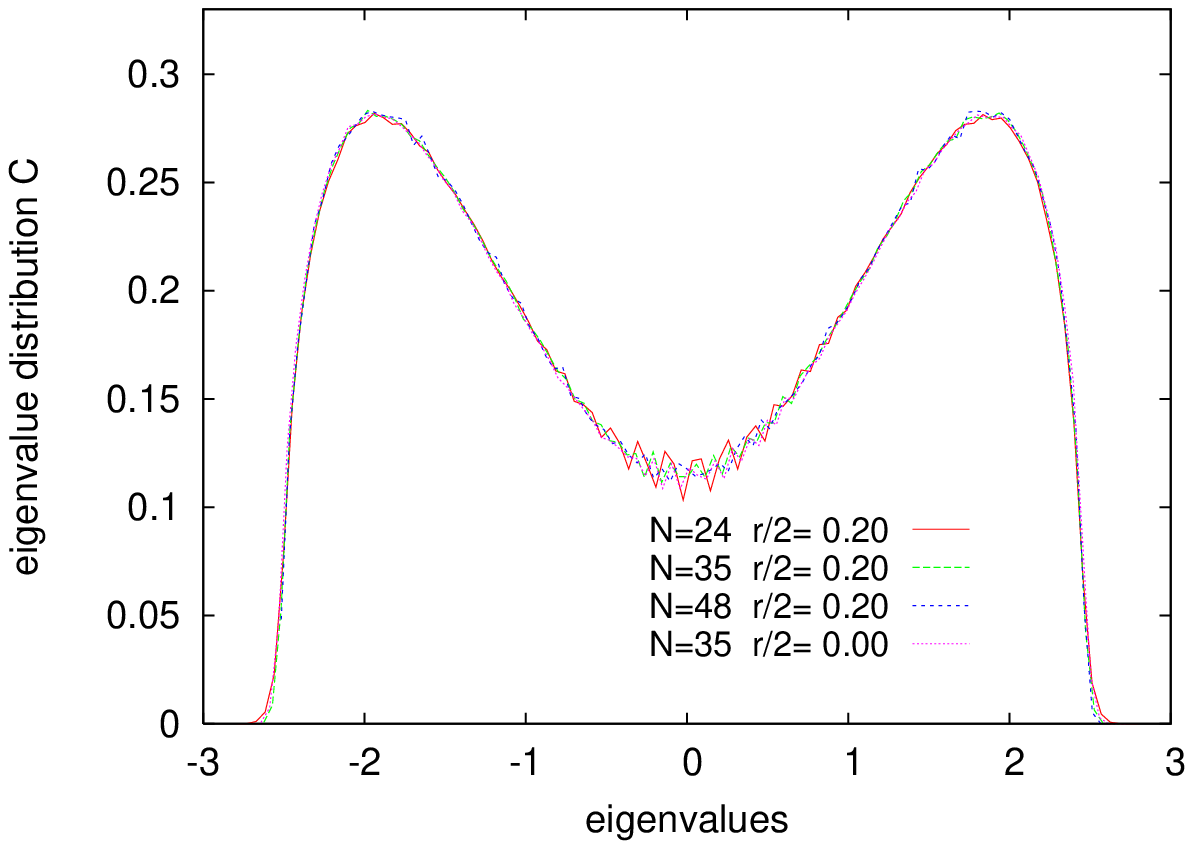}
\caption{Effect of the mass term $\frac{r}{2} Tr X^2_a$ in the Yang-Mills action. Fixing $\frac{r}{2}=0.20$ for different values of $N$ we plot the eigenvalues for matrix $X_a$ (left) and $\mathbb{C}=\sigma_aX_a$ (right). We compare with the $r=0$ ($\tau=0$) case. 
The eigenvalue distribution for the configurations $X_a$ is fit by the parabola (\ref{parabola}) as in the pure Yang-Mills model (thick line). We conclude that the mass term has no effect in the Yang-Mills matrix action.} 
\end{center}
\end{figure}

\paragraph{Matrix phase.}
In this phase as we have mentioned before, the dependence on $N$ drops
out if we concentrate on $X_a=\alpha
D_a=\frac{\tilde{\alpha}}{\sqrt{N}}D_a$ as our basic operator.  We
observe that the distribution for eigenvalues coincides, to our
numerical accuracy, with that of the pure Yang-Mills model and there
is no observable dependence on the parameters $\alpha$, $\tau$ and
matrix size $N$. The data is well fit by the parabolic distribution
(\ref{parabola}) with $R=2.0$. We also observe that the eigenvalue
distribution of the commutator shows no observable deviation from that
of the pure Yang-Mills model. Figure 18 shows the case for
$\tau=0.15$, $\tilde{\alpha}=1$ for different matrix size.

In our simulations we observe that the spectrum of the matrix
$\mathbb{C}=\sigma_aX_a$ and the operator ${\mathbb{D}}$ are not
symmetric, but rather distorted with respect to the pure Yang-Mills
case; see figure 19(a) for $\tau=0.15$ and fixed value
$\tilde{\alpha}=1$ and figure 20 (a) and (b) for ${\mathbb{C}}$ and
figure 20 (c) for ${\mathbb{D}}$. As we see from the graphs the
asymmetry is due to the Myers term. We attribute this asymmetry to a
residual dependence on the Myers term which drops out as
$\frac{1}{\sqrt{N}}$, noting that graphs correspond to $\alpha=0, 0.122$ and 
$\alpha=0.169$ for $\tilde\alpha=0, 0.6$ and $\tilde\alpha=1.0$ respectively.

The operator $\mathbb{D}$ has a continuous spectrum and turning on the
Meyrs term again gives an observable asymmetry to its spectrum, see
figure 19(b) for $\tilde{\alpha}=1.0$ and fixed value $\tau=0.15$ and
figure 20 (c) for a range of parameter values.  It is only for
$\tau\geq \frac{2}{9}$ that the asymmetry survives the large $N$ limit
since for all $\tau < \frac{2}{9}$ increasing $N$ for fixed $\alpha$ drives
across the transition line and into the fuzzy sphere phase.

From the distributions we can see that the transitions that takes place in the model is the one which goes from a non-commutative fuzzy sphere as geometrical background to a solid 3-ball with fixed radius $R=2$. 

\begin{eqnarray}
X_a^2=\frac{N^2-1}{4} \alpha^2\phi^2 \;\;\;\longrightarrow \;\;\; X_a^2\le 4
\end{eqnarray}
Or if we rescale to the $D_a$ we get a collapse of a sphere to a point.

This transition occurs when $\tilde{\alpha}$ reaches the critical
value $\tilde{\alpha}_*$ following the critical in the phase diagram
depicted in figure 3. The transition line is defined by the critical
curve (\ref{critline}) whenever $\tau$ is in the range $0\le\tau<2/9$.
The model is unstable for $\tau<0$. However, when $\tau<0$ and
$\tilde{\alpha}>\tilde{\alpha}_*(\tau)$, the fuzzy sphere is a local
minimum of the model with negative action which goes to $-\infty$ as
$N\rightarrow\infty$ and tunneling out of this well becomes
impossible.  

In figure 21 we track the evolution of the eigenvalue distribution of
$X_a$ and matrix ${\mathbb{C}}$ as a function of $\tau$ along the line
$\tilde\alpha=5.0$ starting in fuzzy sphere. We can see the fuzzy
sphere disappears as we cross the critical value $\tau=2/9$. We
observe that $\tau=\frac{2}{9}$ actually falls in the matrix phase.
We find that the transition tracks the line $\tau=\frac{2}{9}$ for
different $\tilde\alpha$. However, as the end point of this line is
approached, i.e.  $\tilde\alpha=12\left(\frac{4}{107 +
  51\sqrt{17}}\right)^{1/4}\sim 4.02$ we observe that in a small
neighbourhood of this value the transition actually occurs at
$\tau_*<\frac{2}{9}$ see figure 3. In figure 22 we see that there
appears to be some deviation from the predictions of (\ref{phisol})
where one would expect that for $\tilde{\alpha}=4.02$ the transition
would occur at $\tau=\frac{2}{9}\sim0.222$, however we find the
transition at $\tau=0.213\pm0.002$. For this transition $<S>$ shows a
clear discontinuity consistent with the predictions from
(\ref{phisol}) and we expect there is still a divergent specific heat
though this is more difficult to detect.

Figure 23 shows the history of the expectation value of the
action $<S>/N^2$ for $N=24,35$, $\tilde{\alpha}$ and $\tau=0.24$ from
different initial configurations; cold ($D_a=L_a$), zero and random
start. When starting from cold configuration we observe the system
goes through metastable states. All reach the same energy level and no
decay was observed after $2^{19}$ Monte Carlo steps or sweeps, defined
as the step in which all the matrix elements are updated according to
the Metropolis algorithm.

\paragraph{{\bf Massive Yang-Mills model}}
Since, in the model (\ref{actionXalphatilde}) the limit $\alpha=0$ 
removes the quadratic and Meyrs term we treat this the massive deformation 
of this case separately.

For $\alpha=0$ we therefore also study the model 
\begin{eqnarray}
S[X]=N\;Tr \left(-\frac{1}{4}[X_b,X_b]^2+\frac{r}{2} X^2_a\right).
\end{eqnarray}
We study the eigenvalue distributions for matrix $X_a$ and matrix 
$\mathbb{C}=\sigma_aX_a$ for different values of the parameter $r$ and 
matrix size $N$. We find that the phenomenology 
for this model is similar to that of $\tau=0$ described above.
The eigenvalue distribution for the matrix 
configurations $X_a$ and small $r$ are fit by the parabola (\ref{parabola}).

We also observe that the eigenvalues for the matrix $\mathbb{C}$
distribute as in the pure Yang-Mills model. See figure 24.  
For values $\frac{r}{2}<0$ the matrix action is not bounded from below.

\paragraph{{\bf Comment on the Massive Myers model with complex Meyrs coupling.}} For completeness let us consider the following action
\begin{equation}\label{CSmm}
S=N\Tr \left(\frac{i\alpha}{3}\epsilon_{abc}X_a[X_b,X_b]+\frac{r}{2}X^2_a\right).
\end{equation} 
The model is exactly solvable (see Hoppe \cite{hoppe}) and is equivalent 
to a 2-matrix Yang-Mills model with massive deformation.
For real $\alpha$ the model is not stable with no ground state. It may however
be possible to localise configurations in a well by suppressing tunneling 
in the large $N$ limit. We observe that such wells exist for large positive 
$r$.

\section{Conclusions}

We have performed numerical simulations of a simple two parameter
3-matrix model with energy functional given by (\ref{mu-action}). In
particular we studied the phase diagram as a function of the two
parameters $\tau$ and $\tilde\alpha$. Earlier studies \cite{longpaper}
looked at the case $\tau=0$ and found that it exhibited an exotic
phase transition at $\tilde\alpha_*={(\frac{8}{3})}^{3/4}$. It was
argued, based on an effective potential calculation that the model
should have a line of phase transitions dividing the
$(\tau,\tilde\alpha)$ plane in two and predicting the coexistence
curve. We find that for $0<\tau<\frac{2}{9}$ the coexistence curve is
predicted well by the theoretical expressions.  The coexistence curve
asymptotes to the line $\tau=\frac{2}{9}$ and appears to deviate
slightly from the prediction of the effective potential
(\ref{phisol}). This special value of $\tau$ corresponds to the value
where the action is a complete square, see eq. (\ref{F2}). We examine
the eigenvalue distributions of different operators. In particular we
look at $\mathbb{C}=\sigma_aX_a$ which detects the $SU(2)$
representation content of the configurations $X_a$. We find the IRR of
dimension $N$ has fluctuations around a lower $S$ than any other
configuration in the parameter range corresponding to the fuzzy sphere
phase. Also, in this range the spectrum of ${\mathbb C}$ has the two
peaks corresponding to $\alpha \phi\sigma_a L_a$ with $L_a$ the
$SU(2)$ IRR of dimension $N$. In the parameter range corresponding to
the matrix phase ${\mathbb C}$ has a continuum spectrum.  We also
study the Dirac operator $\mathbb{D}=\sigma_a[X_a,\cdot]$ and find
that in the fuzzy sphere phase its spectrum is (as expected) a
shifted, cutoff version of the commutative sphere.

We found evidence for two distinct transition types in the emergent
geometry scenario. For $0<\tau<\frac{2}{9}$ we found that, as the
transition is approached from the fuzzy sphere phase with fixed
$\tau$, the model has a divergent specific heat with critical exponent
$\alpha=\frac{1}{2}$. However for $\tilde\alpha >\tilde\alpha_*
=12\left(\frac{4}{107+51\sqrt{17}}\right)^{1/4}\sim4.02 $, and
crossing the transition at fixed $\tilde\alpha$, there appear to be no
critical fluctuations; the transition is one with a continuous, but
non-differentiable internal energy, and a discontinuous specific
heat. In all cases the transition is from a fuzzy sphere to a matrix
phase. We study the matrix phase in detail and find that a useful
description of this phase is in terms of fluctuations about a
background of commuting matrices whose eigenvalues are concentrated
within a sphere of radius $R=2.0$. This is consistent with the
estimates of \cite{FilevOConnor} who performed a 2-loop analysis and
estimated $R\sim1.8$

We find that, though it is possible for configurations other than the
irreducible fuzzy sphere to be present in the model, they never
correspond to the true ground state of the system. Such
configurations, were they present, would be easily detected by the
matrix ${\mathbb{C}}$. Furthermore for $\tau=\frac{2}{9}$ we find that
$<S>$ when trapped in a fuzzy sphere configuration is larger 
than that for fluctuations in the matrix phase, and the fuzzy sphere 
is not a true ground state of the system.  Also, for
$\tau>\frac{2}{9}$ we observe decays from the fuzzy sphere to the
matrix phase, i.e. the fuzzy sphere is a meta-stable configuration for
the system, with observable decay. In contrast for the critical line
$\tau=\frac{2}{9}$ we observe no decay of the fuzzy sphere to the true
ground state. We infer that the barrier is sufficiently high in this
case that the limit of $N\rightarrow\infty$ prevents tunneling out of
the local minimum corresponding to the fuzzy sphere.

A natural further stage in the study undertaken here is to include the
effect of Fermions. However, the most interesting case involves either
complex actions or fluctuating signs in the Fermionic sector
\cite{sign} both of which lead to significant numerical difficulties.

In $d=10$ the pure Yang-Mills model has received significant attention
in a scenario of emergent gravity \cite{steinemgravity}.  It appears
from our study, that for finite $N$, all such configurations will be
meta stable. In the large $N$ limit tunneling will be suppressed and
these states may become stable. unfortunately, reliable numerical
simulations in such a situation are more difficult.

\paragraph{{\bf Acknowledgements}}
R.D.B. was supported by a Marie-Curie Fellowship from the Commission of the European Communities. Currently he acknowledges support from Conacyt Mexico. D.O'C. thanks the Perimeter Institute for hospitality where part of this work was performed.

\end{document}